\documentclass[graybox]{svmult}
\usepackage{amssymb}
\usepackage{mathptmx}       
\usepackage{helvet}         
\usepackage{courier}        
\usepackage{type1cm}        
\usepackage{makeidx}        
\usepackage{graphicx}       
\usepackage{multicol}       
\usepackage[bottom]{footmisc}
\makeindex             

\begin{document}

\title*{Supernovae from massive stars}
\author{Marco Limongi}
\institute{Marco Limongi \at INAF - Osservatorio Astronomico di Roma, Via Frascati 33, 00078 Monteporzio Catone (Roma) - Italy, \email{marco.limongi@oa-roma.inaf.it}}
\maketitle

\abstract{
Massive stars, by which we mean those stars exploding as core collapse supernovae, play a pivotal role in the evolution of the Universe. Therefore, the understanding of their evolution and explosion is fundamental in many branches of physics and astrophysics, among which, galaxy evolution, nucleosynthesis, supernovae, neutron stars and pulsars, black holes, neutrinos and gravitational waves. In this chapter, the author presents an overview of the presupernova evolution of stars in the range between 13 and 120 $\rm M_\odot$, with initial metallicities between [Fe/H]=-3 and [Fe/H]=0 and initial rotation velocities $\rm v=0,~150,~300~km/s$. Emphasis is placed upon those evolutionary properties that determine the final fate of the star with special attention to the interplay among mass loss, mixing and rotation. A general picture of the evolution and outcome of a generation of massive stars, as a function of the initial mass, metallicity and rotation velocity, is finally outlined.
}

\section{Introduction}
\label{sec:introduction}

Massive stars, i.e. those stars evolving through all the stable nuclear burning stages and eventually exploding as core collapse supernovae, play a fundamental role in the evolution of the Universe. They provide most of the mechanical energy input into the interstellar medium through strong stellar winds and supernova explosions and therefore induce star formation and mixing of the interstellar matter. Because they are hot and luminous, they generate most of the ultraviolet ionizing radiation, power the far-infrared luminosities of galaxies through the heating of the dust and contribute significantly to the integrated luminosity of the unresolved galaxies. During their hydrostatic evolution, as well as during the supernova explosion, they synthesize most of the elements, especially those necessary to life. The interiors of massive stars constitute invaluable laboratories where to study physical phenomena not seen elsewhere in the Universe, like, e.g., the neutrino burst occurring few seconds prior their explosion. Thus, the understanding of the evolution and the explosion of massive stars is fundamental for the interpretation of many astrophysical evidences and in general for the correct understanding of the evolution of the Universe.

In this chapter an overview of the evolution of stars in the mass range between 13 and 120 $\rm M_\odot$ from the main sequence phase up to the presupernova stage will be presented. It will be also outlined a general picture of the global properties and outcome of a generation of massive stars for various metallicities and initial rotation velocities. The specific results and conclusions presented in this chapter, however, must not be considered as conclusive but, on the contrary, one possible and plausible result of the stellar evolution. The reason is that they depend on many details of the computation of the stellar models, among which the most relevant are (1) the treatment of convection; (2) the adopted mass loss rate; (3) the treatment of both the rotation driven mixing and the angular momentum transport; (4) the adopted input physics (equation of state, opacities, nuclear cross sections, etc.). On the contrary, the general properties of the presupernova evolution of a massive star, addressed in the following sections, can be considered now well established and well accepted by the astronomical community.
There are, however, also many other excellent papers and reviews on this subject, that the author recommends to the reader for further insights \cite{2003ApJ...591..288H,2002RvMP...74.1015W,2012ARA&A..50..107L,2012IAUS..279....1N,2008EAS....32..233L,2008AIPC.1016...91L,1998ApJ...502..737C,1986nce..conf....1W,2000ApJS..129..625L,1996snai.book.....A,1988PhR...163...13N}.

\section{Massive stars: distinctive features}
\label{sec:features}

Stars are self-gravitating objects of hot plasma in hydrostatic equilibrium, losing energy from the surface, in the form of photons, and/or from the center, in the form of neutrinos. The equilibrium is guaranteed by the balance between the pressure gradient and the force of gravity - the pressure being provided by a combination of radiation, ideal gas and partially or totally degenerate electrons. Since the star loses energy, because it is hotter than the environment and/or because of the neutrino losses, it heats up and shrinks at the same time, according to the virial theorem, i.e., a fraction of the gravitational energy gained by the contraction goes into internal energy while the remaining fraction replaces the energy lost \cite{2012sse..book.....K}. When the central temperature is high enough, thermonuclear fusion reactions take place, nuclear energy supplies the energy lost, and gravitational contraction halts. As the nuclear fuel is exhausted, gravitational contraction and heating start again until the next nuclear fuel is ignited. Therefore, from this point of view, the life of a star can be envisioned as the progressive gravitational contraction of a self gravitating ball of gas, punctuated by occasional delays when the nuclear burning supplies the energy lost in the form of radiation and/or neutrinos.

By properly combining the equations of the conservation of momentum (hydrostatic equilibrium) and of the conservation of mass and by assuming an equation of state for an ideal gas, the temporal evolution of the central temperature as a function of the central density, for a star of a given mass, can be simply estimated: $T_c \propto \rho_c^{1/3} M^{2}$. This means that a contracting star with a mass M and with a constant composition, in which the energy generation and neutrino losses are negligible, supported by an ideal gas pressure, will increase its central temperature following the relation $T_c \propto \rho_c^{1/3}$. This relation will hold until one of the above assumptions is violated, i.e., by the ignition of thermonuclear fusion or by the onset of degeneracy. Low mass stars \index{low mass stars} ($\rm M\lesssim 8~M_\odot$) reach a point where the internal structure is fully supported by the electron degeneracy and the contraction of the degenerate core is slowed down. These stars end their life as white dwarfs. More massive stars \index{massive stars}, on the contrary, never experience a significant electron degeneracy in the core during all their nuclear burning stages, therefore they evolve to higher and higher temperatures, fusing heavier and heavier elements until an iron core is formed. The iron core becomes unstable and through a sequence of events the star explodes as a core collapse supernova. The presupernova evolution of these stars and their final fate as core collapse supernovae \index{core collapse supernovae} are the subjects of the present chapter.

During the progressive contraction and heating, the core of a massive star achieves progressively high temperatures. Therefore all the nuclear burning stages, i.e., H-, He-, C-, Ne-, O- and Si-burning, are activated and go to completion until the nuclear statistical equilibrium is achieved. The nuclear burning stages following the H and the He ones (the so called advanced burning stages \index{advanced burning stages}) are characterized by an increasing number of efficient nuclear reactions. In general any given advanced burning stage is activated by few key reactions that release light particles (i.e., protons, neutrons and alpha particles). Because of the high temperatures, these light particles can be captured by almost all the isotopes present in the hot plasma and therefore huge number of nuclear reactions are promptly activated. The general properties of the advanced burning stages in massive stars are discussed in details in many papers and excellent reviews \cite{2002RvMP...74.1015W,1998ApJ...502..737C,2008EAS....32..233L,1998npa..conf...27T,1986nce..conf....1W}.

An important aspect of the evolution of massive stars is the neutrino losses \index{neutrino losses}. In very hot environments there are enough energetic photons in the Planck distribution having energies in excess of $m_e c^2$, $m_e$ being the electron rest mass. As a consequence the creation of positron-electron pairs is activated and proceeds at a high rate \cite{1996ApJS..102..411I}. The created pairs ($e^+ e^-$), however, quickly recombine and give back two photons per pair. Due to the very short annihilation time, the pair creation and annihilation quickly come to an equilibrium. However, in this continuous back and forth exchange, there is a small one-way leakage, because a small fraction of ($e^+ e^-$) recombinations result in a ($\nu_e \bar{\nu_e}$) pair, instead of two photons. Since, for the typical physical conditions of these phases, the neutrino opacity is extremely low, the neutrinos exit the star at the speed of ligth without interacting with the matter and therefore they constitute an efficient energy loss mechanism. The neutrino emission from pair production starts to become efficient when the central temperature exceeds $\rm \sim 8 \cdot 10^8~K$, i.e., at the beginning of core C burning, and progressively increases all along the further evolutionary phases up to the onset of the iron core collapse. This is the reason why the advanced evolutionary phases of massive stars are usually referred to as "neutrino dominated" phases. The dramatic increase of the neutrino emission implies a dramatic increase of the total luminosity, i.e. the sum of the photon and neutrino luminosities ($L_{tot}=L_\gamma+L_\nu$). The almost constant nuclear energy $E_{nuc}$ provided by each advanced nuclear burning stage coupled to the dramatic increase of the total luminosity (mainly the neutrino luminosity, because the photon luminosity remains essentially constant during these phases) implies a dramatic acceleration of the stellar evolution. The lifetime of any given nuclear burning stage, in fact, can be simply estimated as $\tau_{nuc} \sim E_{nuc} M/L_{tot}$. The typical nuclear burning lifetimes of C, Ne, O and Si burning are $\rm \sim 10^4,~\sim 0.5,~\sim 0.4$ and $\rm 10^{-2}$ years, respectively.

Because of the tremendous shortening of the advanced burning lifetimes, in the presence of convection, the nuclear burning timescales may become comparable to the mixing turnover ones and therefore, in this case, the chemical composition cannot be fully homogenized within the whole convective zone. This interaction between local nuclear burning and convective mixing requires a proper numerical treatment and, perhaps, constitutes one of the most challenging aspects of the computation of the presupernova evolution of massive stars \cite{1998ApJ...502..737C,2008EAS....32..233L}.

\begin{figure}[htbp]
\centering
\includegraphics[scale=.40]{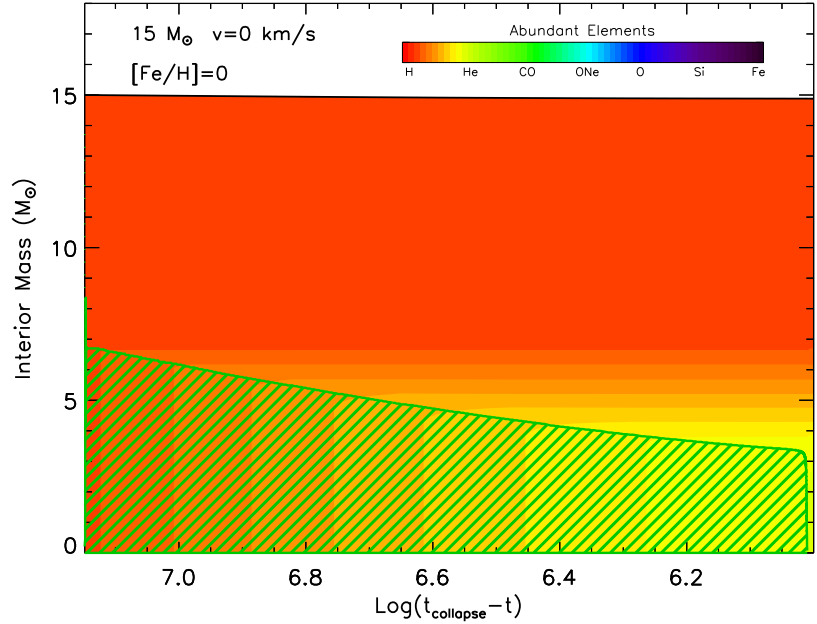}
\subfigures
\caption{Convective and composition history of a non rotating solar metallicity $\rm 15~M_\odot$ model. The convective zones are marked by green shaded areas, while the chemical composition follows the color codes reported in the upper right color bar. The quantity in the $x$-axis is the logarithm of the remaining time to the collapse in years. The quantity reported in the $y$-axis is the interior mass coordinate in solar masses.}
\label{kipphburn15a000}       
\end{figure}
\begin{figure}[htbp]
\samenumber
\centering
\includegraphics[scale=.40]{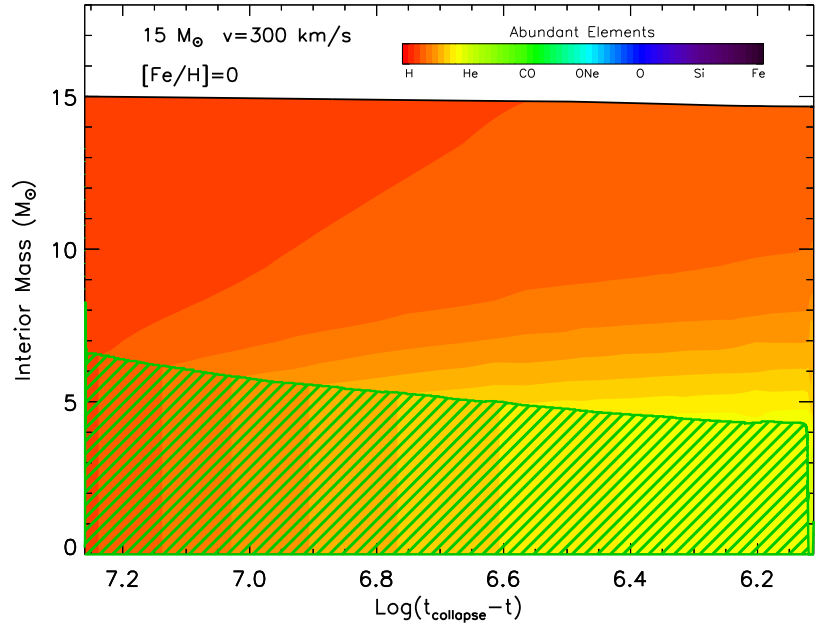}
\subfigures
\caption{Same as Figure \ref{kipphburn15a000} but for a model with initial rotation velocity $\rm v=300~km/s$.}
\label{kipphburn15a300}       
\end{figure}

 
\section{Presupernova Evolution: Overview}

\subsection{Core H burning}
\label{sec:hburn}
Core H burning is the first stable and long lasting nuclear burning stage.
In massive stars with initial metallicity $\rm Z \gtrsim 10^{-7} $, it occurs 
at temperatures of the order of $\rm \sim 4$ to $\sim 6\cdot 10^{7}~K$ and therefore it is powered by the CNO cycle.
The high sensitivity of the CNO cycle on the temperature implies a high nuclear energy flux and hence 
the formation of a convective core that reaches its maximum extension at the very beginning of core H burning.
The main source of opacity in the core is due to the electron scattering, therefore as
the burning proceeds the radiative gradient progressively decreases and therefore the convective core reduces as well.
Such a reduction leaves a region of variable composition enriched by the partial CNO processed material (Figure \ref{kipphburn15a000}).
When the central H mass fraction drops below $10^{-7}$ the convective core vanishes, the interior zones undergo an overall gravitational contraction and then the H burning shifts in a shell. The core H burning lifetimes range between $\rm \sim 10^7~yr$ and $\rm \sim 10^6~yr$ in the mass range $\rm 13-120~M_\odot$, see, e.g. \cite{2013ApJ...764...21C}.

\begin{figure}[htbp]
\centering
\includegraphics[scale=.30]{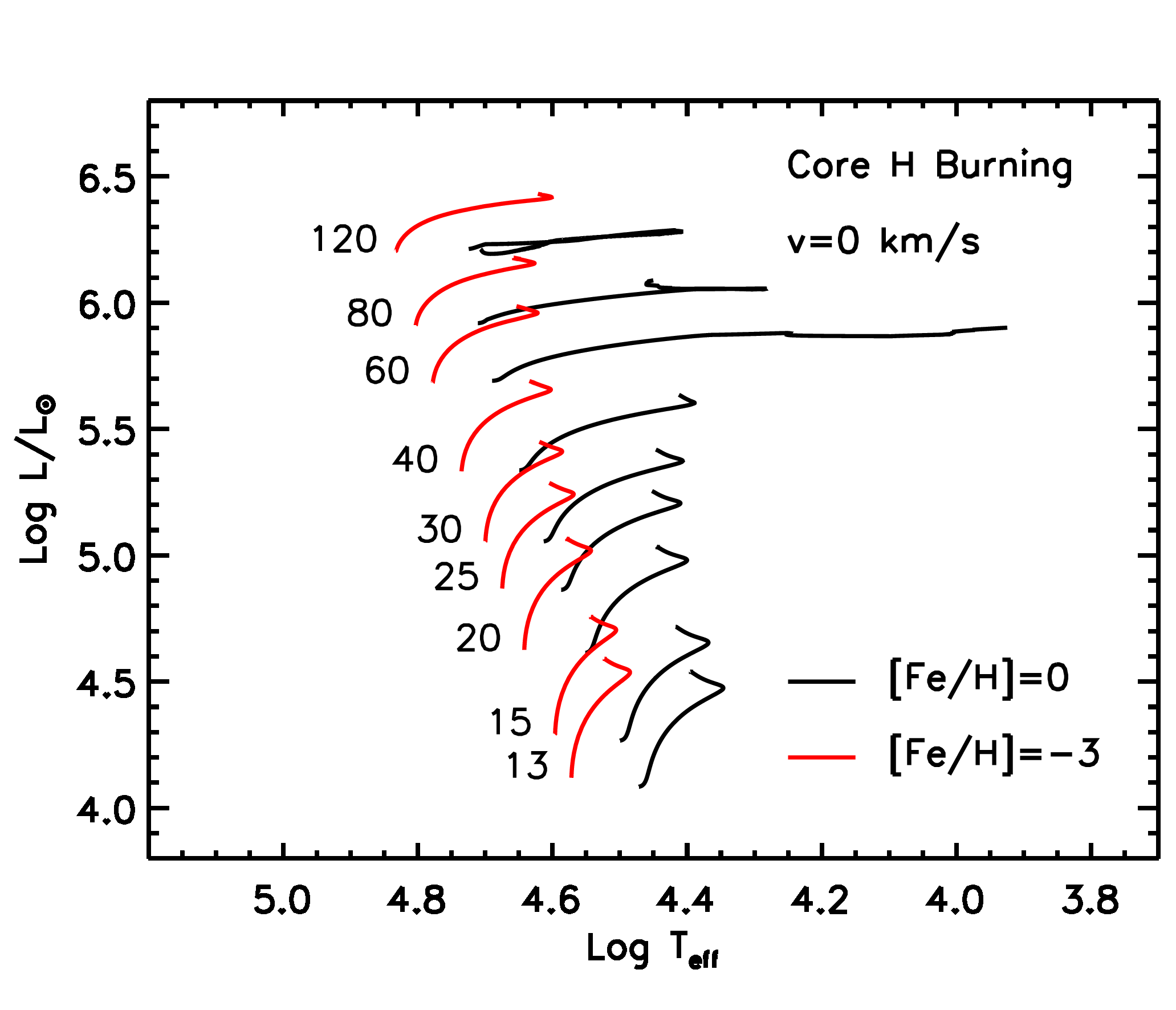}
\caption{Hertzsprung-Russell (HR) diagram of non rotating models of metallicities [Fe/H]=0 (black lines) and [Fe/H]=-3 (red lines) during the core H burning phase. Solar metallicity models with initial mass $\rm M>60~M_\odot$ become WR during this phase.}
\label{hrhburnall}       
\end{figure}

The interior properties of the star during this phase determine the evolutionary path in the HR diagram. 
In general, during core H burning the luminosity progressively increases while the effective temperature progressively
decreases. Figure \ref{hrhburnall} shows the evolutionary path of core H burning solar metallicity massive stars in the HR diagram. For this set of models the minimum mass entering the region corresponding to the O-type stars ($\rm T_{eff} \ge 31500~K$) is $\rm \sim 13~M_\odot$. The fraction of core H burning lifetime spent as a O-type star \index{O-type star} increases with the mass and ranges in this case between $\sim 0.15$ and $\sim 0.8$ for the 13 and the $\rm 120~M_\odot$ models respectively.  

Mass loss is quite efficient during this phase and scales directly with the mass (i.e., luminosity). In the more massive solar metallicity models it ranges typically between $\rm 10^{-6}$ and $\rm 10^{-5}~M_\odot/yr$ \cite{2001A&A...369..574V,2000A&A...362..295V}. This value is high enough to induce a substantial reduction of the total mass of the star and hence to expose to the surface the zones partially modified by the core H burning. When this happens the star becomes a Wolf-Rayet star \index{Wolf-Rayet} (WR)  \cite{2007ARA&A..45..177C} and evolves toward higher effective temperatures (see also \cite{2006ApJ...647..483L} for our definitions of the various WR stages). In the present set of models, solar metallicity non rotating stars with initial mass larger than 60 $\rm M_\odot$ become WR stars already during core H burning (Figure \ref{hrhburnall}). This limit, however, is sensitive to a number of uncertainties among which the efficiency of the stellar wind. 
\begin{figure}[htbp]
\centering
\includegraphics[scale=0.35]{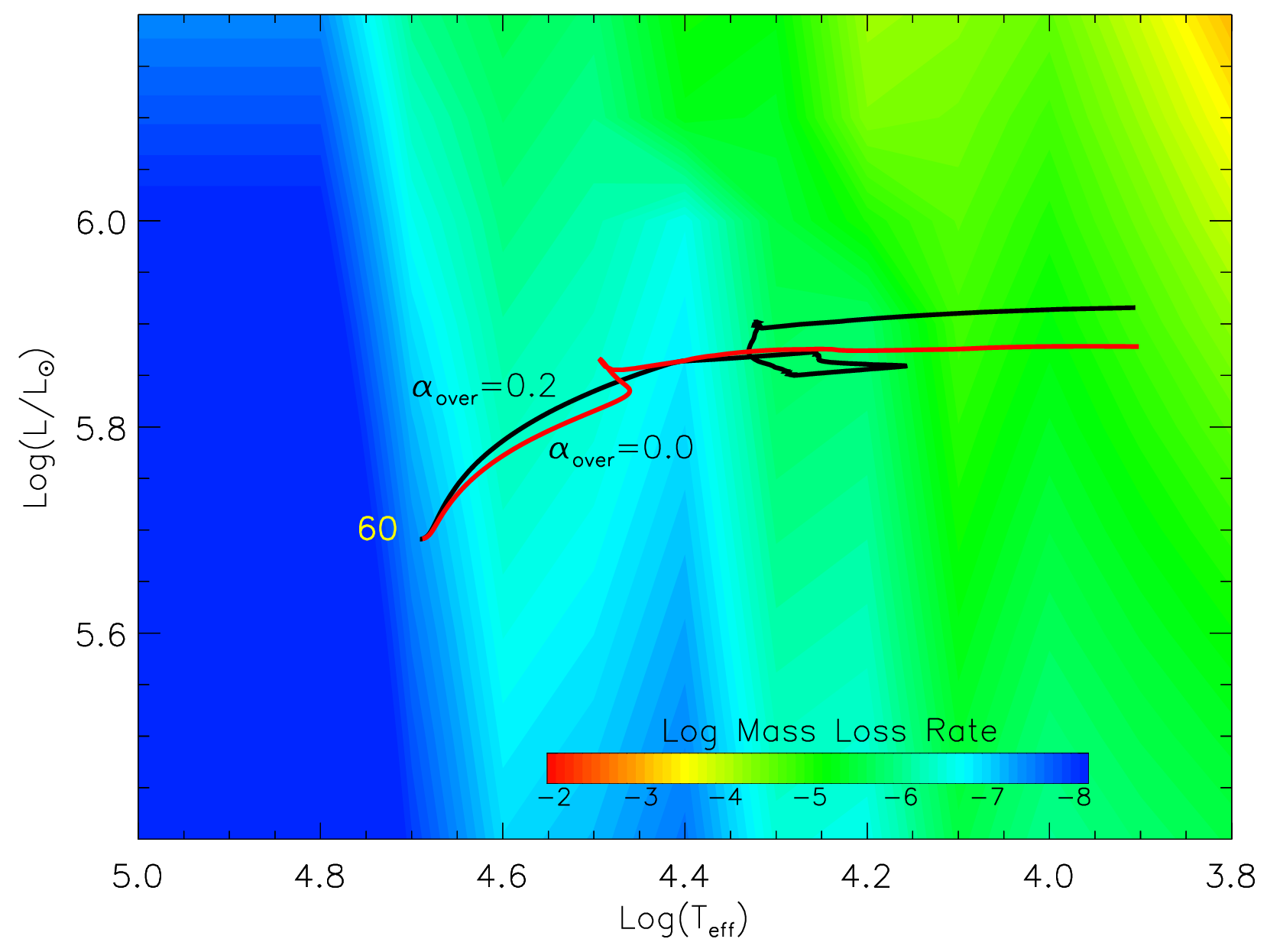}
\caption{Evolutionary tracks in the HR diagram of two non rotating $\rm 60~M_\odot$ models of solar composition during core H burning. The black line refers to the model computed assuming $\rm 0.2~H_P$ of core overshooting while the red line refers to the model computed without any convective core overshooting. The mass loss rate corresponding to the various regions of the HR diagram follows the color codes reported in the color bar.}
\label{mpunto_60}       
\end{figure}
The mass loss depends mainly on both the luminosity and the effective temperature, i.e. on the evolutionary path of the star in the HR diagram, and therefore, ultimately, on the interior properties of the star. For example, the size of the convective core, that in turn depends on the amount of convective core overshoot adopted in the calculations, may influence, even significantly, the total mass of the star at core H depletion. Figure \ref{mpunto_60} shows the evolutionary tracks of two non rotating models of solar metallicity with initial mass $\rm M=60~M_\odot$, computed with and without overshooting. The evolutionary track of the star computed with overshooting is much more extended toward lower effective temperatures. This implies a higher mass loss and therefore a final mass at core H depletion ($\rm M_{H-dep}\sim 35~M_\odot$), much lower than the one of the corresponding model computed without overshooting ($\rm M_{H-dep}\sim 50~M_\odot$). 

\begin{figure}[htbp]
\centering
\includegraphics[scale=.30]{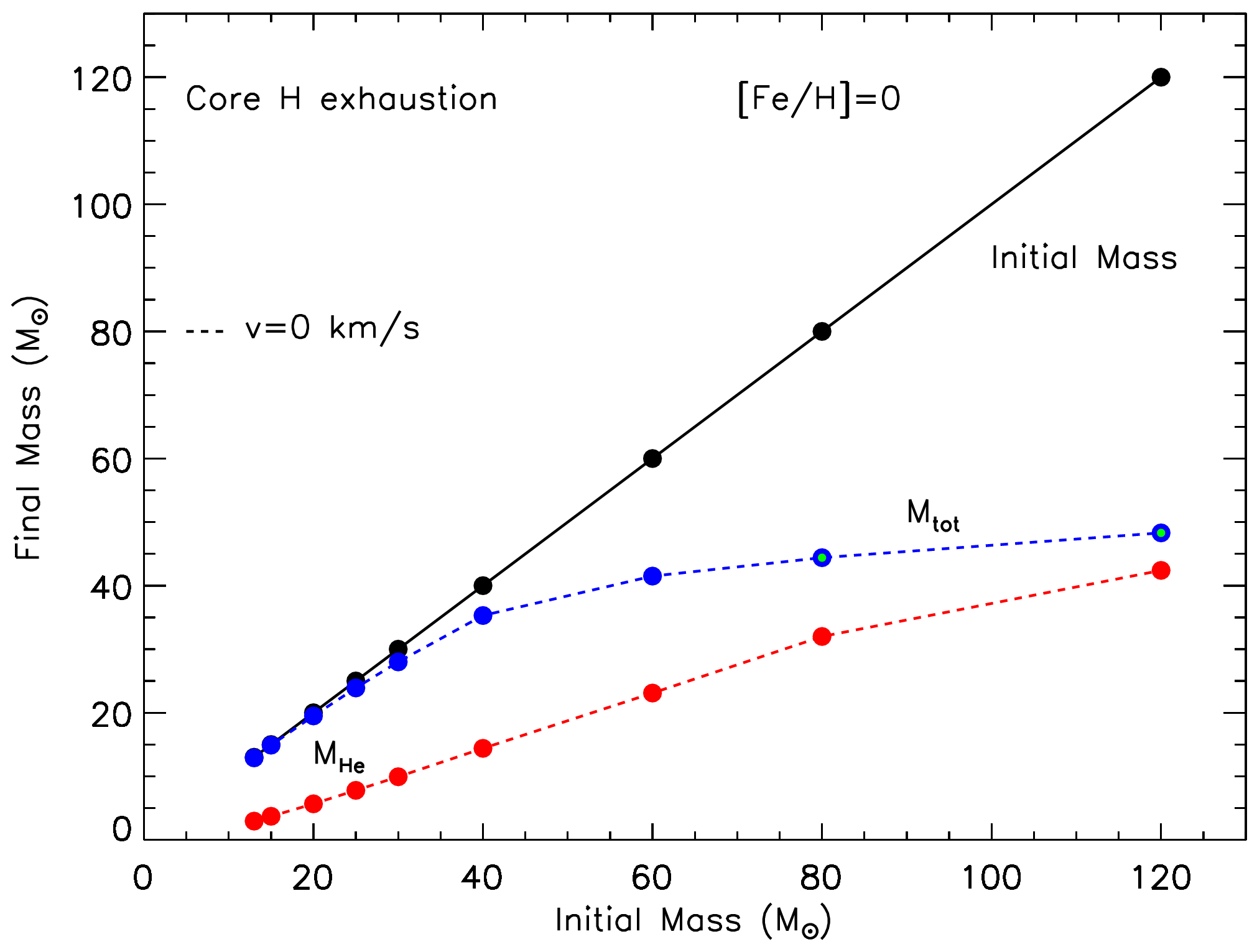}
\caption{Total mass (blue line) and He core mass (red line) of solar metallicity non rotating models at core H depletion as a function of the initial mass.}
\label{masshecore}       
\end{figure}

\begin{figure}[htbp]
\centering
\includegraphics[scale=.30]{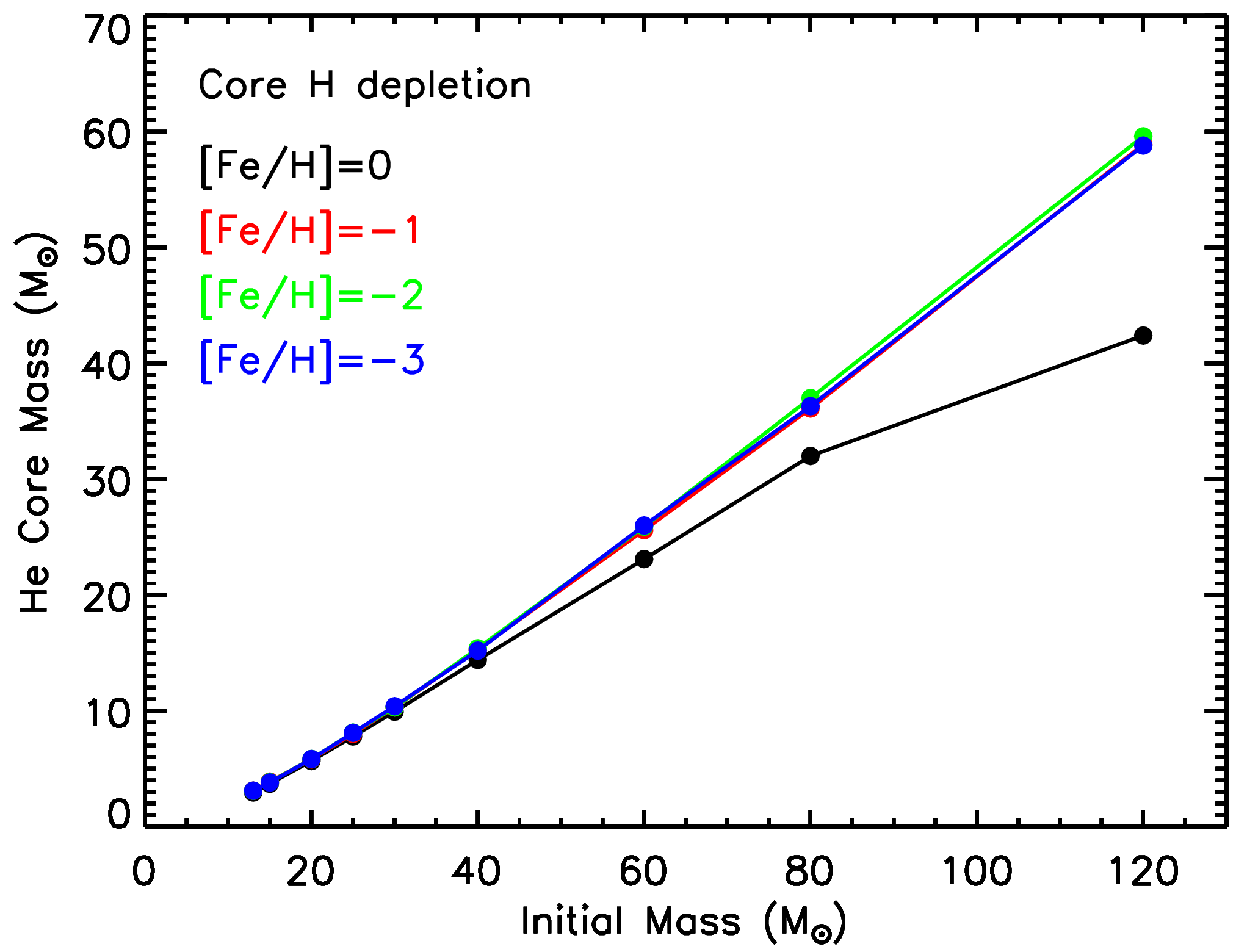}
\caption{He core mass at core at core H depletion as a function of the initial mass for non rotating models with initial metallicities [Fe/H]=0 (black line), [Fe/H]=-1 (red line), [Fe/H]=-2 (green line) and [Fe/H]=-3 (blue line).}
\label{masshecore2}       
\end{figure}

The He core at core H depletion depends on the size of the H convective core - the larger the size of the H convective core, the larger the zone where H is converted into He and therefore the larger the final He core mass at core H depletion. The maximum size of the H convective core, in general, increases with the mass of the star, therefore the He core at core He depletion will increase with the mass of the star as well. However such a relation may depend on the efficiency of mass loss. In the more massive stars, in fact, mass loss is strong enough to reduce substantially the total mass and therefore to induce a reduction of the H convective core during the core H burning phase. As a consequence, in this case, the He core at core He depletion is smaller than it would be in the case of reduced, or absent, mass loss. Figures \ref{masshecore} and \ref{masshecore2} shows this effect for solar metallicity models with initial mass larger than 60 $\rm M_\odot$. As mentioned above, these stars become WR during the core H burning phase.

As the metallicity decreases, the stars become more compact and hotter due to the reduction of the opacity of the matter. This implies bluer evolutionary tracks in the HR diagram (Figure \ref{hrhburnall}). 
The mass loss scales with the initial metallicity as $\rm \dot{M}\sim Z^{0.85}$ \cite{2001A&A...369..574V,2000A&A...362..295V} therefore at low metallicities, i.e. $\rm [Fe/H]\leq -1$, the stars during the core H burning phase evolve essentially at constant mass. The reduction of the initial metallicity also implies a reduction of the abundance of the CNO nuclei and therefore an increase of the core H burning temperature. This would lead to larger convective cores and hence larger He cores at core H depletion \cite{1986MNRAS.220..529T}. However, in models with convective core overshooting, this is only a mild effect. Figure \ref{masshecore2} shows that for stars with initial mass $\rm M<40~M_\odot$ the He core at core H depletion is essentially independent on the initial metallicity. On the contrary, for stars above this limiting mass there are sizable  differences between models with solar metallicity and models with metallicities $\rm [Fe/H]\leq -1$. These differences are due to the influence of mass loss on the size of the H convective core (discussed above), that is sizable at solar metallicity and becomes progressively negligible at lower metallicities.

The inclusion of rotation has essentially two main effects on the evolutionary path of a massive star in the HR diagram during core H burning: (1) the lowering of the effective gravity, due to both the centrifugal force and the angular momentum transport, makes the track redder; (2) the rotation driven mixing (see below) a) increases the size of the H depleted zone, making the track brighter and cooler (like the convective core overshooting), and b) enriches the radiative envelope with He, reducing the opacity and making the track brighter and bluer \cite{2000A&A...361..101M,2013ApJ...764...21C,2012RvMP...84...25M,2000ARA&A..38..143M}. By the way, a changing of the path in the HR diagram also implies a changing of the mass loss rate which is, in this case, an indirect effect of the stellar rotation. Depending on the initial mass, initial metallicity and initial rotation velocity, one effect may prevail on the others. 
\begin{figure}[t]
\centering
\includegraphics[scale=.24]{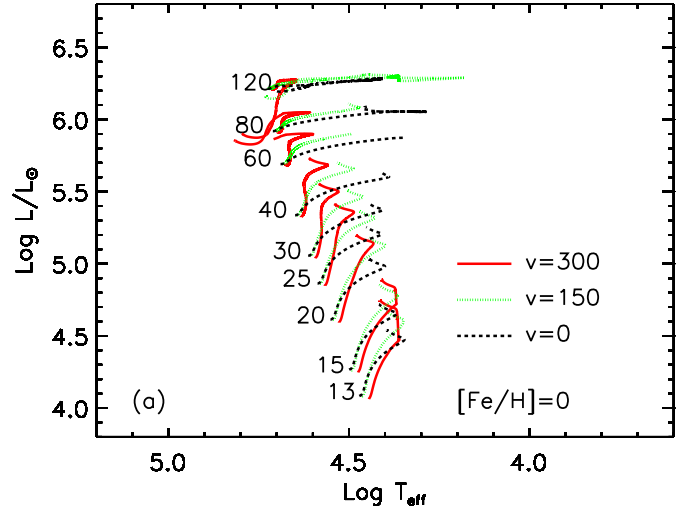}
\subfigures
\caption{HR diagram of models during core H burning with initial velocities v=0 km/s (black lines), v=150 km/s (green lines) and v=300 km/s (red lines) and with initial metallicity [Fe/H]=0. }
\label{hrhburnallz0}       
\end{figure}
\begin{figure}[t]
\centering
\samenumber
\subfigures
\includegraphics[scale=.24]{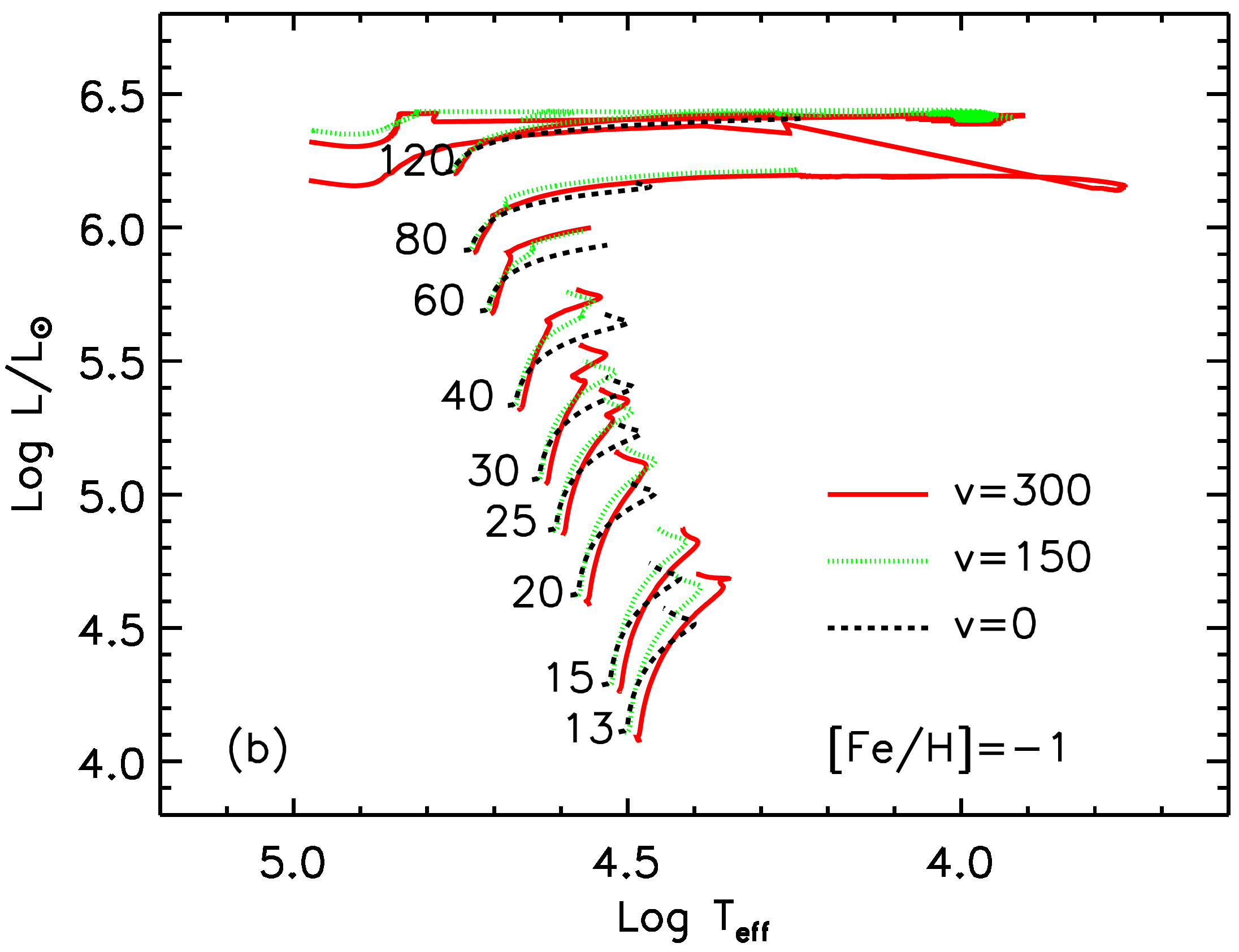}
\caption{Same as Figure \ref{hrhburnallz0} but for metallicity [Fe/H]=-1. }
\label{hrhburnallz1}       
\end{figure}
\begin{figure}[t]
\centering
\samenumber
\subfigures
\includegraphics[scale=.24]{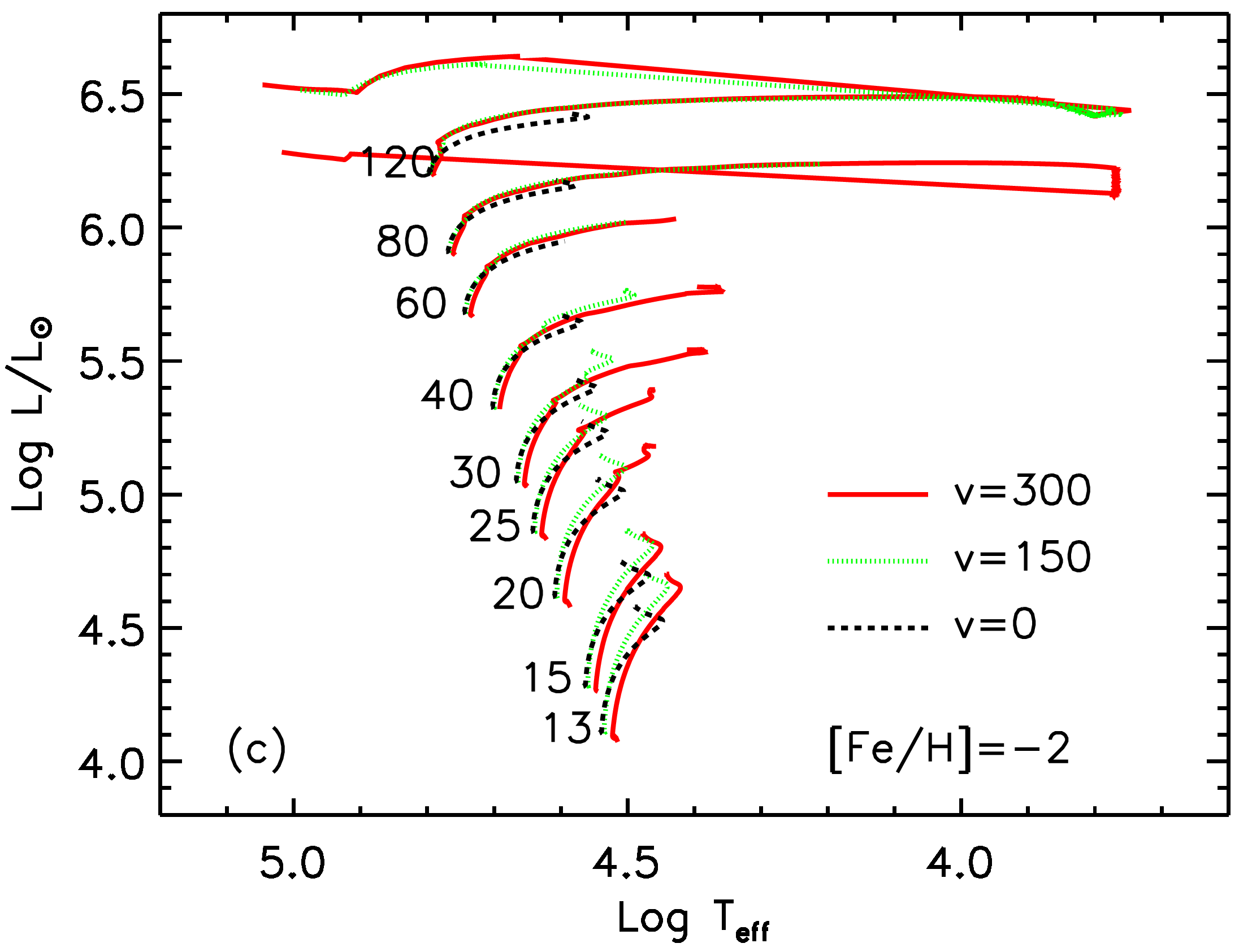}
\caption{Same as Figure \ref{hrhburnallz0} but for metallicity [Fe/H]=-2. }
\label{hrhburnallz2}       
\end{figure}
\begin{figure}[t]
\centering
\samenumber
\subfigures
\includegraphics[scale=.24]{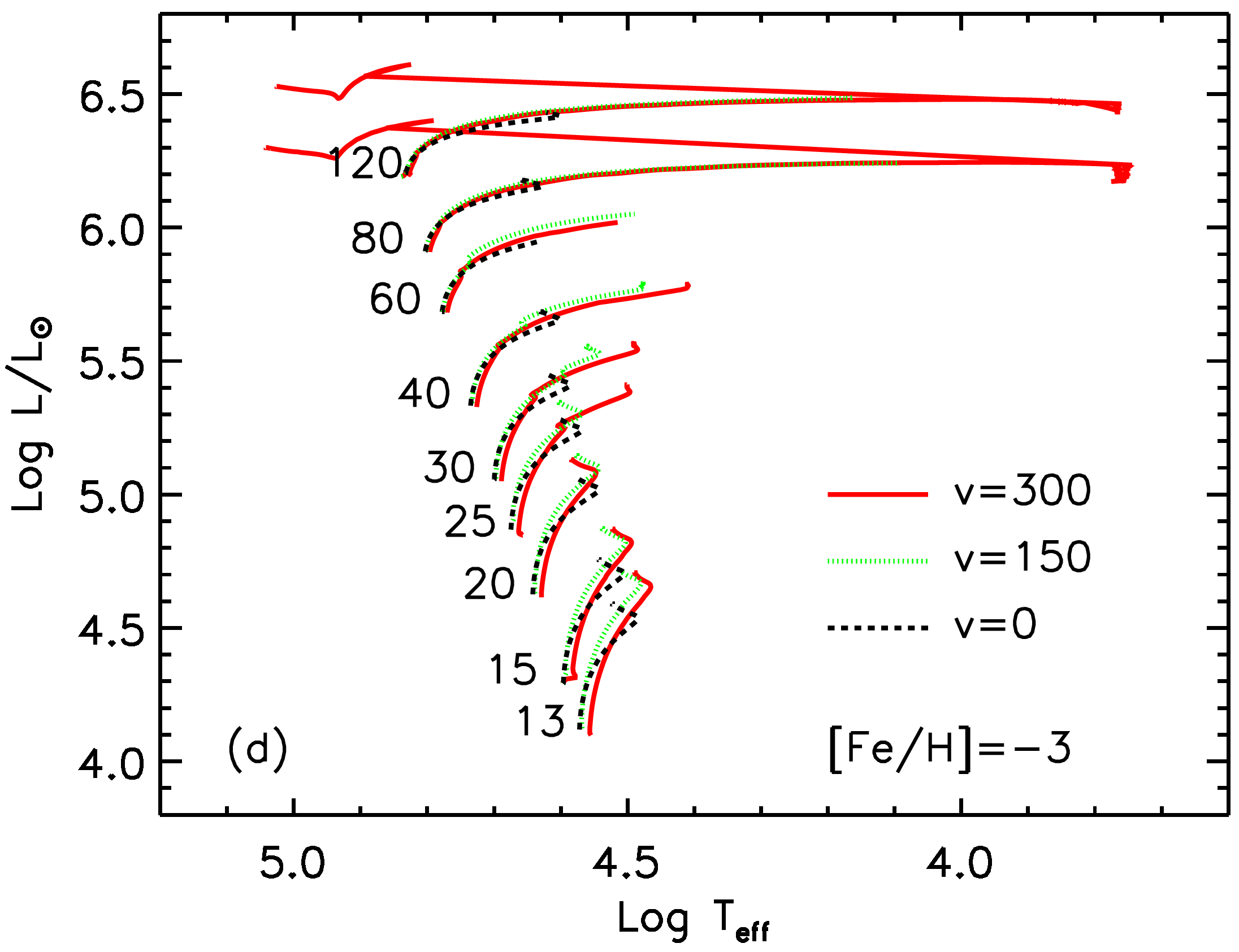}
\caption{Same as Figure \ref{hrhburnallz0} but for metallicity [Fe/H]=-3. }
\label{hrhburnallz3}       
\end{figure}
Figures \ref{hrhburnallz0}, \ref{hrhburnallz1}, \ref{hrhburnallz2}, \ref{hrhburnallz3} show that for lower metallicities the evolutionary tracks of rotating models appear broader and more extended toward lower effective temperatures compared to those of the non rotating stars, therefore in this case the lowering of the effective gravity and the increase of the H depleted zone due to the rotation driven mixing are the dominant effects. On the contrary, for higher metallicities, the evolutionary tracks of rotating models are brighter and bluer compared to those of the non rotating stars, clearly demonstrating that for these models the main effect of rotation is the He enrichment of the radiative envelope.  
Since, in general, the mass loss increases with increasing the luminosity and with decreasing the effective temperature, at solar metallicity the indirect effect of the inclusion of rotation is that of a global reduction of the total mass of the star during core H burning. As a consequence, the minimum mass entering the WR stage during this phase decreases from 80 $\rm M_\odot$, for non rotating stars, to 60 $\rm M_\odot$, for stars initially rotating with v=300 km/s. For metallicities corresponding to $\rm [Fe/H]\leq -1$ this effect is negligible because of the strong reduction of the mass loss with decreasing the metallicity. However, the more massive rotating stars ($\rm M > 60~M_\odot$) at metallicities lower than solar undergo a remarkable redward excursion in the HR diagram, which is followed by a bluer evolution. This occurrence is due to the fact that, during the redward evolution, these stars approach their Eddington luminosity \index{Eddington luminosity}, lose a substantial amount of mass and become WR stars. Therefore, summarizing, the minimum mass entering the WR stage ($\rm M_{min}^{WR}$), during the core H burning phase, decreases with increasing the metallicity and with increasing the initial rotation velocity. At solar metallicity $\rm M_{min}^{WR}$ is $\rm \sim 80~M_\odot$ and $\rm \sim 60~M_\odot$ for the non rotating and the rotating ($\rm v_{ini}=300~km/s$) models respectively. At metallicities corresponding to $\rm [Fe/H]\leq -1$, $\rm M_{min}^{WR} \sim 80~M_\odot$ for the rotating models ($\rm v_{ini}=300~km/s$) - no model enters the WR stage in the non rotating case. 


\begin{figure}[htbp]
\centering
\includegraphics[scale=.3]{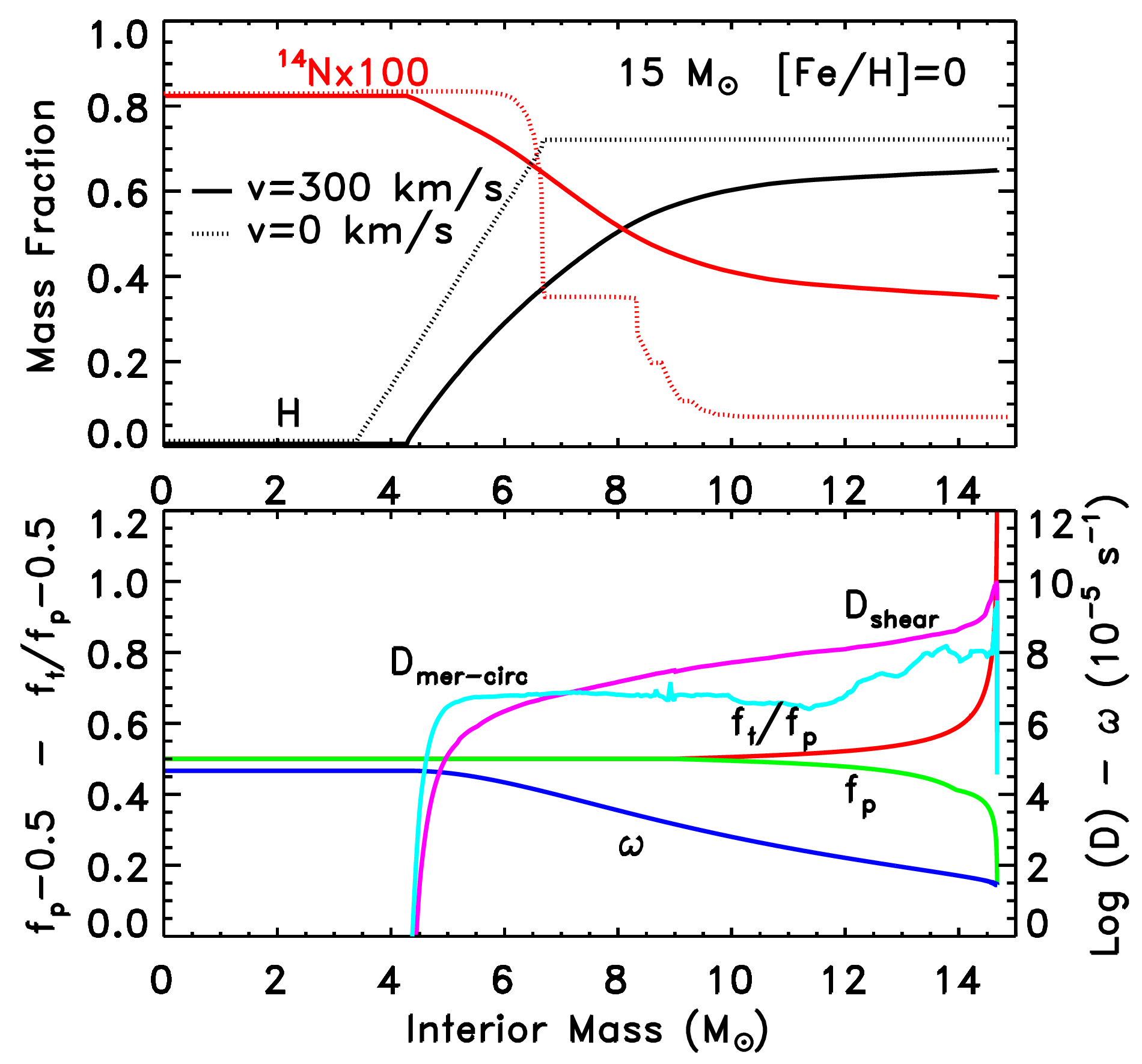}
\caption{Upper panel: interior profiles of H (black lines) and $\rm ^{14}N$ (red lines) mass fractions of two $\rm 15~M_\odot$ models with initial solar composition at core H depletion - the solid and dotted lines refer to the rotating ($\rm v=300~km/s$) and non rotating models, respectively. Lower panel: interior profiles of selected quantities of the $\rm 15~M_\odot$ rotating model at core H depletion: the angular velocity (blue line - right $y$-axis); the diffusion coefficients corresponding to the meridional circulation (cyan line - right $y$-axis) and shear turbulence (magenta line - right $y$-axis); the two form factors $f_P$ (green line - left $y$-axis) and $f_T/f_P$ (red line - left $y$-axis) \cite{1970stro.coll...20K}}
\label{rot15a300chim}       
\end{figure}

The transport of both the angular momentum and the chemical species (rotation driven mixing) during core H burning is essentially due to (1) convection, (2) meridional circulation \index{meridional circulation} and (3) shear turbulence \index{shear turbulence} \cite{2000A&A...361..101M,2013ApJ...764...21C,2012RvMP...84...25M,2000ARA&A..38..143M,2000ApJ...528..368H}. Convection is obviously working within the convective zones while meridional circulation and shear turbulence, operate in the radiative ones. As mentioned above, during core H burning, rotation induced mixing drives a continuous slow ingestion of fresh fuel (H) into the H convective core, as well as a slow mixing of the freshly synthesized H-burning products (mainly $\rm ^{4}He$ and $\rm ^{14}N$) into the radiative envelope of the star (Figure \ref{kipphburn15a300}). Such a mixing is essentially due to the meridional circulation, which dominates at the base of the radiative envelope, i.e., close to the convective-radiative interface, and to the shear turbulence, which is, on the contrary, more efficient in the outer layers (Figure \ref{rot15a300chim}). The main effects of this rotation induced mixing \index{rotation induced mixing} are (1) the increase of the core H burning lifetime; (2) the increase of the He core mass at core H depletion and (3) the enrichment of the $\rm ^{14}N$ surface abundance.

\begin{figure}[h]
\centering
\includegraphics[scale=.3]{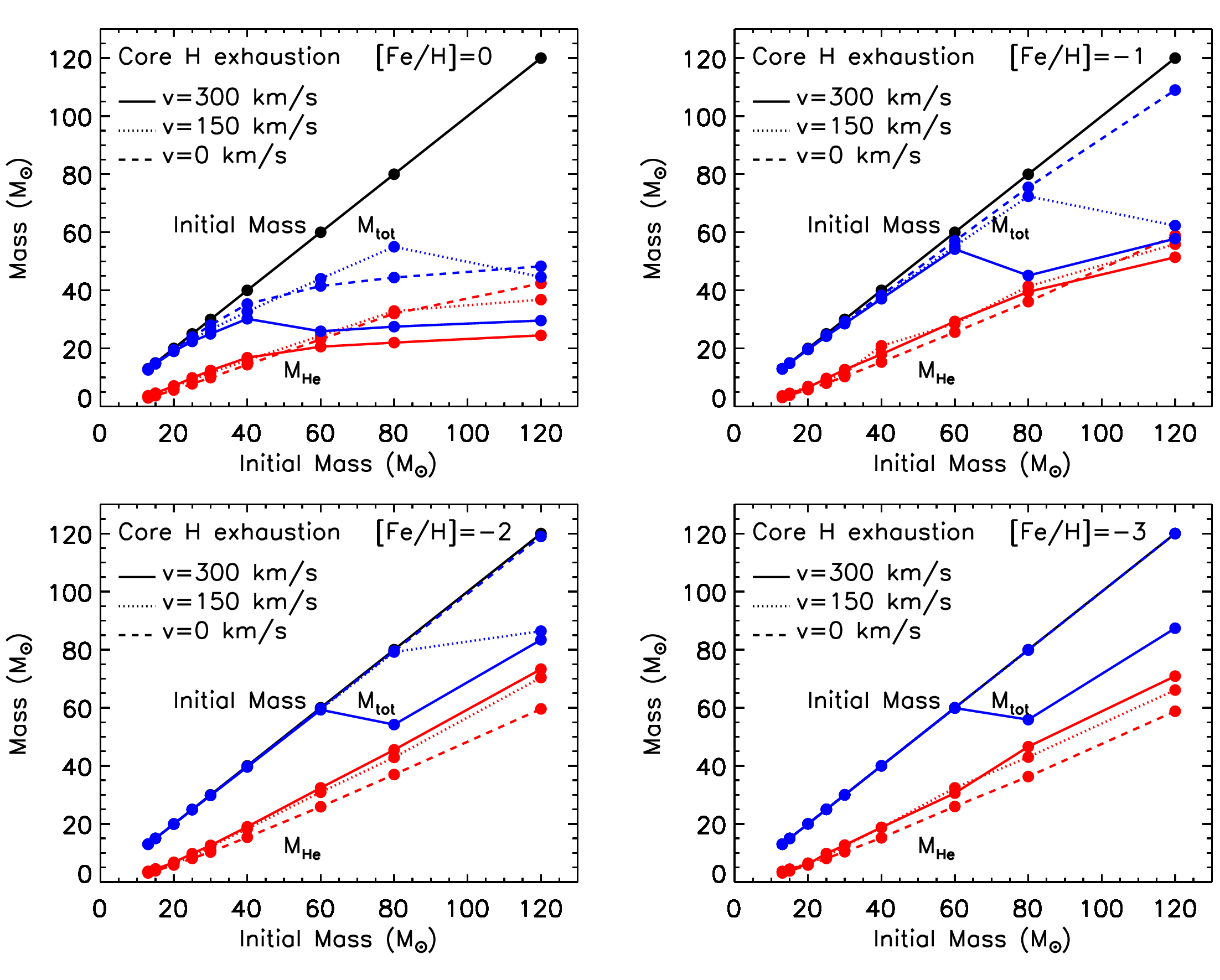}
\caption{Total mass (blue lines) and He core mass (red lines) at core H exhaustion as a function of the initial mass, for models with metallicity [Fe/H]=0 (upper left panel), [Fe/H]=-1 (upper right panel), [Fe/H]=-2 (lower left panel) and [Fe/H]=-3 (lower right panel). The dashed, dotted and solid lines refer to models with initial rotation velocity $\rm v=0,~150,~300~km/s$, respectively. }
\label{massHburncompare}       
\end{figure}

Figure \ref{massHburncompare} shows the He core mass at core H exhaustion for the various metallicities, for rotating and non rotating models. In general, for the reasons mentioned above, the He core mass at core H depletion increases with increasing the initial rotation velocity. This effect is more pronounced at lower metallicities because the efficiency of the rotation driven instabilities increases as the metallicity reduces due to the following reasons: (1) the diffusion timescale of the rotation induced mixing is proportional to $\rm \Delta R^2/D_{mix}$, where $\rm \Delta R$ is the region where the diffusive mixing operates and $\rm D$ is the diffusion coefficient; (2) the stars become more and more compact as the initial metallicity progressively reduces. However, an opposite behavior is found in those rotating models where the mass loss is significantly enhanced by rotation that the H convective core is substantially reduced during the core H burning phase and therefore the resulting He core is smaller than the one of the corresponding non rotating model (see, e.g., the fast rotating solar metallicity models with $\rm M > 40~M_\odot$ and the fast rotating models with $\rm M > 60~M_\odot$ and with metallicity $\rm [Fe/H]\leq -1$ shown in Figure \ref{massHburncompare}). 

\begin{figure}[htbp]
\centering
\includegraphics[scale=.28]{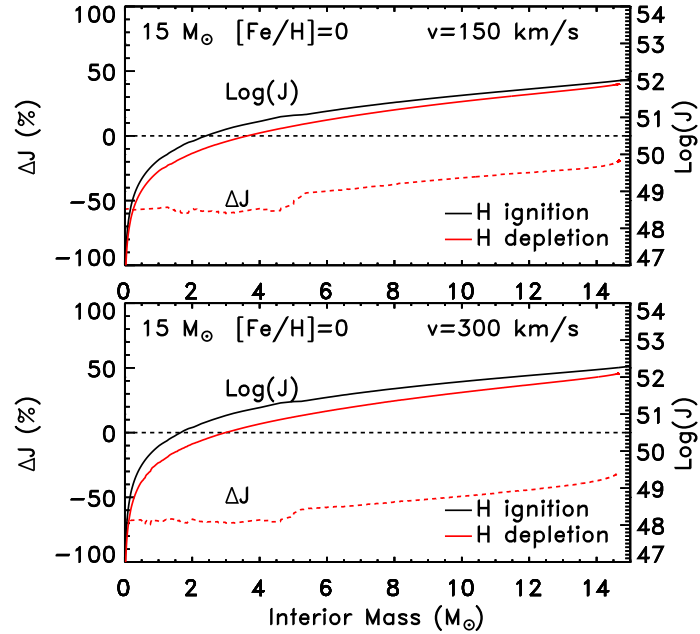}
\subfigures
\caption{Angular momentum as a function of the enclosed mass for a 15 $\rm M_\odot$ model with initial metallicity [Fe/H]=0 and with initial rotation velocities $\rm v=150~km/s$ (upper panel) and $\rm v=300~km/s$ (lower panel). The black and red solid lines refer to the stage corresponding to the core H ignition and core H depletion, respectively (their scale is on the right $y$-axis). In each plot it is also reported the percentage variation of the angular momentum between the core H ignition and core H depletion (red dotted lines - left $y$-axis).}
  \label{momanghburna1}
\end{figure}
\begin{figure}[htbp]
\centering
\samenumber
\includegraphics[scale=.28]{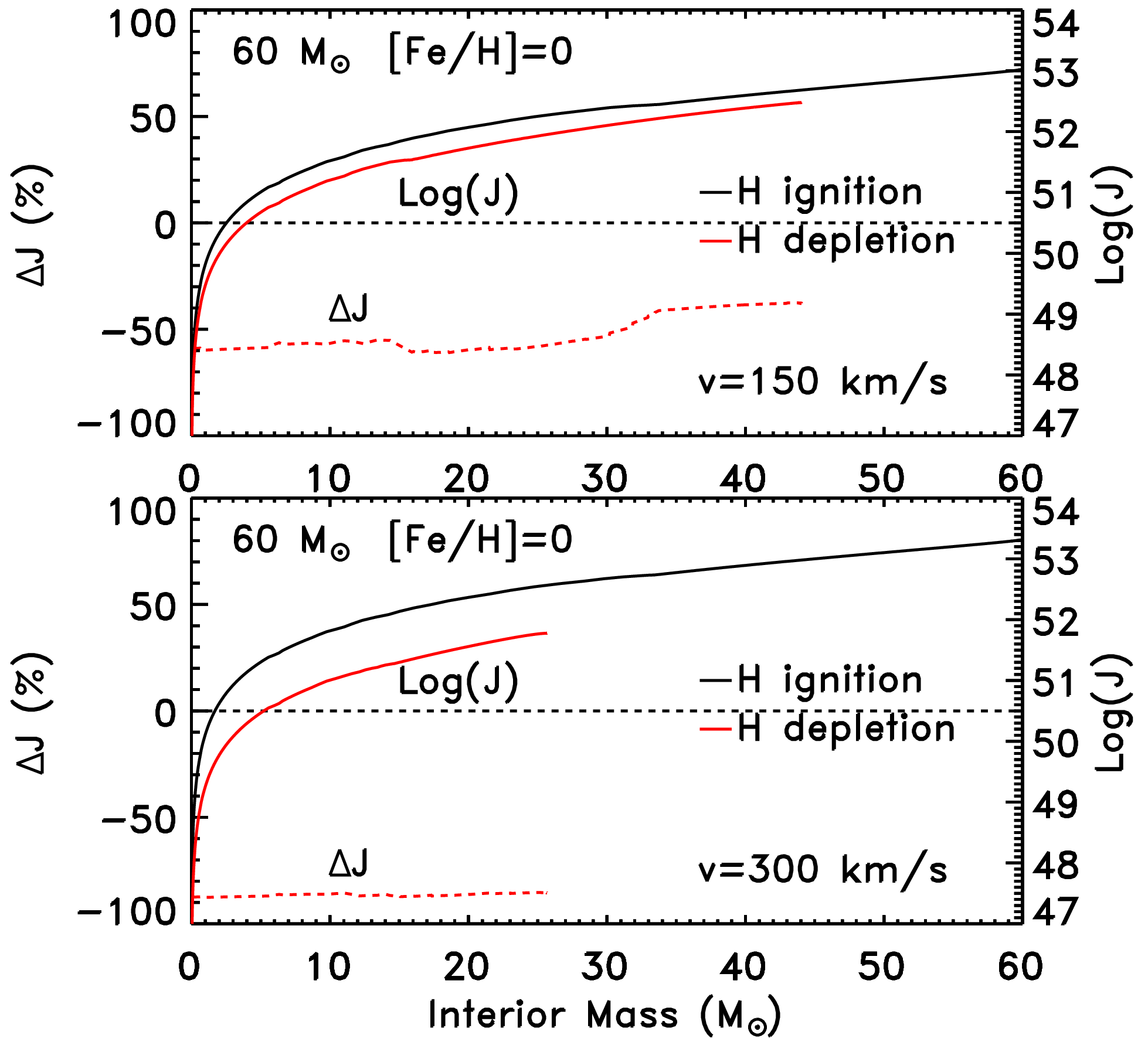}
\subfigures
\caption{Same as Figure \ref{momanghburna1} but for a $\rm 60~M_\odot$ model.}
  \label{momanghburna2}
\end{figure}
\begin{figure}[htbp]
\centering
\samenumber
\includegraphics[scale=.28]{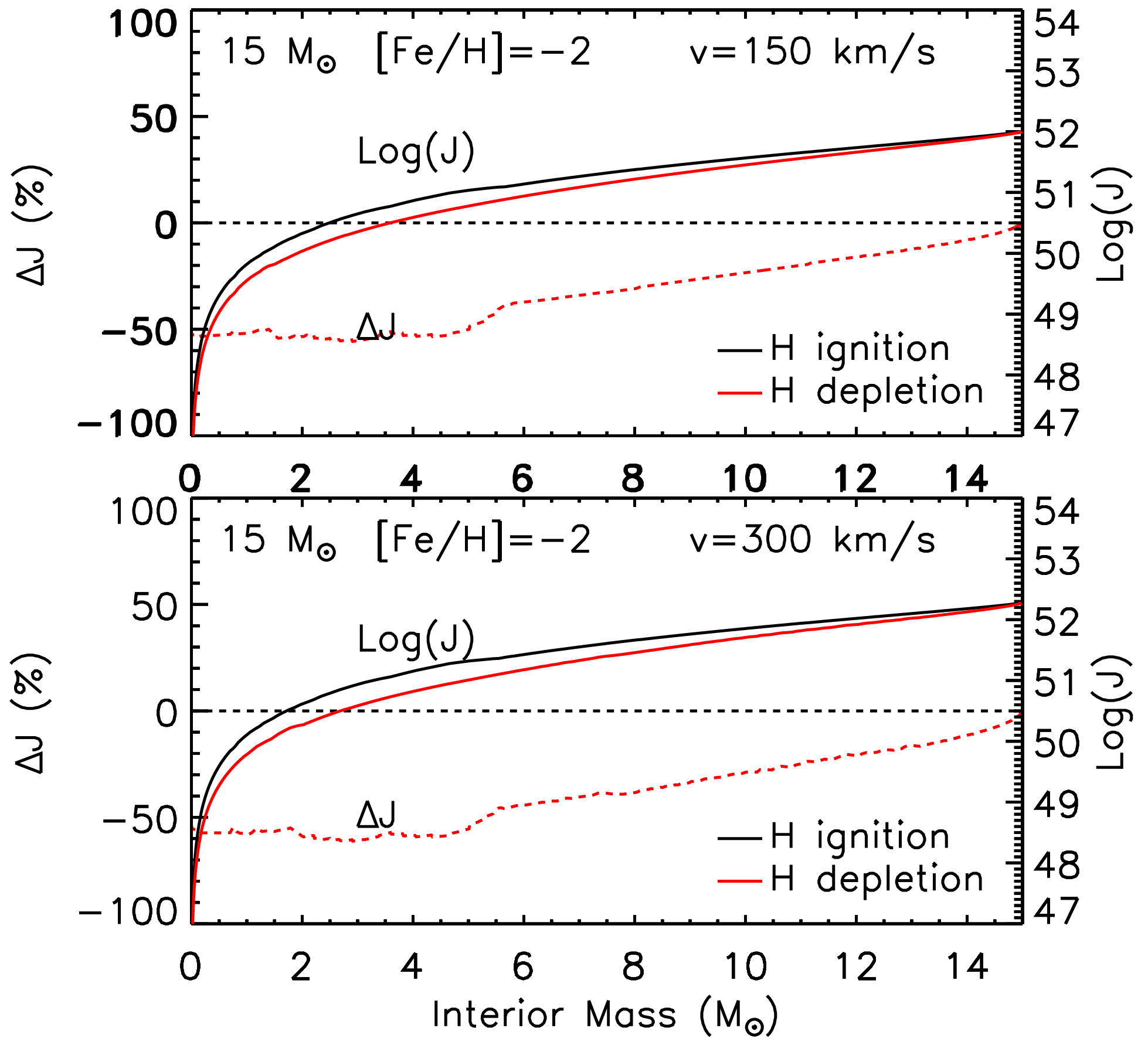}
\subfigures
\caption{Same as Figure \ref{momanghburna1} but for an initial metallicity [Fe/H]=-2.}
  \label{momanghburnc1}
\end{figure}
\begin{figure}[htbp]
\centering
\samenumber
\includegraphics[scale=.28]{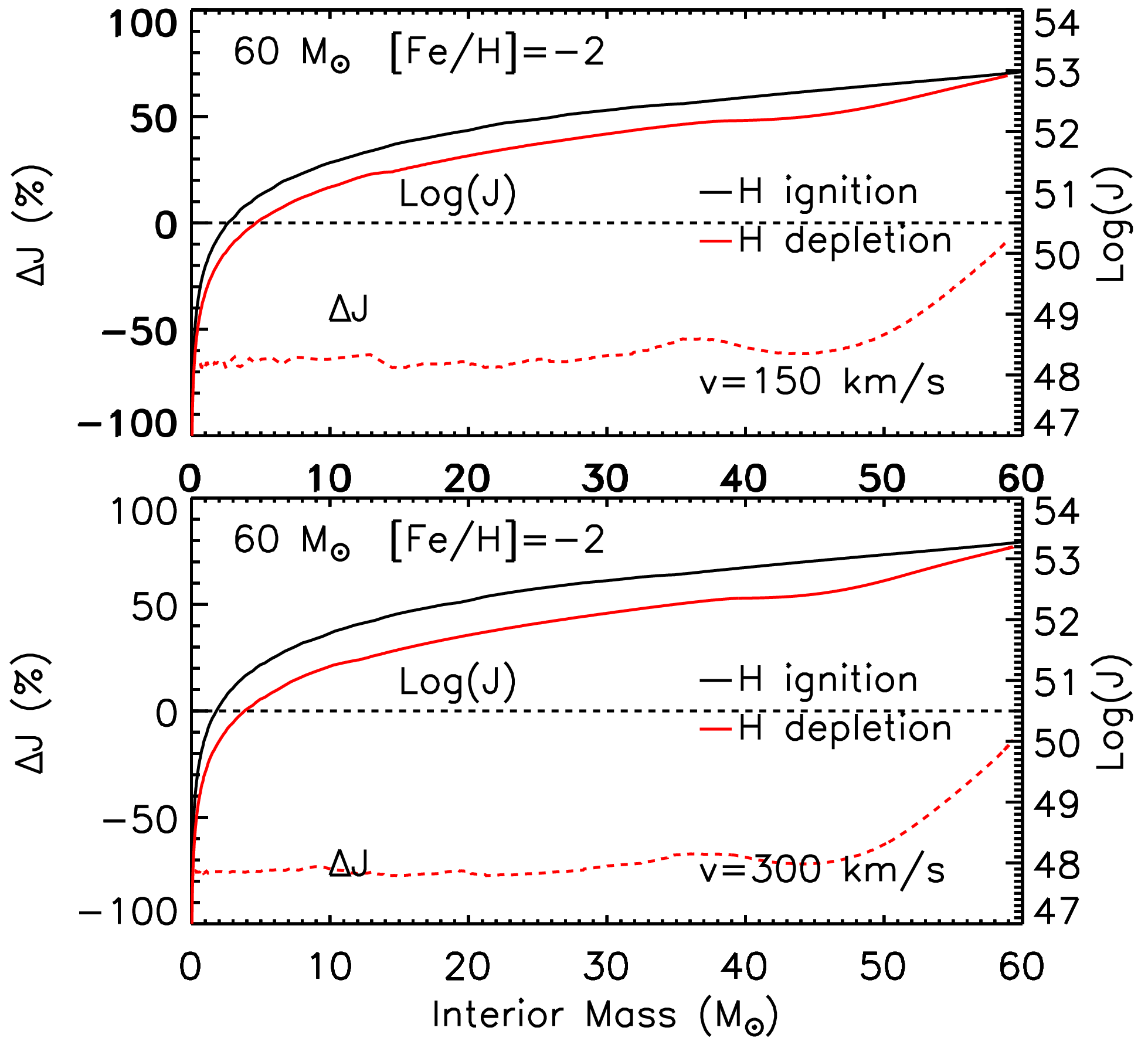}
\subfigures
\caption{Same as Figure \ref{momanghburna2} but for an initial metallicity [Fe/H]=-2.}
  \label{momanghburn2}
\end{figure}

The total mass at core H depletion, in general, scales inversely with the initial rotation velocity. 
However, there may be some exceptions to this general rule because this quantity depends on the complex interplay between the mass loss, which as mentioned above depends on a number of factors, and the efficiency of the rotation driven mixing. For example, at solar metallicity both the 60 and the 80 $\rm M_\odot$ non rotating models lose more mass than their corresponding models with $\rm v_{ini}=150~km/s$. The reason for such a non monotonic behavior is that the evolutionary tracks of the non rotating models extend toward lower effective temperatures in the HR diagram compared to those of the rotating ones (with $\rm v_{ini}=150~km/s$) and therefore they undergo a stronger mass loss.
 
Another important property of the rotating models during core H burning, worth mentioning, is the efficiency of the angular momentum transport. Due to the interplay among convection, meridional circulation and turbulent shear, the angular momentum is in general transported from the innermost zones toward the more external ones and, then, it is eventually lost by stellar wind. As a result of these combined effects, solar metallicity massive stars lose from $\sim 30\%$ to $\sim 90\%$ of their initial total angular momentum during core H burning (Figures \ref{momanghburna1} and \ref{momanghburna2}). In general, the larger the mass and the larger the initial rotation velocity the larger the angular momentum loss. 
At metallicities  corresponding to $\rm [Fe/H]\leq -1$ the total angular momentum loss reduces dramatically because of the strong reduction of the mass loss - for $\rm [Fe/H]\leq -2$ the total angular momentum is essentially conserved in all the stars (Figures \ref{momanghburnc1} and \ref{momanghburnc2}). Note, however, that in the cores of all these models the angular momentum reduces by $\sim 60-90\%$ in low and high masses, respectively, regardless of the initial metallicity. This is due the fact that the reduction of the angular momentum in the core is essentially due to the convection that is the most efficient mechanism for angular momentum transport and that, in turn, is almost independent of the initial metallicity.

\subsection{Core He burning}

\begin{figure}[h]
\centering
\includegraphics[scale=.25]{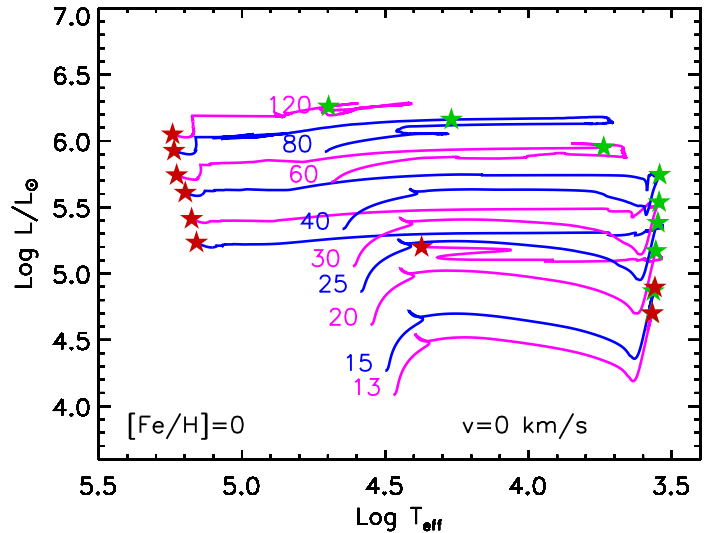}
\subfigures
\caption{HR diagram of solar metallicity non rotating models from the MS phase up to the core He depletion stage. The green stars mark the beginning of core He burning while the red stars refer to the core He depletion stage.}
\label{hrheburninga000}       
\end{figure}

\begin{figure}[h]
\samenumber
\centering
\includegraphics[scale=.25]{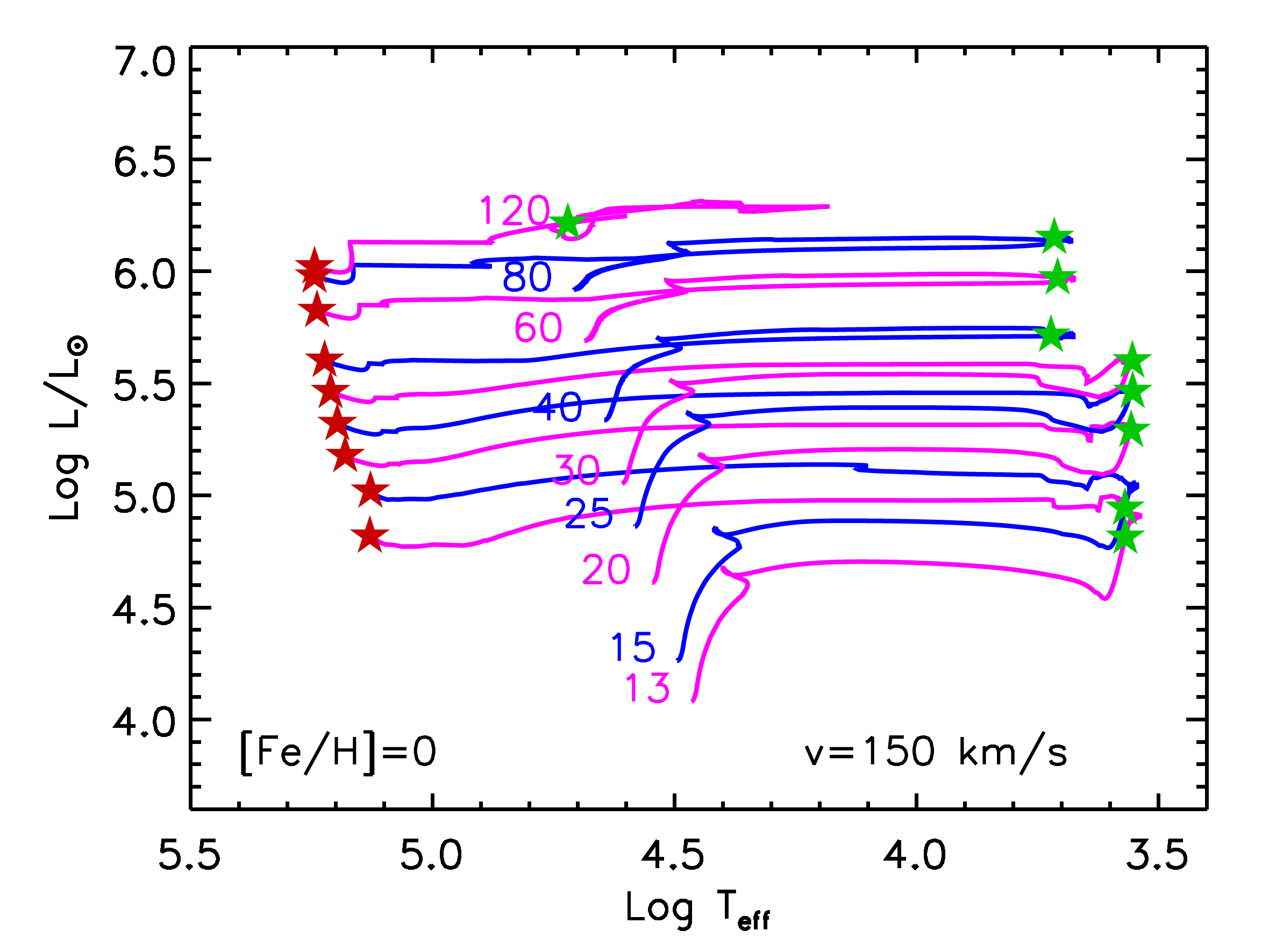}
\subfigures
\caption{Same as Figure \ref{hrheburninga000} but for models with initial rotation velocity $\rm v=150~km/s$.}
\label{hrheburninga150}       
\end{figure}

\begin{figure}[h]
\samenumber
\centering
\includegraphics[scale=.25]{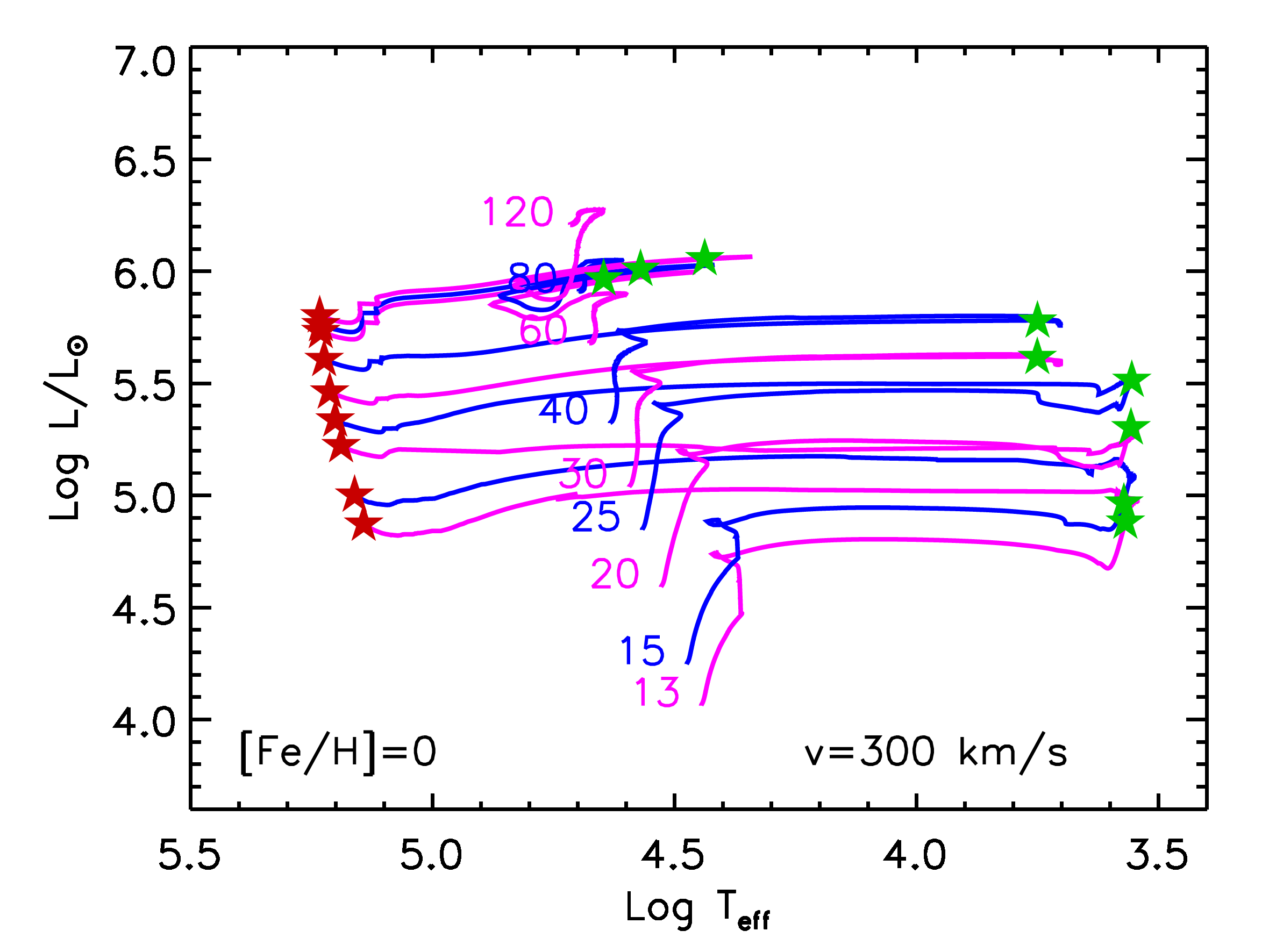}
\subfigures
\caption{Same as Figure \ref{hrheburninga000} but for models with initial rotation velocity $\rm v=300~km/s$.}
\label{hrheburninga300}       
\end{figure}

\begin{figure}[h]
\samenumber
\centering
\includegraphics[scale=.25]{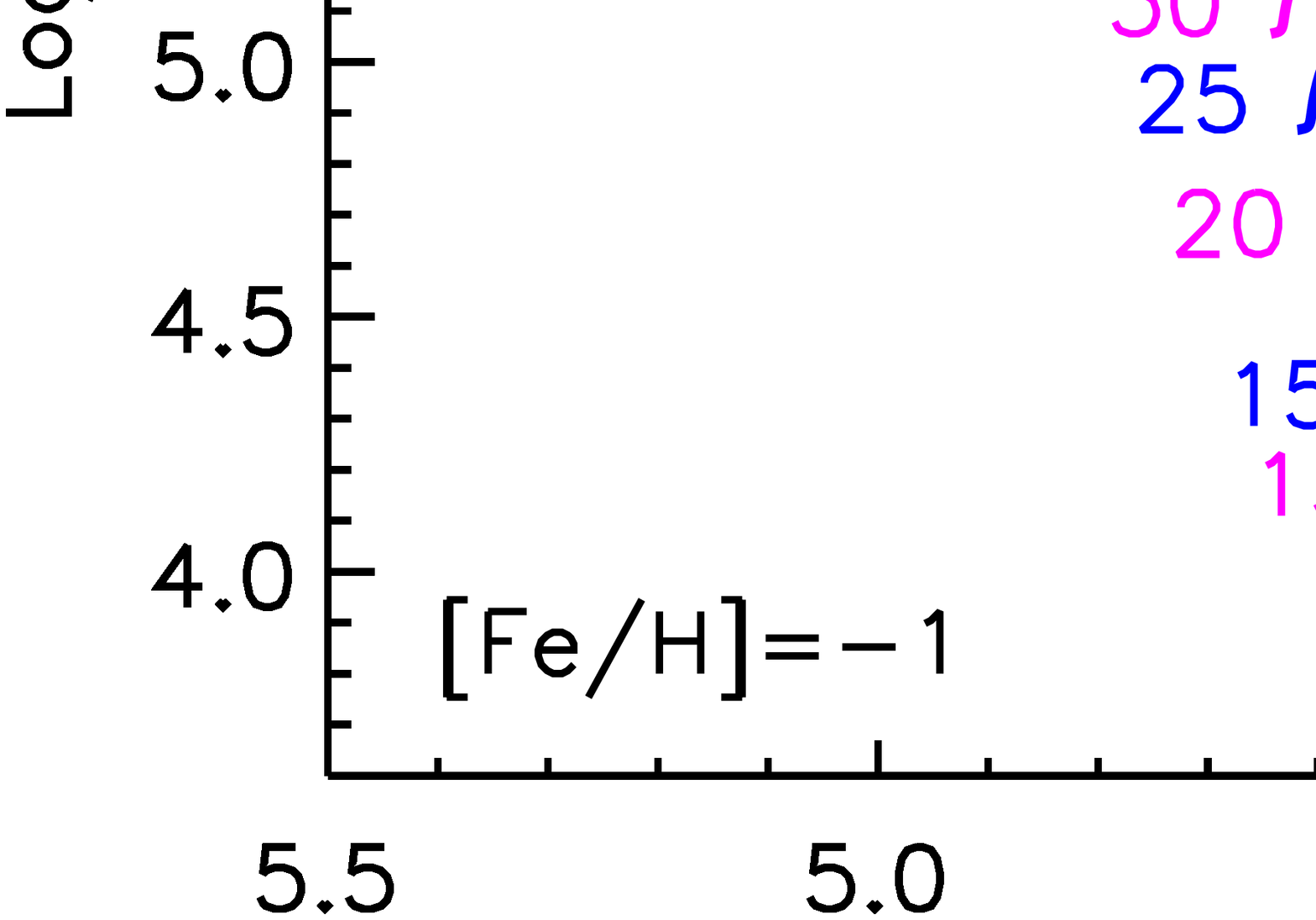}
\subfigures
\caption{Same as Figure \ref{hrheburninga000} but for models with initial metallicity [Fe/H]=-1.}
\label{hrheburningb000}       
\end{figure}

\begin{figure}[h]
\samenumber
\centering
\includegraphics[scale=.25]{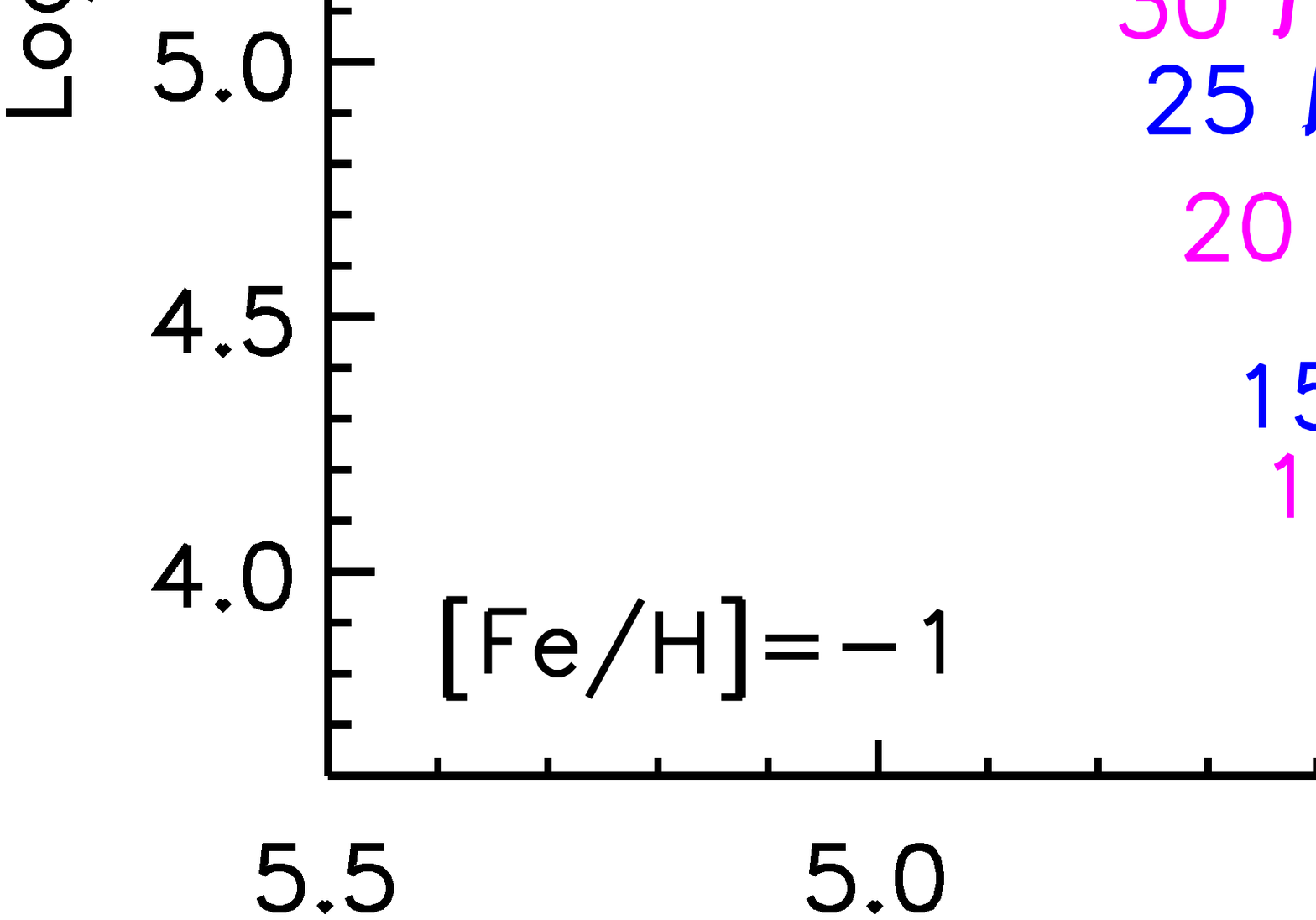}
\subfigures
\caption{Same as Figure \ref{hrheburningb000} but for models with initial rotation velocity $\rm v=150~km/s$.}
\label{hrheburningb150}       
\end{figure}

\begin{figure}[h]
\samenumber
\centering
\includegraphics[scale=.25]{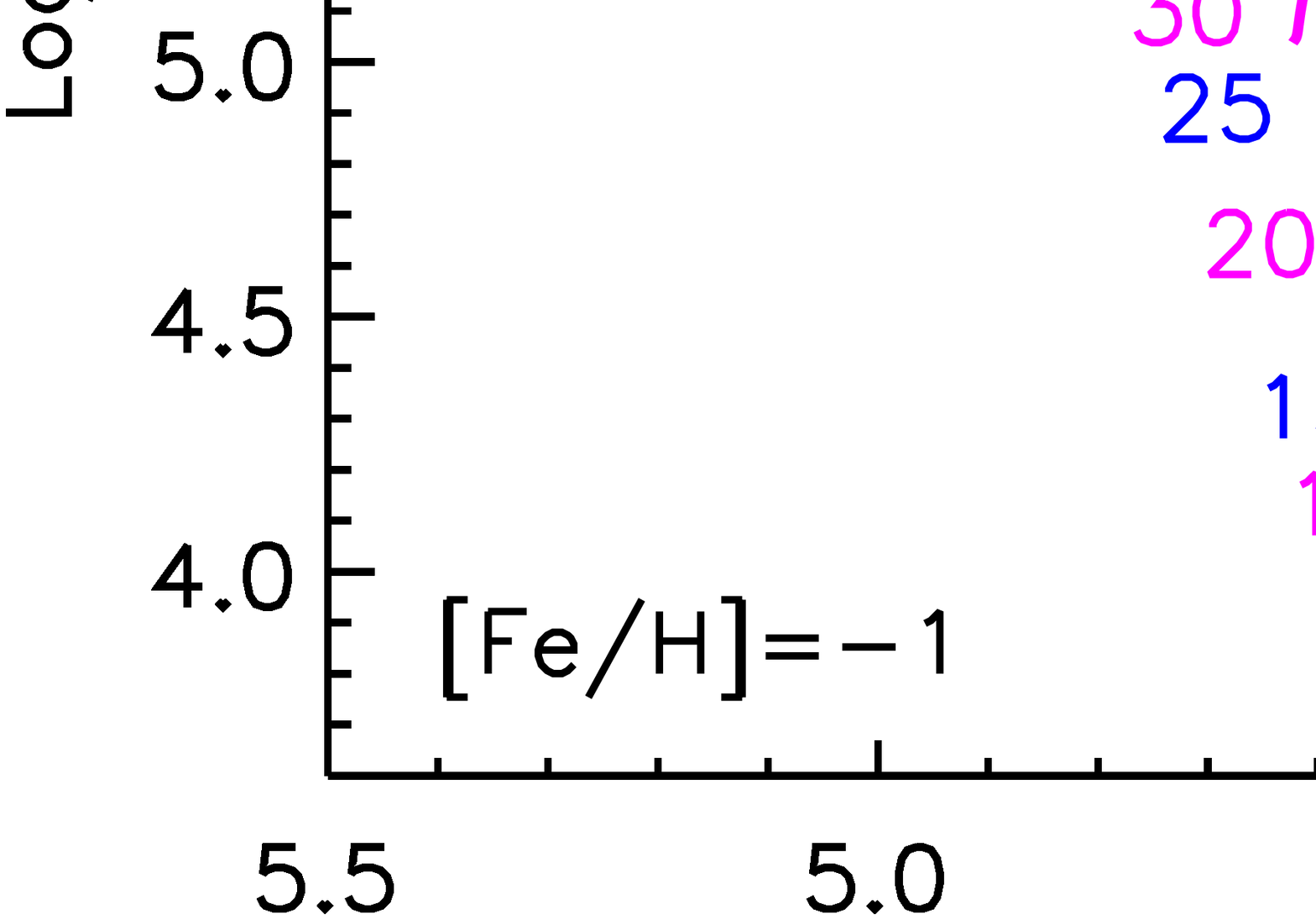}
\subfigures
\caption{Same as Figure \ref{hrheburningb000} but for models with initial rotation velocity $\rm v=300~km/s$.}
\label{hrheburningb300}       
\end{figure}

\begin{figure}[h]
\samenumber
\centering
\includegraphics[scale=.25]{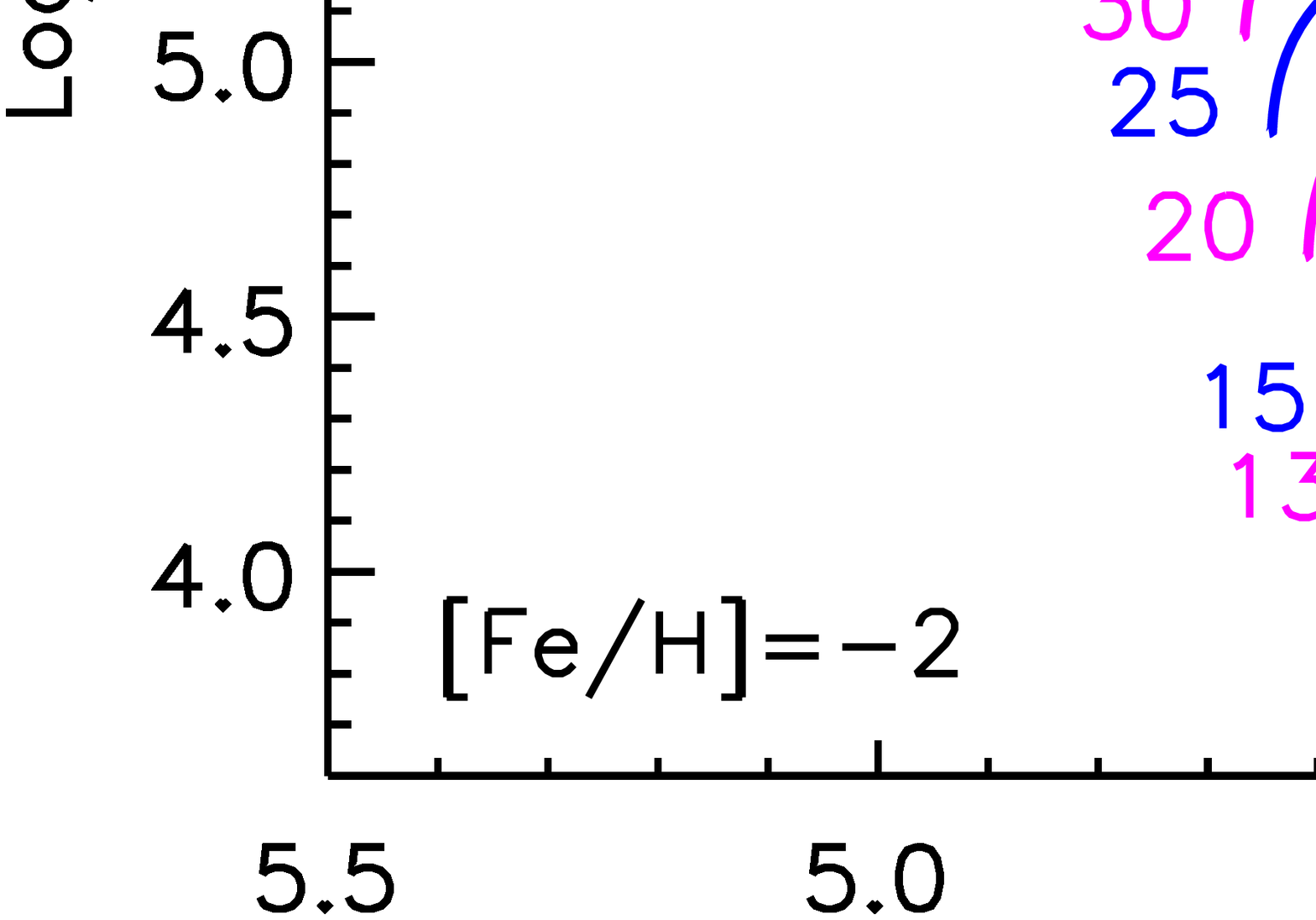}
\subfigures
\caption{Same as Figure \ref{hrheburninga000} but for models with initial metallicity [Fe/H]=-2.}
\label{hrheburningc000}       
\end{figure}

\begin{figure}[h]
\samenumber
\centering
\includegraphics[scale=.25]{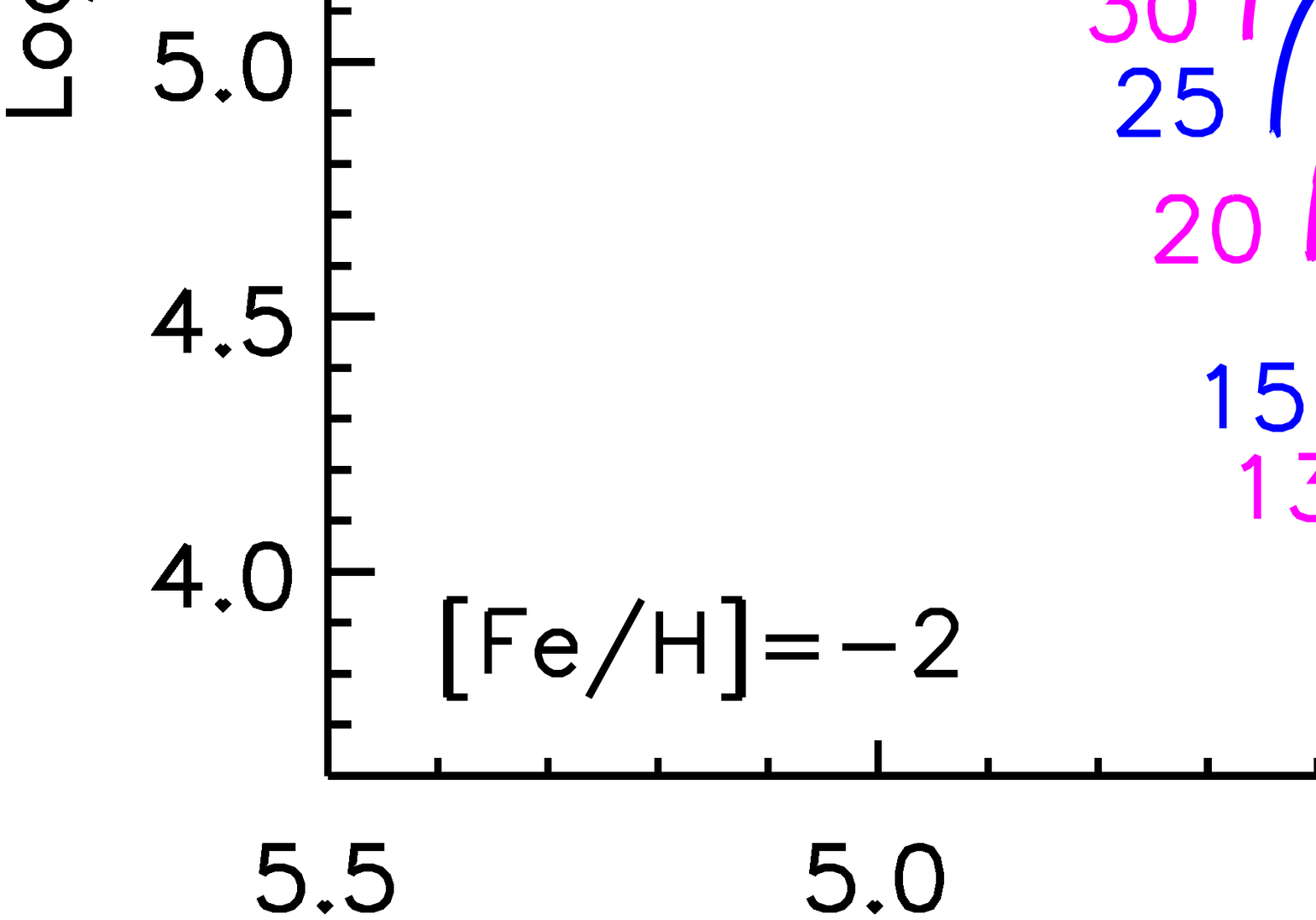}
\subfigures
\caption{Same as Figure \ref{hrheburningc000} but for models with initial rotation velocity $\rm v=150~km/s$.}
\label{hrheburningc150}       
\end{figure}

\begin{figure}[h]
\samenumber
\centering
\includegraphics[scale=.25]{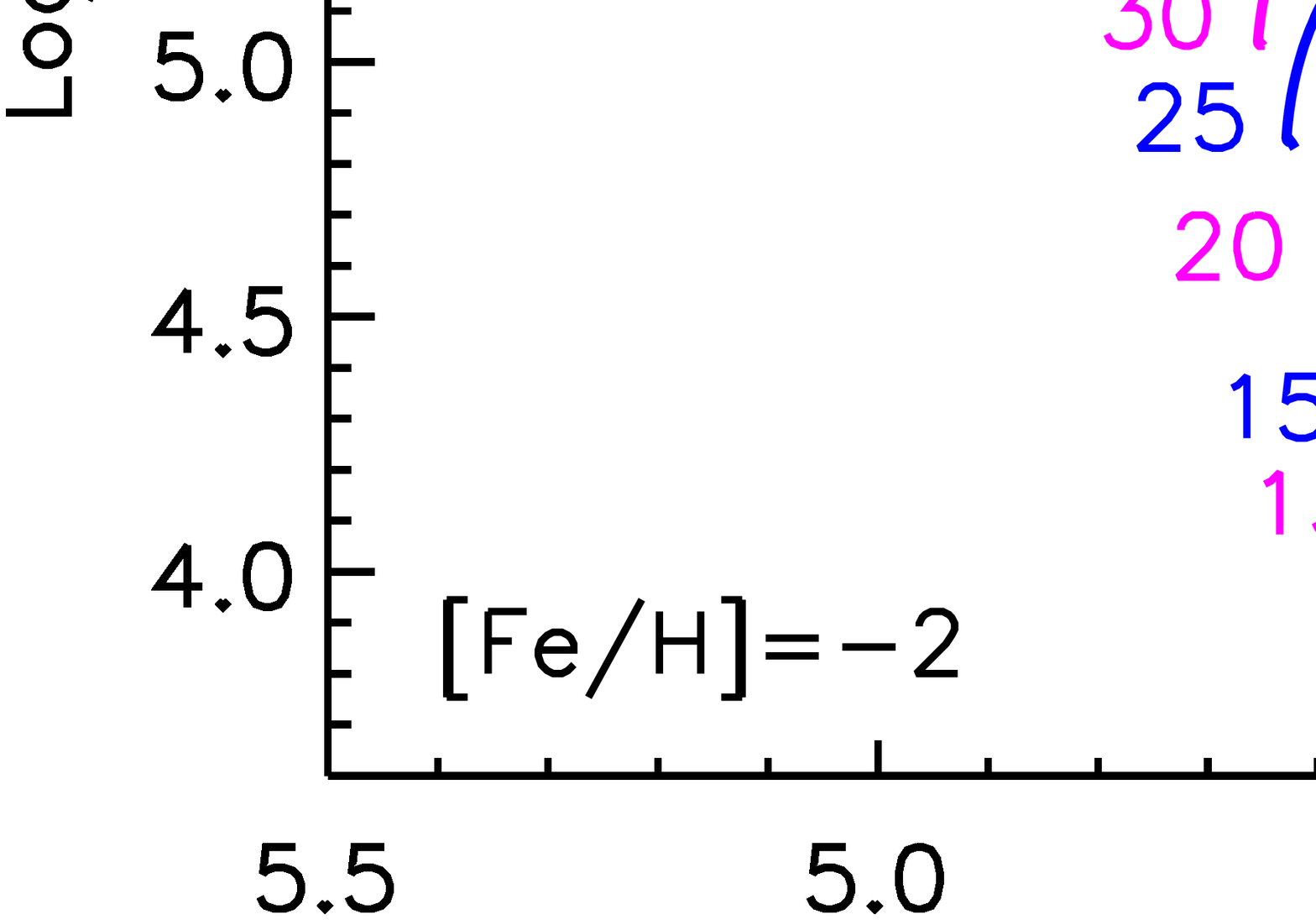}
\subfigures
\caption{Same as Figure \ref{hrheburningc000} but for models with initial rotation velocity $\rm v=300~km/s$.}
\label{hrheburningc300}       
\end{figure}

\begin{figure}[h]
\samenumber
\centering
\includegraphics[scale=.25]{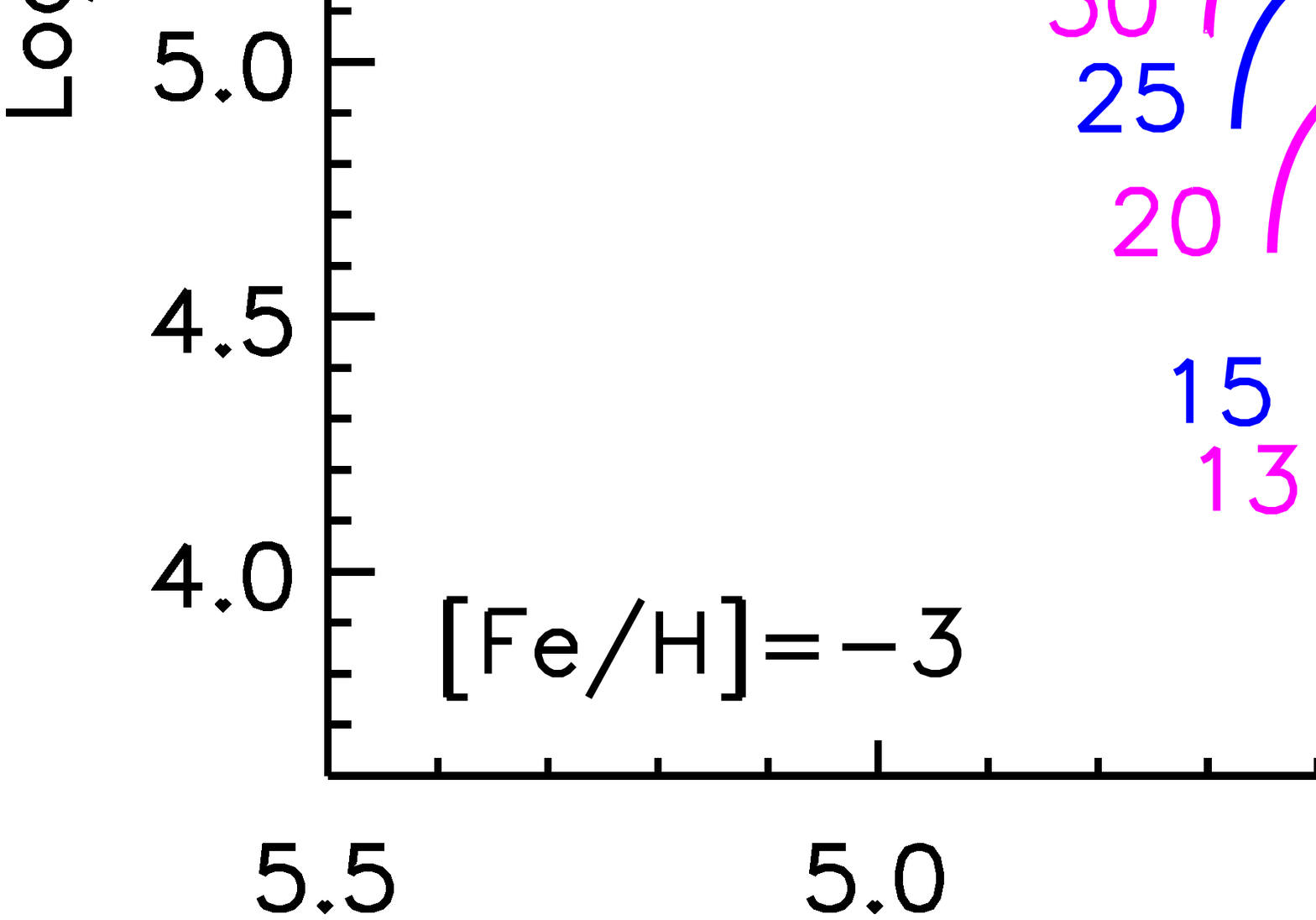}
\subfigures
\caption{Same as Figure \ref{hrheburninga000} but for models with initial metallicity [Fe/H]=-3.}
\label{hrheburningd000}       
\end{figure}

\begin{figure}[h]
\samenumber
\centering
\includegraphics[scale=.25]{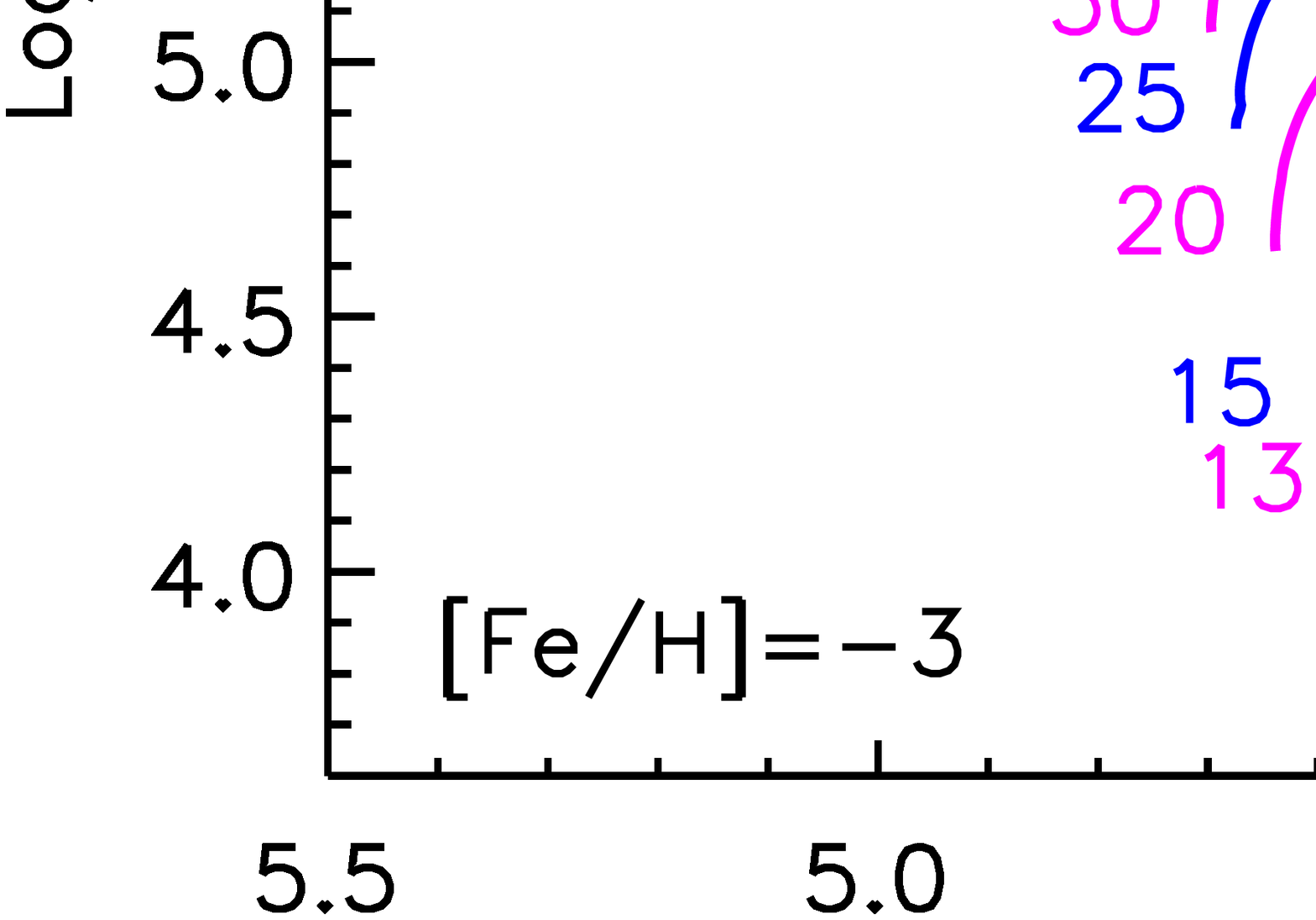}
\subfigures
\caption{Same as Figure \ref{hrheburningd000} but for models with initial rotation velocity $\rm v=150~km/s$.}
\label{hrheburningd150}       
\end{figure}

\begin{figure}[h]
\samenumber
\centering
\includegraphics[scale=.25]{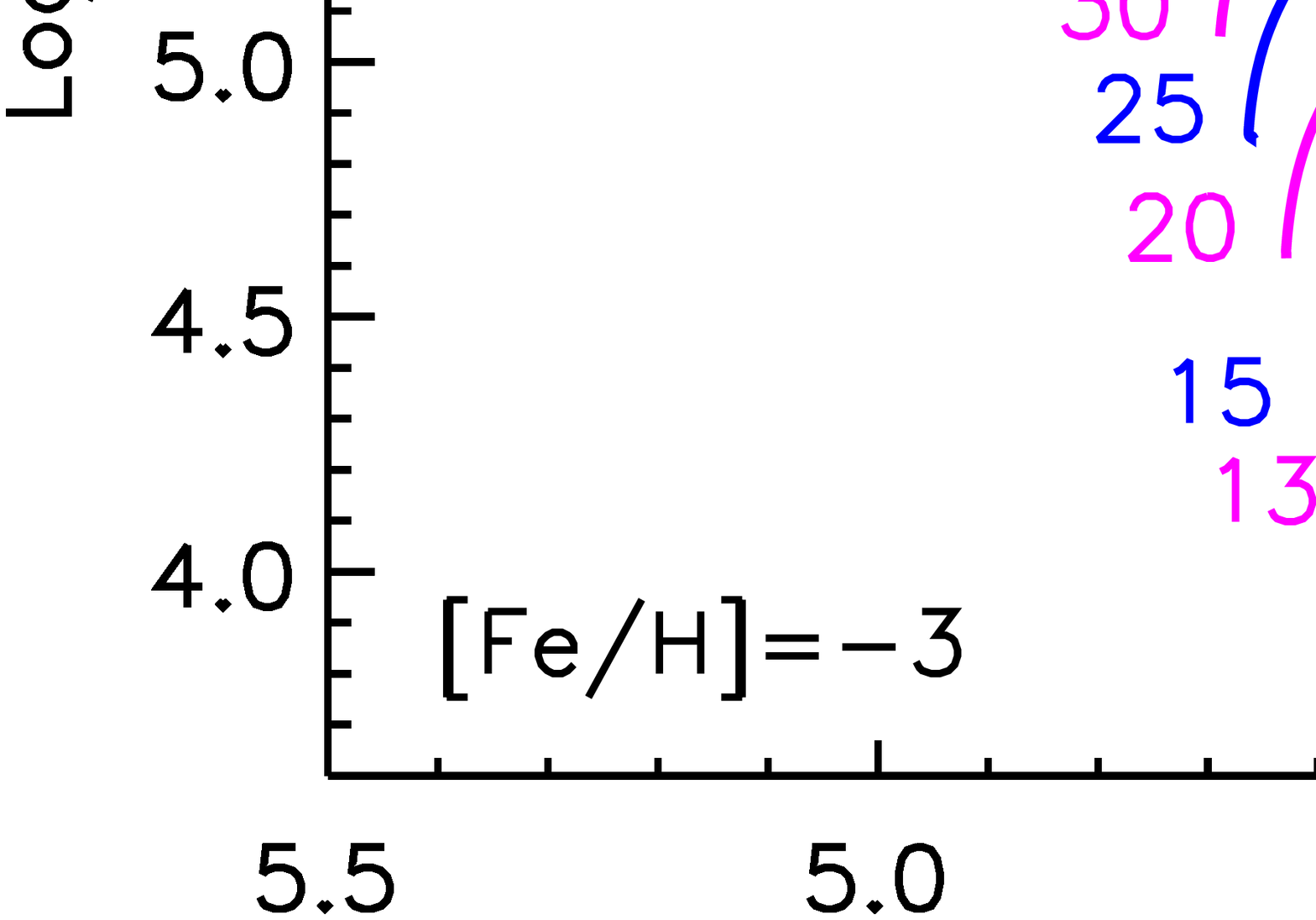}
\subfigures
\caption{Same as Figure \ref{hrheburningd000} but for models with initial rotation velocity $\rm v=300~km/s$.}
\label{hrheburningd300}       
\end{figure}

After core H depletion, the H burning shell progressively shifts outward in mass and progressively increases the mass of the He core.
During this phase all the models move toward the red side of the HR diagram, at constant luminosity, while the He core progressively contracts and heats up until the He burning reactions are activated. The timescale of such a transition cannot be determined with precision on the basis of first principles because it depends on the efficiency of the chemical mixing in the region of variable composition left by the receding convective core during core H burning (Figure \ref{kipphburn15a000}) that, in turn, is still highly uncertain. This region, in fact, becomes unstable according to the Schwarzschild criterion \index{Schwarzschild criterion} and stable according to the Ledoux \index{Ledoux criterion} one, because of the stabilizing effect of the $\mu$ gradient ($\nabla_\mu$). This zone is usually referred to as a "secmiconvective" region \index{semiconvective region} \cite{1958ApJ...128..348S,1970MNRAS.151...65S,1985A&A...145..179L}. If the Schwarzschild criterion for convection is adopted, this zone is mixed very efficiently, i.e. on a dynamical timescale, and the redward excursion in the HR diagram occurs on a nuclear timescale, i.e. during core He burning. As a consequence we would expect, in this case, that the region between the main sequence (MS) and the red giant branch (RGB) would be well populated. On the contrary, if the Ledoux criterion for convection is adopted, the mixing is suppressed and the redward evolution occurs on much faster Kelvin-Helmoltz timescales. In this case we would expect very few stars, i.e. a gap, in the HR diagram between the MS and the RGB. The problem is that the Ledoux criterion is not that robust because $\nabla_\mu$, that stabilizes the zone against convection, may be reduced, and even destroyed, by any kind of stochastic turbulence, favoring the onset of convection that, in turn, would reduce even more $\nabla_\mu$. This is clearly an unstable situation. The physics of this phenomenon has been studied by a number of authors. \cite{1966PASJ...18..374K}, for example, showed that these small perturbations would rise on thermal timescales, because they are due to heat exchange processes, with the net result of mixing eventually the whole region. Nevertheless, at present, the precise values of such timescales as well as the real efficiency of the chemical mixing operating in this region are still unknown. As a consequence the only guidance we have to treat this region comes from the observations, that clearly show the presence of a gap between the MS and the RGB in the color-magnitude diagrams of massive stars populations in the Milky Way and the in the Magellanic Clouds \cite{2003ARA&A..41...15M}. This implies that, at most, a small efficiency of the mixing in the semiconvective region is allowed by the observations. After a proper calibration of the semiconvection, core He burning begins when the star is still a blue supergiant (BSG) \index{blue supergiant} or has become a red supergiant (RSG) \index{red supergiant} depending on the initial mass, initial metallicity and initial rotation velocity. The green stars in Figures \ref{hrheburninga000} to \ref{hrheburningd300},
mark the location in the HR diagram corresponding to the beginning of core He burning for all the models.



At solar metallicity, all the models start core He burning as RSGs (Figures \ref{hrheburninga000}, \ref{hrheburninga150}, \ref{hrheburninga300}) with the exceptions of those stars, i.e. the most massive ones, that become WR (1) already during core H burning or (2) during the redward excursion, before the beginning of core He burning, due to the approaching to their Eddington luminosity. Stars that start core He burning as RSGs become cool enough that the dust driven wind become efficient \cite{2005A&A...438..273V} and a phase of strong mass loss begins. The central He mass fraction corresponding to the onset of the dust driven wind is crucial in determining the amount of mass which is lost during the remaining core He burning and therefore in determining whether the star remains a RSG or become a WR (BSG) star. If the star enters the dust driven wind phase at late stages of core He burning, the remaining core He burning lifetime will allow a very small amount of mass loss and therefore the star will remain a RSG all along the subsequent evolution. On the contrary, if the star enters the dust production regime at an early stage of core He burning, there is enough time for the star to loose a substantial amount of mass favoring the evolution from a RSG to a BSG-WR configuration. The transition mass between these two evolutionary paths depends on the initial rotation velocity. For the set of solar metallicity models discussed in this chapter the minimum mass entering the WR stage is $\rm \sim 17~M_\odot$ in the non rotating case. Such a value decreases to $\rm \sim 13~M_\odot$ for models with initial rotation velocities of $\rm v\geq 150$ km/s. 

The number of RSGs and WRs at core He depletion, decreases progressively as the metallicity reduces due to the dramatic reduction of the mass loss and to the increase of the compactness of the stars. For metallicities corresponding to $\rm [Fe/H]<-1$ all the non rotating stars in the range $\rm 13-120~M_\odot$ skip the redward excursion in the HR diagram and remain BSG during all the core He burning phase. The inclusion of rotation favors, at all the metallicities, a redward evolution for the lower mass models, as well as an approach to the Eddington luminosity for the higher mass ones. Therefore, for these low metallicities, the inclusion of rotation determines an increase of the number of both the RSGs and the WRs at core He depletion - the WR/RGB ratio increasing with increasing the initial rotation velocity. The red stars in Figures \ref{hrheburninga000} to \ref{hrheburningd300},
mark the location of the models at core He depletion and therefore provide the mass intervals for RSGs, BSGs and WRs at core He depletion.

\begin{figure}[htpb]
\centering
\includegraphics[scale=.30]{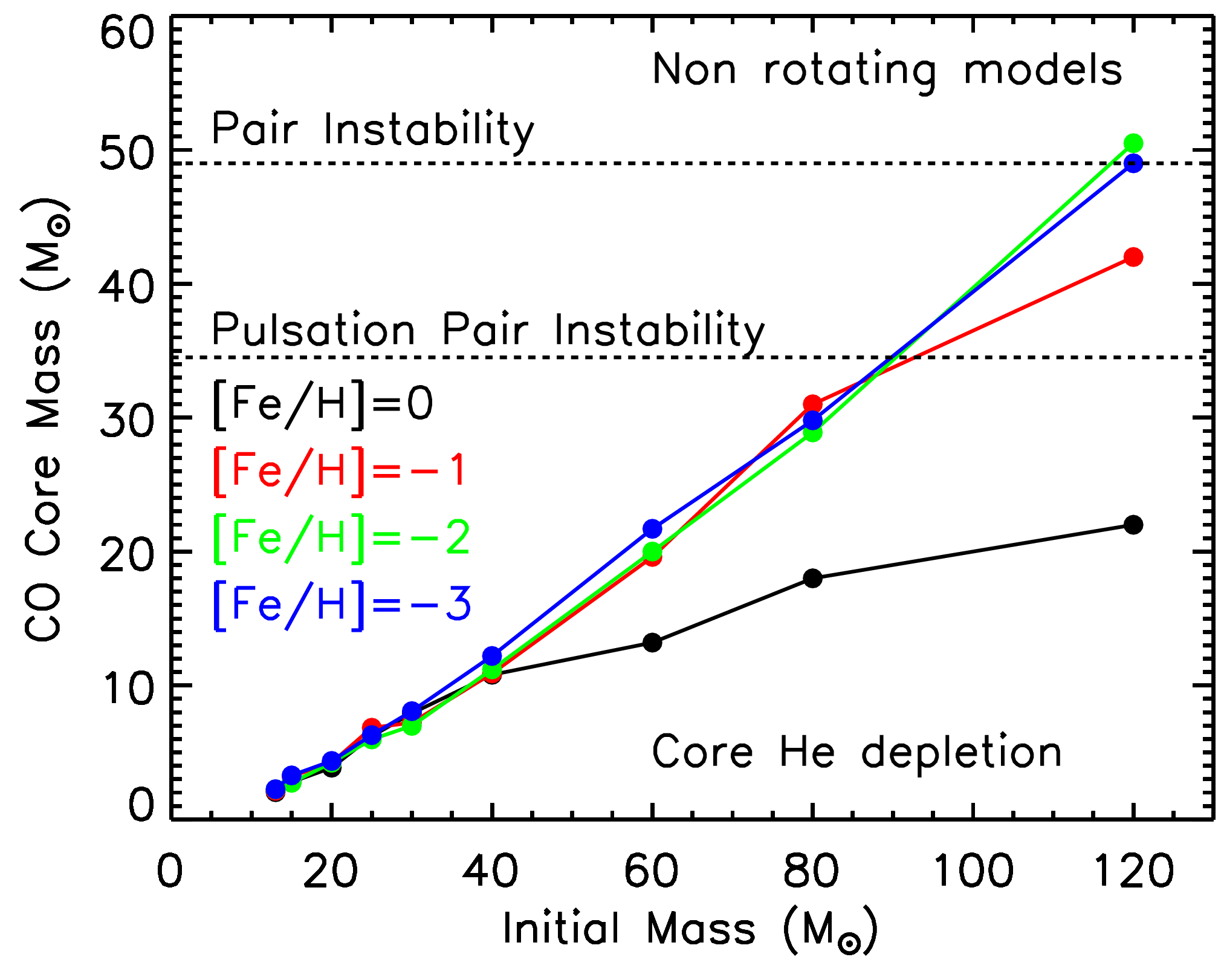}
\caption{CO core mass at core He depletion as a function of the initial mass, for non rotating models. The various metallicities follow the color codes reported in the Figure. Also shown is the CO core mass limit above which pulsation pair instabilities and pair instabilities are expected according to \cite{2002ApJ...567..532H}.}
\label{masscocoreZ}       
\end{figure}

\begin{figure}[htbp]
  \centering
  \includegraphics[scale=.30]{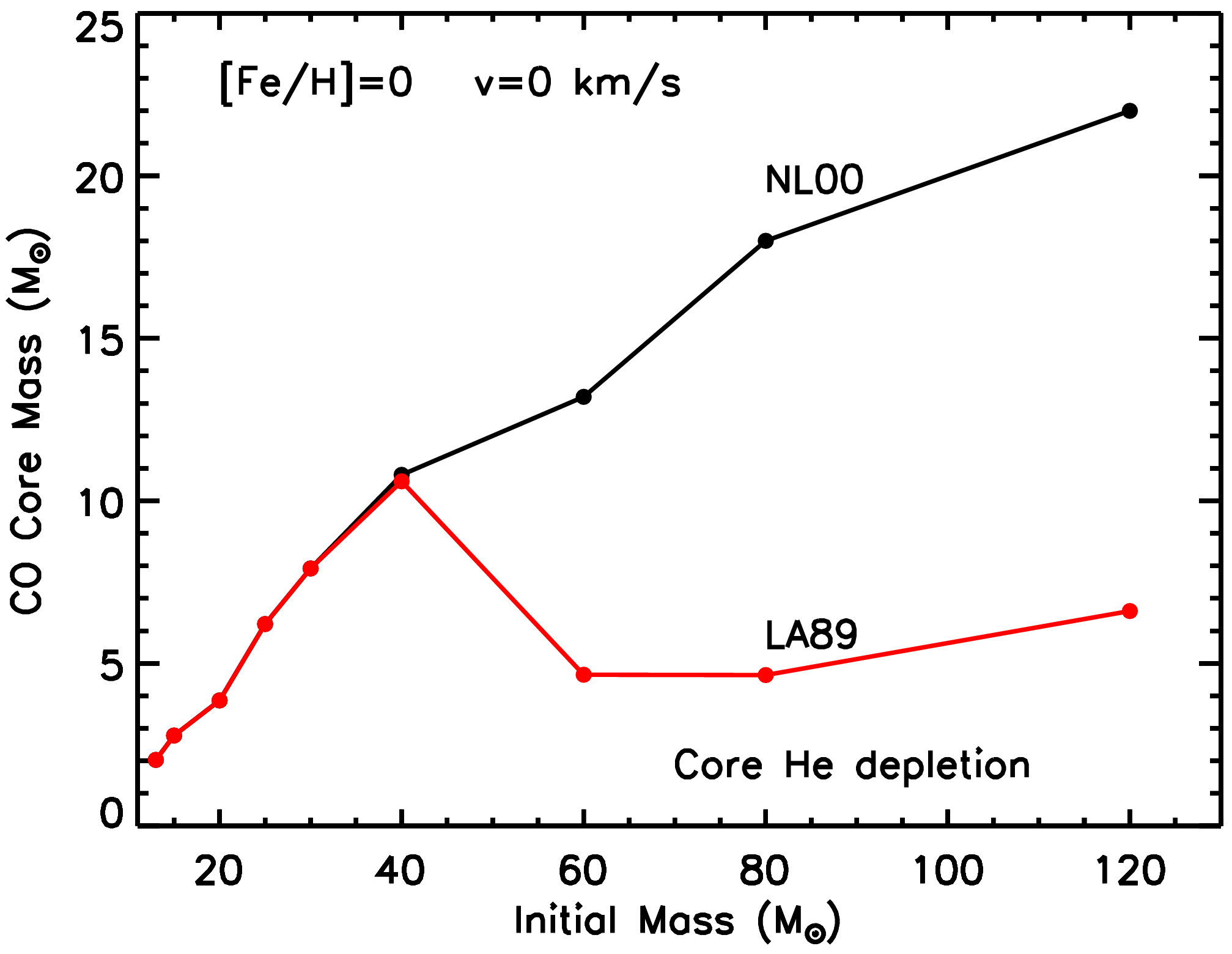}
  \caption{CO core mass at core He depletion as a function of the initial mass, for solar metallicity non rotating models, computed by adopting two different prescriptions of the mass loss rate during the WNE/WC phases, i.e., the one provided by \cite{2000A&A...360..227N} (NL00 - black line) and the one proposed by \cite{1989A&A...220..135L} (LA89 - red line).}
  \label{mconl00la89}       
\end{figure}

Core He burning lasts $\rm \sim 10^6-10^5~yr$, in the mass range 13-120 $\rm M_\odot$. It 
occurs in a convective core whose mass size in general increases or, at most, remains constant; then it vanishes when the central He mass fraction drops below $\sim 10^{-8}$. Such a behavior is typical of the stars for which the mass loss in not strong enough to uncover the He core and eventually to reduce its mass (as already mentioned above, Figures \ref{hrheburninga000} to \ref{hrheburningd300} provides the minimum mass entering this phase as a function of the initial mass, initial metallicity and initial rotation velocity). 
Stars for which the mass loss is high enough ($\rm \dot{M}\sim 10^{-5}-10^{-4}~M_\odot/yr$) to progressively reduce the He core, on the contrary, enter the WNE \index{WNE} stage and eventually, if the total mass is reduced below the mass coordinate corresponding to the maximum extension of the He convective core, may become WC \index{WC} stars. In these stars, the evolutionary properties during the remaining core He burning stage is mainly driven by the actual size of the He core. In particular, as the He core progressively reduces in mass because of the mass loss, the star tends to behave as a star of a lower mass (i.e., a star having the same actual He core), essentially by reducing its central temperature. Such an occurrence has the following effects: (1) the He convective core shrinks progressively in mass and leaves a region of variable chemical composition; (2) the core He burning lifetime increases; (3) the total luminosity progressively decreases, i.e., the star in the HR diagram moves downward; (4) the $\rm ^{12}C$ mass fraction at core He exhaustion becomes larger than it would be without mass loss; (5) the CO core at core He exhaustion is smaller than it would be without mass loss and resembles that of a star having a similar final He core mass (regardless on the initial mass of the star). 
In the present set of non rotating models, the CO core mass, in general, increases with the initial mass at all the metallicities (Figure \ref{masscocoreZ}). 
The sensitivity of the CO core mass on the initial metallicity is negligible for stars with initial mass $\rm M_{ini} \leq 40~M_\odot$ (as for the He core, Figure \ref{masshecore2}). Above this limit, on the contrary, the $\rm M_{CO}-M_{ini}$ relation is shallower at solar metallicity compared to the corresponding ones at lower metallicities. The reason for this different behavior is that solar metallicity models with $\rm M_{ini} > 40~M_\odot$ become WNE/WC stars and therefore their He cores are significantly reduced by the mass loss - this limit the increase of the CO core; on the contrary, low metallicity models evolve during this phase essentially at constant mass because of the highly reduced efficiency of the mass loss and therefore none of them become a WNE/WC star. One of the consequence of such an occurrence is that, in the most massive low metallicity non rotating models, the CO core mass may increase even above the limit for the onset of the pair instability \index{pair instability} \cite{2002ApJ...567..532H}, as in the case of the $\rm 120~M_\odot$ models with $\rm [Fe/H] \leq -1$. \cite{2002ApJ...567..532H} identify two different outcomes driven by the pair instabilities, i.e. a black hole formation, in the case of the pulsation pair instabilities occurring in stars with He cores in the range $\rm \sim 40-63~M_\odot$, and the complete disruption of the star, in the case of the pair instabilities occurring in stars with He cores more massive than $\rm \sim 63~M_\odot$ (i.e. the so called pair instability supernovae). According to these He cores limits, the two $\rm 120~M_\odot$ models with $\rm [Fe/H] \leq -2$ enter the pair instability regime, while the $\rm 120~M_\odot$ model with $\rm [Fe/H] = -1$ undergoes pulsation pair instabilities (Figure \ref{masscocoreZ}). Let us however remember that we cannot follow the evolution after the onset of these instabilities, therefore we cannot identify with precision of certainty the evolutionary paths of these stars. The $\rm M_{CO}-M_{ini}$ relation, however, is highly sensitive to the mass loss rate during the WNE/WC stages. As an example, Figure \ref{mconl00la89} shows the CO core mass at core He exhaustion as a function of the initial mass, for solar metallicity non rotating models computed adopting two different prescriptions of the mass loss rate during the WNE/WCO stages, i.e., the one provided by \cite{2000A&A...360..227N} (NL00, that we consider the reference one in this paper), and the one proposed by \cite{1989A&A...220..135L} (LA89) and widely adopted in the past. 
This older mass loss rate is significantly higher than the NL00 one (by $\rm \sim 0.2-0.6~dex$) and it scales as $\rm M^{2.5}$ (where $\rm M$ is the actual mass). In this way all the very massive star models tend to converge toward a similar, quite small, mass. Since this strong reduction of the total mass occurs essentially during core He burning, all the LA89 models will develop a quite similar, low mass, CO core. Figure \ref{mconl00la89} clearly shows that while in the NL00 case the CO core mass preserves a clear trend with the initial mass, in the case of the LA89 all the models develop a significantly smaller CO core, i.e., similar to that of the lower mass models. For example, in this case, a $\rm 60~M_\odot$ develops a CO core similar to that of the $\rm 20~M_\odot$. Since the evolution of a massive star after core He burning is mainly driven by both the CO core mass and its chemical composition (see below), in this case a $\rm 60~M_\odot$ star will behave, in the following evolution, as a $\rm 20~M_\odot$ star. It is clear, therefore, that the mass loss during the WNE/WC stages is fundamental in determining the evolutionary properties of these stars during the more advanced burning stages and also their final fate (see below).
 
In rotating models the CO core mass determination, at core He exhaustion, is complicated by the presence of the rotation driven mixing. As in the core H burning, during core He burning the interplay among convection, meridional circulation and turbulent shear (the last two mechanisms operating in the radiative zone above the He convective core) determines both the angular momentum transport and the mixing of the chemicals. The last phenomenon has essentially the following effects: (1) the increase of the CO core; (2) the reduction of the $\rm ^{12}C$ mass fraction at core He exhaustion; (3) the continuous diffusion of core He burning products, mainly $\rm ^{12}C$, up to the base of the H burning shell that activates a primary $\rm ^{14}N$ production \cite{2007A&A...461..571H,2013ApJ...764...21C}. The importance of these effects, however, is not the same in all the models of the present grid but, on the contrary, it reaches its maximum for the lowest mass models and reduces progressively as the initial mass increases, for any given initial metallicity.  The reason is partly due to the fact that in a grid of models having the same initial equatorial rotation velocity, the ratio $\rm \omega/\omega_{crit}$ [$\omega_{crit}=(2/3)^{3/2}(GM/R_{pole}^3)^{(1/2)})$ in the framework of the Roche model \index{Roche model}] decreases as the initial mass increases, therefore the effects of rotation become progressively less important in the more massive stars. In addition to that, it must be remembered that the evolutionary timescales decrease with the initial mass, and therefore the larger the mass, the faster the evolution, and hence the smaller the timescale over which the rotation driven secular instabilities may operate.

\begin{figure}[htbp]
  \centering
  \includegraphics[scale=.14]{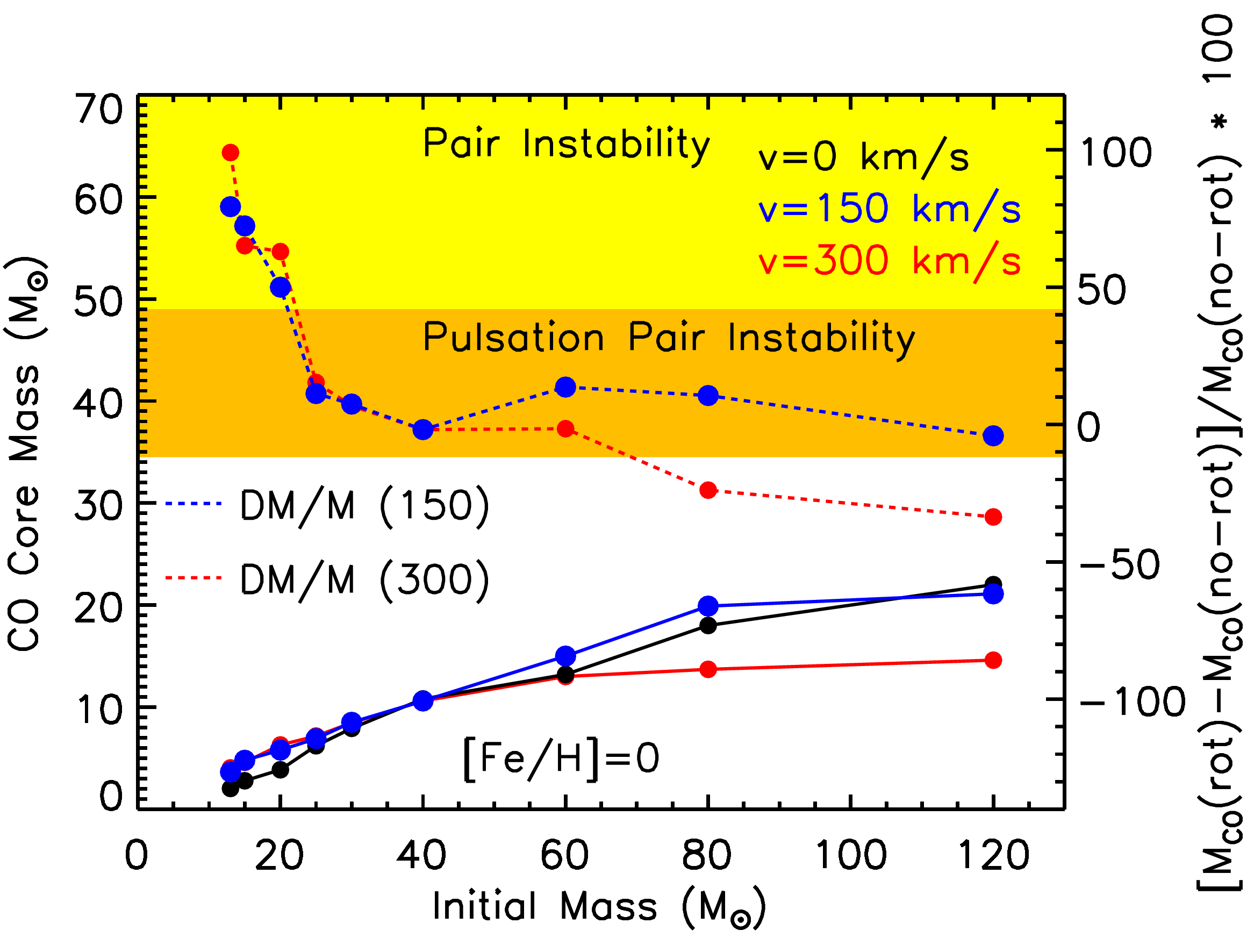}%
  \includegraphics[scale=.14]{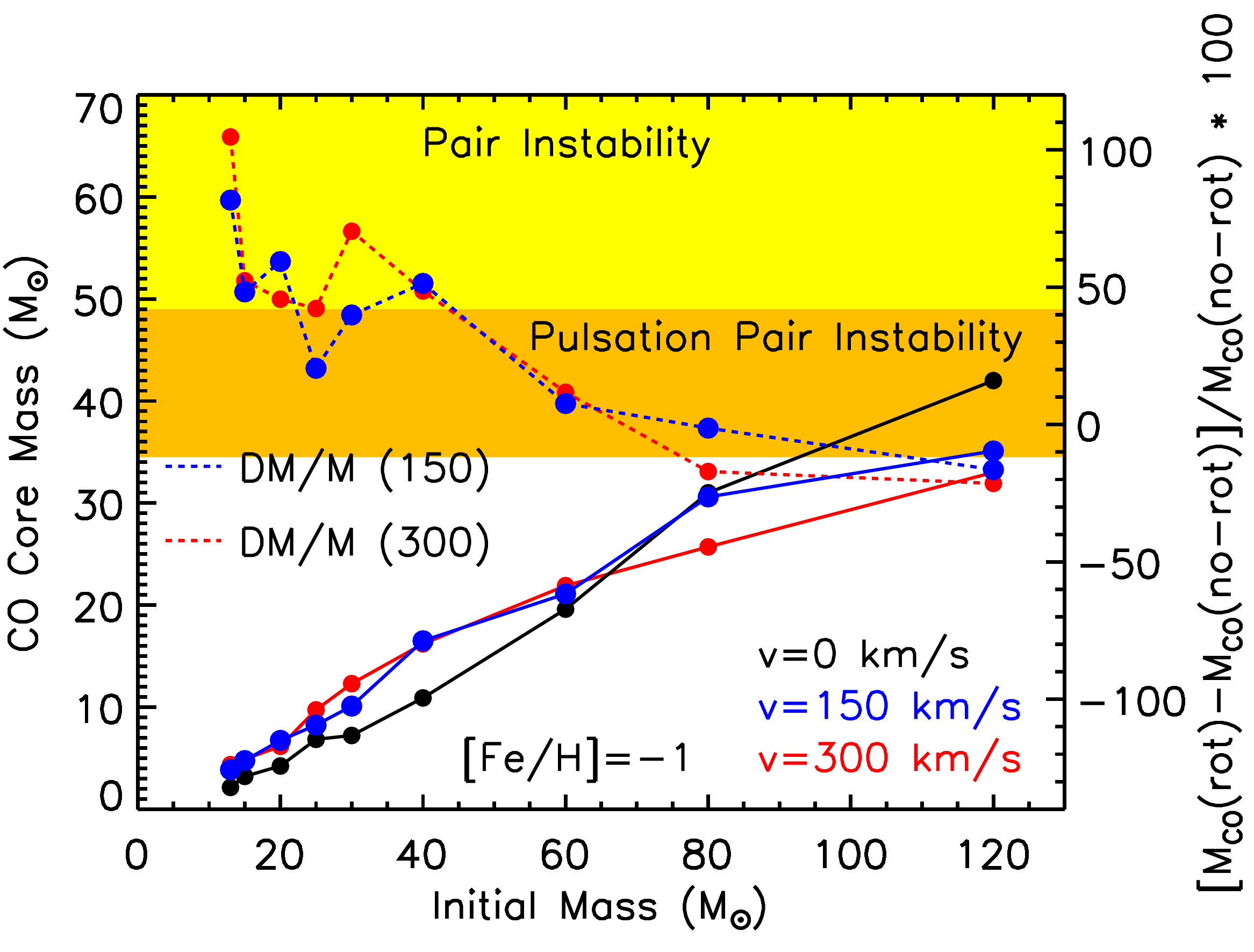}
  \includegraphics[scale=.14]{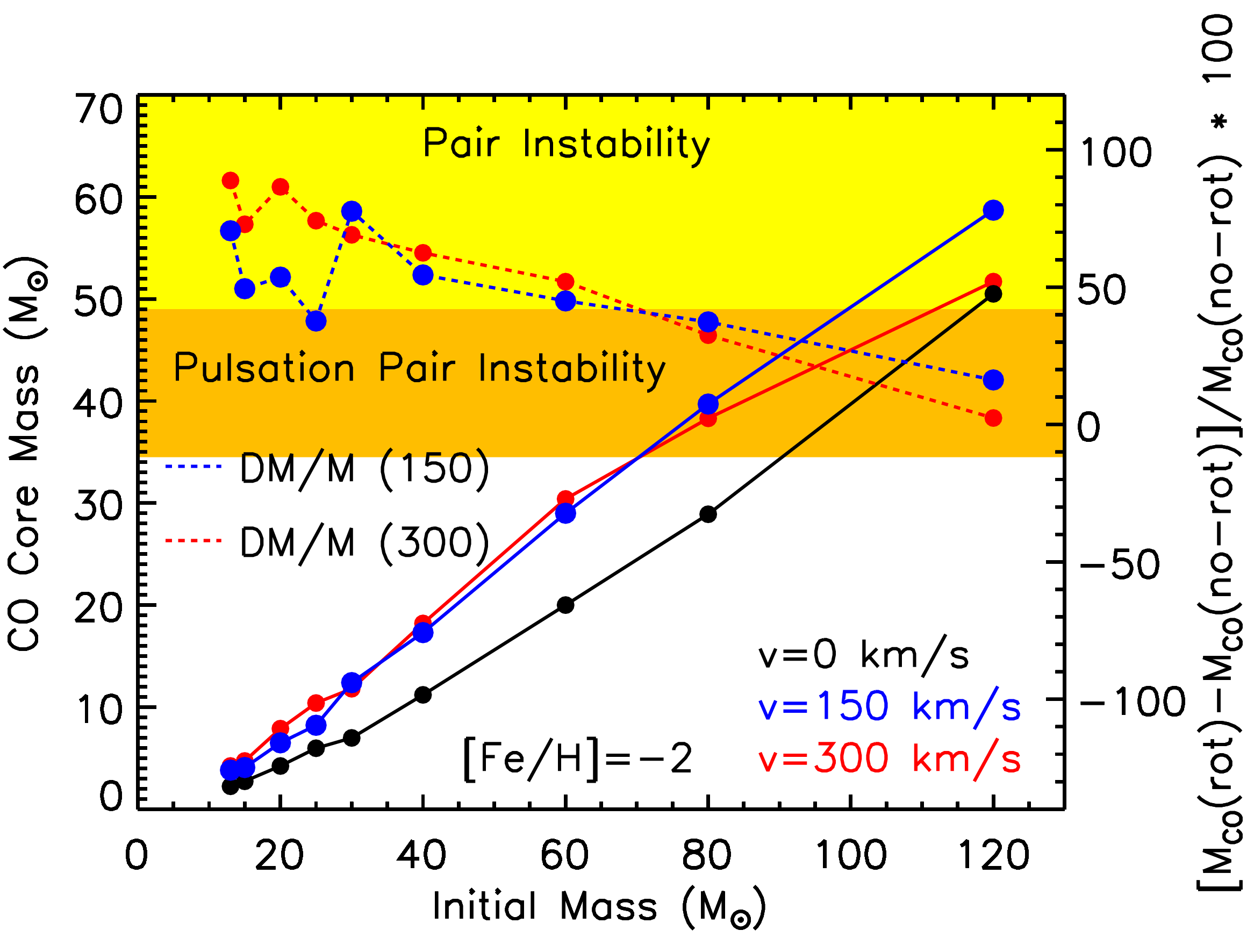}
  \includegraphics[scale=.14]{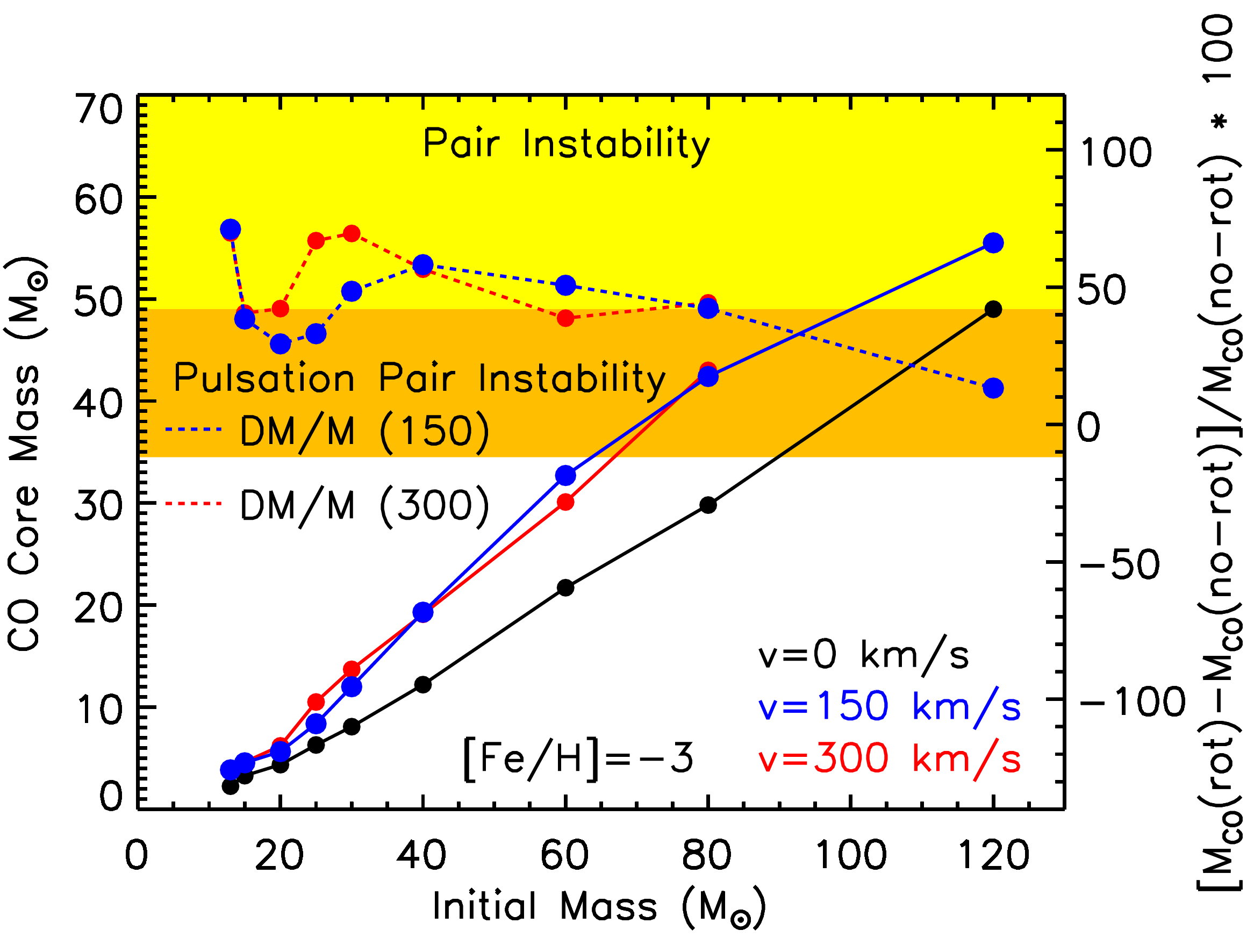}  
  \caption{CO core mass at core He depletion as a function of the initial mass for non rotating (black solid lines) and rotating models (blue solid lines refer to model with $\rm v_{ini}=150~km/s$, red solid lines to models with $\rm v_{ini}=300~km/s$). The models with metallicities [Fe/H]=0, -1, -2 and -3 are shown in the upper left, upper right, lower left and lower right panels, respectively. Also shown in the figures, and reported in the left $y$-axis, is the percentage variation (DM/M) of the CO core between rotating and non rotating models (blue and red dot lines refer to models with $\rm v_{ini}=150~km/s$ and $\rm v_{ini}=300~km/s$, respectively). The orange and yellow zones mark the regions corresponding to CO core mass values for which we expect pulsation pair instabilities and pair instabilities according to \cite{2002ApJ...567..532H}, respectively.}
  \label{masscocoreperc}       
\end{figure}

The net result of the interplay between mass loss and rotation is shown in Figure \ref{masscocoreperc}.
As it is expected, in general, the increase of the CO core mass associated to a given initial rotation velocity, reduces progressively with the initial mass, for any given initial metallicity. The importance of this effect, however, increases with increasing the metallicity because of the increase of mass loss. At solar metallicity, for example, the mass loss is high enough in stars with initial mass $\rm M\geq 40~M_\odot$ to prevent the increase, or even reduce, the size of the CO core with respect to the corresponding non rotating models (upper left panel of Figure \ref{masscocoreperc}). 
As the metallicity decreases, the mass loss decreases as well and therefore the increase of the CO core due to rotation in the more massive stars is not limited anymore. For this reason the most massive ($\rm M_{ini}\gtrsim 60~M_\odot$) rotating models of low metallicities may develop CO cores large enough to enter either the pulsation pair instability or the pair instability regime \cite{2002ApJ...567..532H} (orange and yellow area in Figure \ref{masscocoreperc}, respectively).

The evolution of the angular momentum during core He burning depends, in general, on the structure of the star, i.e., if the star is a RSG or if it is a BSG. In a RSG configuration the star has a He convective core surrounded by a deep convective envelope, these two zones being separated by a thin radiative region. On the contrary, in a BSG supergiant configuration the convective envelope is lacking.
Since in the convective zones the angular momentum transport efficiency is maximal and since during this phase the star loses a substantial amount of mass, the angular momentum loss during this stage depends mainly on the efficiency of the angular momentum transport in the radiative zone and on the presence of the convective envelope. 
\begin{figure}[htbp]
\centering
\includegraphics[scale=.28]{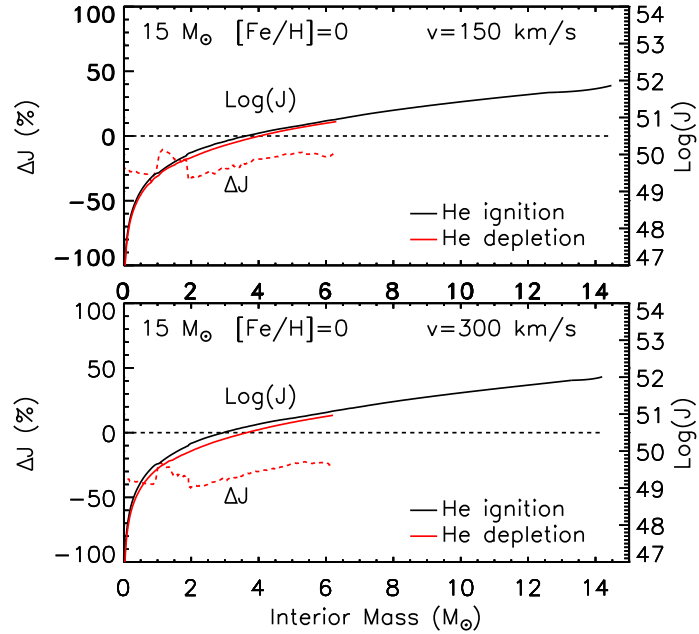}
\subfigures
\caption{Angular momentum as a function of the enclosed mass for a 15 $\rm M_\odot$ model with initial metallicity [Fe/H]=0 and with initial rotation velocities $\rm v=150~km/s$ (upper panel) and $\rm v=300~km/s$ (lower panel). The black and red solid lines refer to the stage corresponding to the core He ignition and core He depletion, respectively (their scale is on the right $y$-axis). In each plot it is also reported the percentage variation of the angular momentum between the core He ignition and core He depletion (red dotted lines - left $y$-axis).}
\label{momangheburna15}       
\end{figure}
\begin{figure}[htbp]
\samenumber
\centering
\includegraphics[scale=.28]{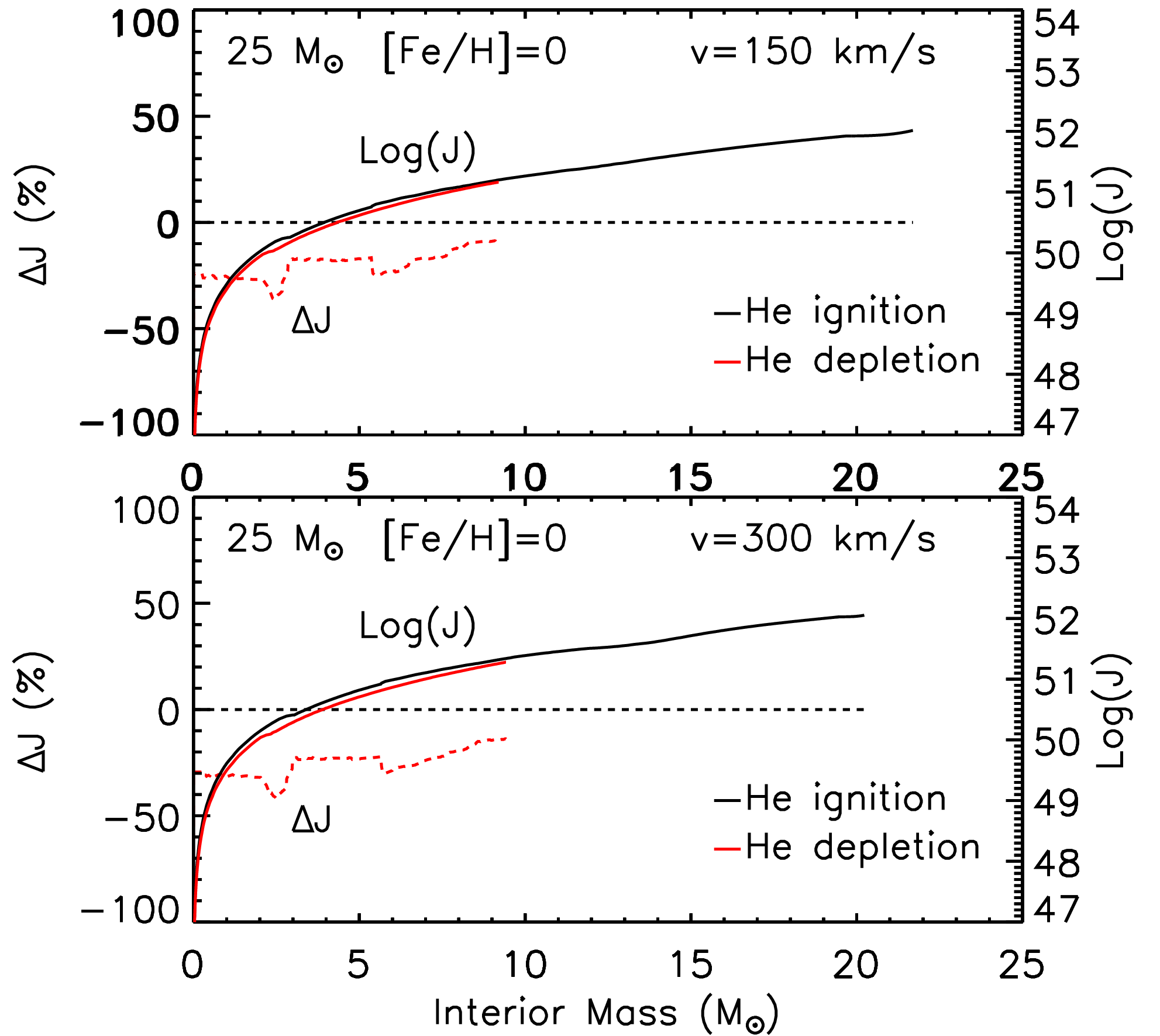}
\subfigures
\caption{Same as Figure \ref{momangheburna15} but for a $\rm 25~M_\odot$ model.}
\label{momangheburna25}       
\end{figure}
\begin{figure}[htbp]
\samenumber
\centering
\includegraphics[scale=.28]{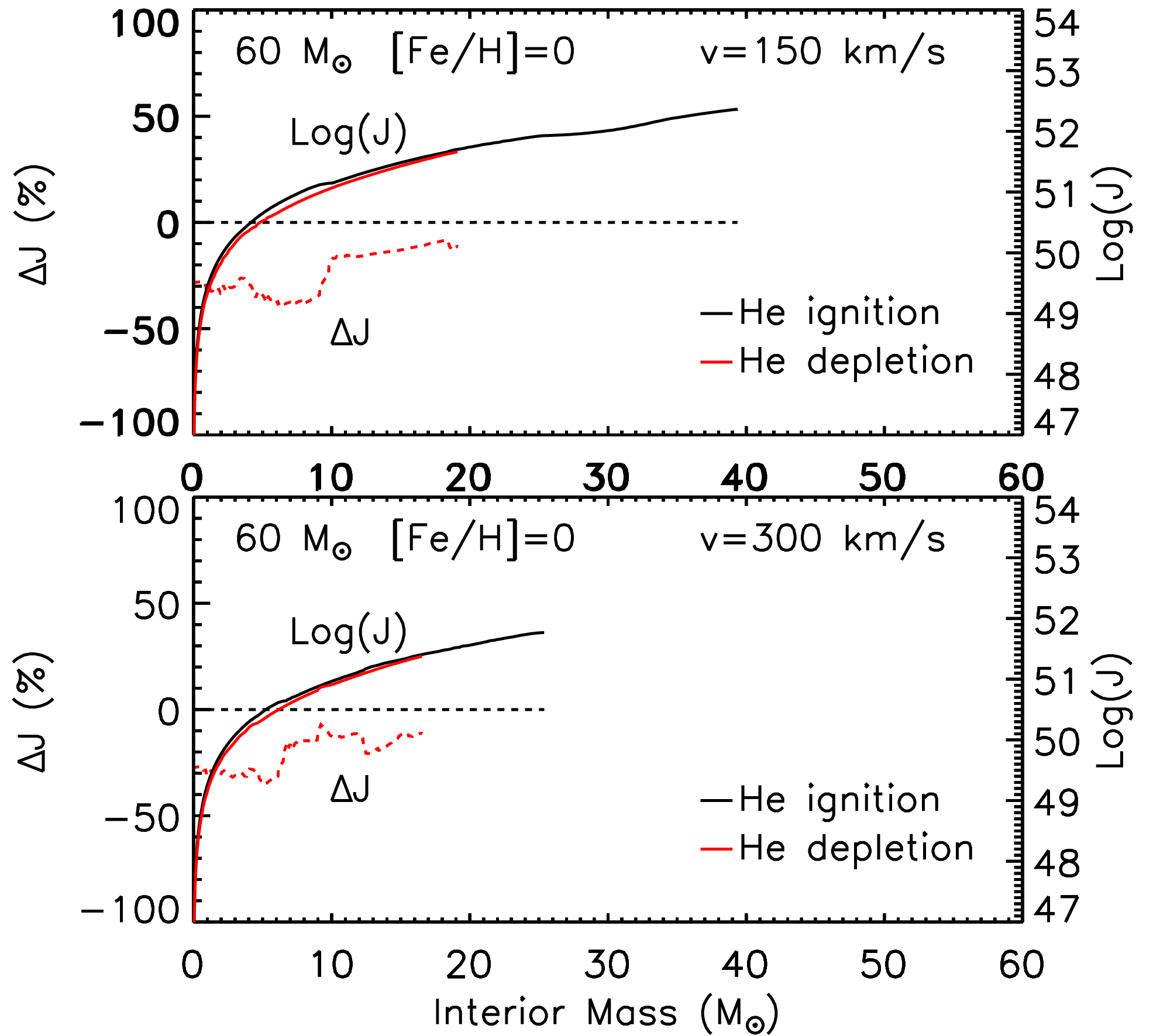}
\subfigures
\caption{Same as Figure \ref{momangheburna15} but for a $\rm 60~M_\odot$ model.}
\label{momangheburna60}       
\end{figure}
\begin{figure}[htbp]
\samenumber
\centering
\includegraphics[scale=.28]{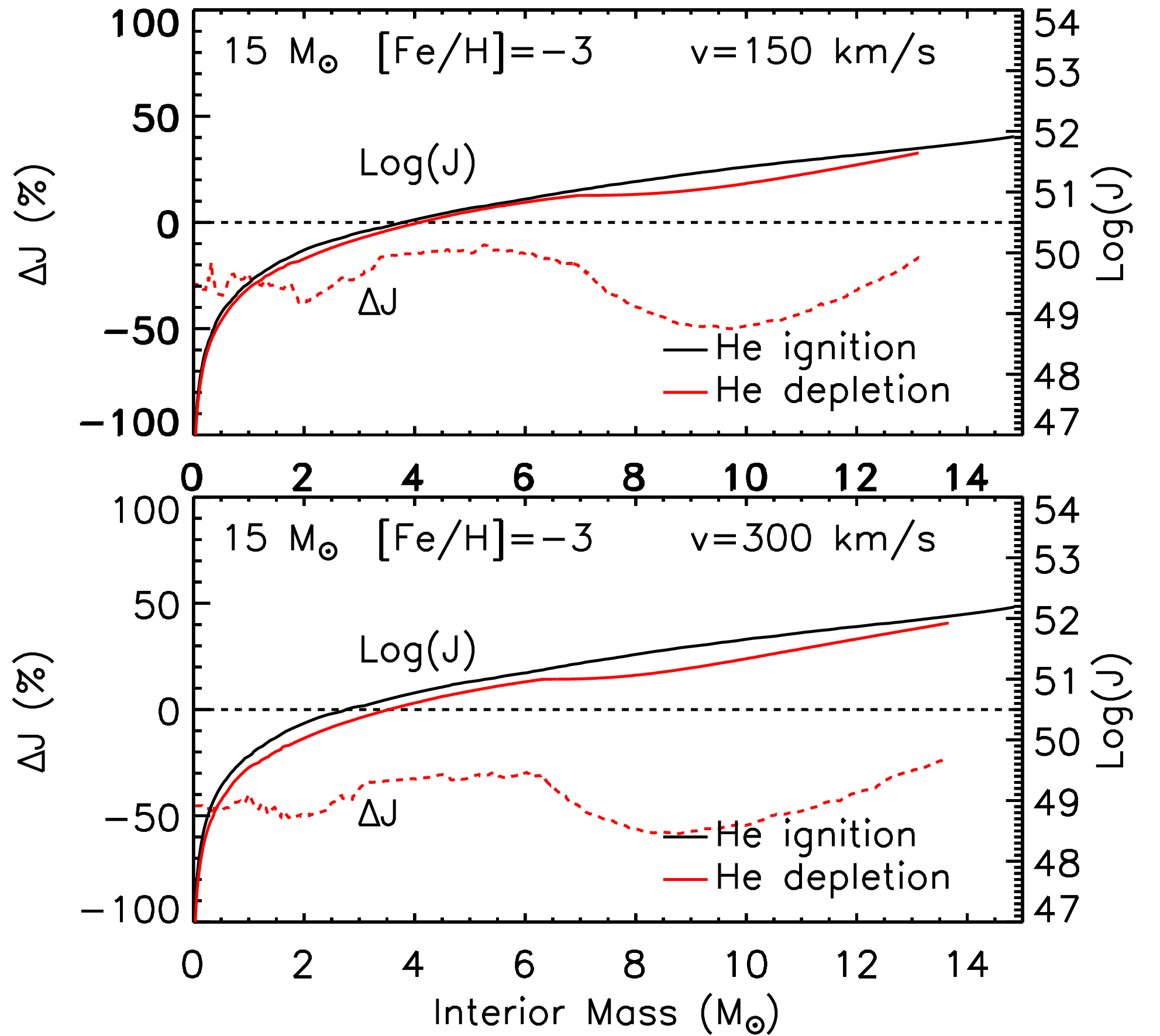}
\subfigures
\caption{Same as Figure \ref{momangheburna15} but for a model with initial metallicity [Fe/H]=-3.}
\label{momangheburnd15}       
\end{figure}
\begin{figure}[htbp]
\samenumber
\centering
\includegraphics[scale=.28]{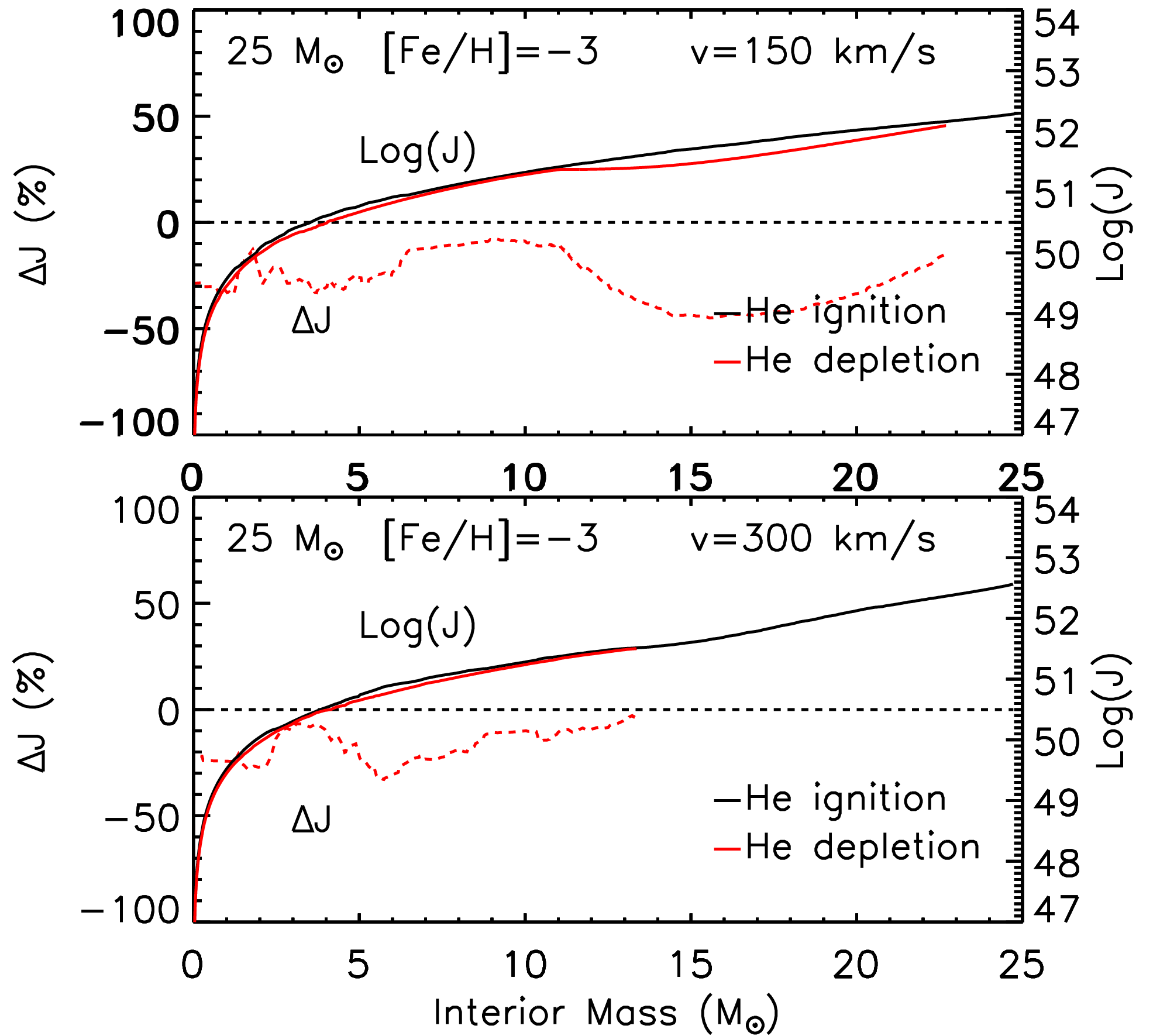}
\subfigures
\caption{Same as Figure \ref{momangheburnd15} but for a $\rm 25~M_\odot$ model.}
\label{momangheburnd25}       
\end{figure}
\begin{figure}[htbp]
\samenumber
\centering
\includegraphics[scale=.28]{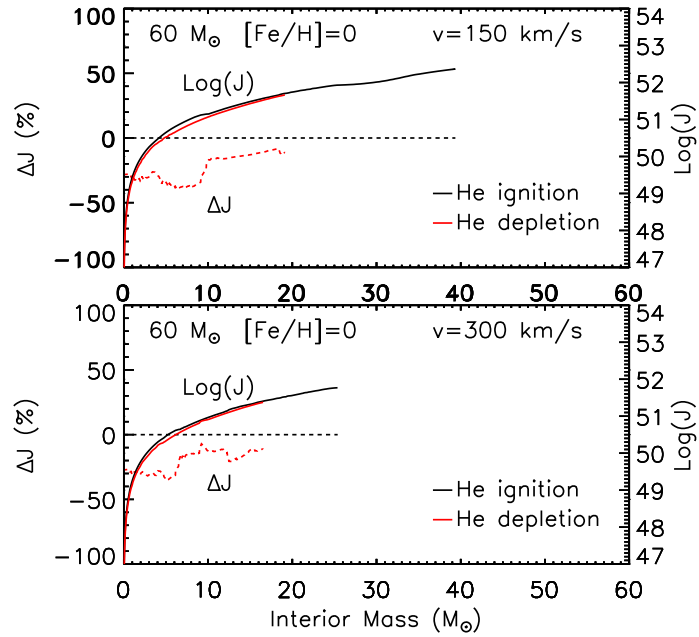}
\subfigures
\caption{Same as Figure \ref{momangheburnd15} but for a $\rm 60~M_\odot$ model.}
\label{momangheburnd60}       
\end{figure}

Figures \ref{momangheburna15} to \ref{momangheburnd60} show that, on average, in all the models of the present grid, the angular momentum in the core reduces by $\sim 20-40~\%$ of the initial value, essentially due to the transport within the He convective core, with the general trend that the higher reduction is obtained in low mass highly rotating models.
On the contrary, the total angular momentum is essentially constant in low metallicity models, due to the strong reduction of the stellar wind, while it reduces to $\sim 30-40~\%$ of the initial value at solar metallicity, the highest reduction occurring, again, in the low mass higher rotating models. Let us remind the reader that, as already mentioned above, for the same initial rotation velocity, the effects of rotation are generally lower in the more massive stars.

After core He depletion, the newly formed CO core begins to contract in order to ignite the following nuclear fuel while the He burning shifts in a shell and drives the formation of a convective zone just above the He burning shell. In stars where the He core remains constant in mass, the He convective shell forms in a region where the He profile is flat and equal to the one left by the H burning, i.e. beyond the He discontinuity that marks the outer edge of the CO core. Once the He convective shell forms it increases in mass until it reaches its maximum extension without, in any case, reaching the tail of the H burning shell. Because of the short lifetimes of the following nuclear burning stages (see below), only a small amount of He is burnt inside the shell before the final explosion of the star. In stars in which the He core is reduced by mass loss, or increased by rotation driven mixing, the He convective shell forms in a region of variable chemical composition, i.e., the one left by the receding He convective core (in stars where the He core is reduced by mass loss) or the one produced the continuous diffusion of He burning products driven by the rotation induced mixing (in rotating stars). In both cases, the He convective shell turns out to be hotter than the one formed in a region of flat He profile; such a different behavior has profound and interesting implications on the production of some specific isotopes \cite{2013ApJ...764...21C,2006ApJ...647..483L}.

\subsection{Advanced nuclear burning stages}

The evolutionary properties of a massive star after core He depletion are mainly controlled by (1) the CO core mass, (2) the $\rm ^{12}C/^{16}O$ ratio left by core He burning and, if the star rotates, (3) both the total and the internal distribution of the angular momentum. The CO core mass, that takes the role of the total mass (Figure \ref{masscocoreperc}), essentially determines the thermodynamic history of the core while $\rm ^{12}C$ and $\rm ^{16}O$ constitute the basic fuel for all the more advanced burning stages up to the formation of the iron core. In the previous section it has been already discussed the dependence of the $\rm ^{12}C$ mass fraction at core He depletion on both the mass loss and the rotation. It is worth to mention, however, that this quantity also depends, in general, on both the treatment of convection during core He burning and the value of the $\rm ^{12}C(\alpha,\gamma)^{16}O$ cross section \cite{2001ApJ...558..903I}. The physics of the convective motions is still poorly known and the determination of the  $\rm ^{12}C(\alpha,\gamma)^{16}O$ cross section at the relevant energies is still affected by large errors \cite{2012PhLB..711...35S}, hence these two ingredients constitute one of the major uncertainties in the computation of massive star models. 

Four major nuclear burning, distinguished by their principal fuel, can be identified during the evolution of a massive star from the core He exhaustion up to the presupernova stage, namely, carbon, neon, oxygen and silicon burning (see \cite{2008EAS....32..233L} for a detailed discussion of the nucleosynthesis occuring during each one of these nuclear burning). 

\begin{figure}[htpb]
\centering
\includegraphics[scale=.40]{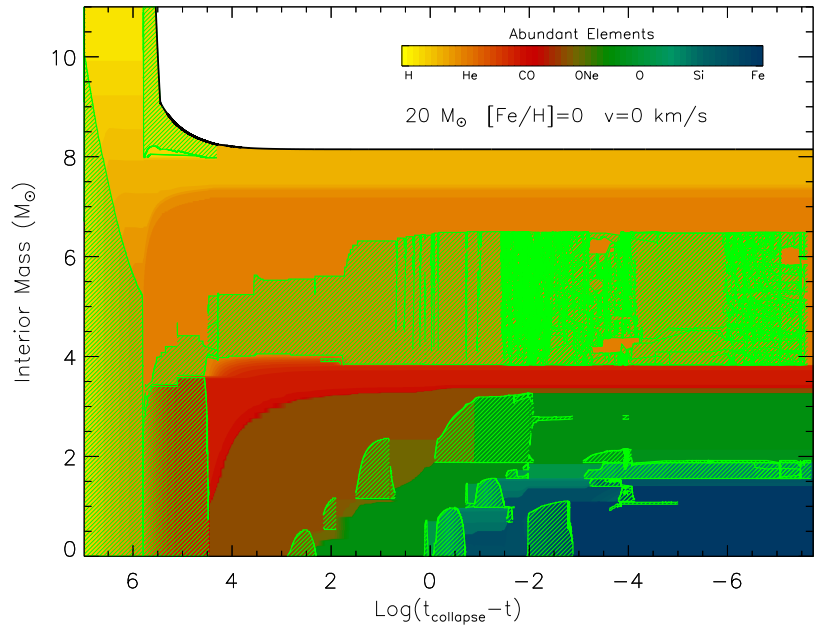}
\subfigures
\caption{Convective and composition presupernova history of a solar metallicity non rotating  $\rm 20~M_\odot$ star. Convective zones are marked by green shaded areas, while the chemical composition is coded as in the upper right color bar. The quantity on the $x$-axis is the logarithm of the residual time to the collapse in years, while the quantity reported on the $y$-axis is the interior mass coordinate in solar masses.
}
\label{convhista000}       
\end{figure}
\begin{figure}[htpb]
\samenumber
\centering
\includegraphics[scale=.40]{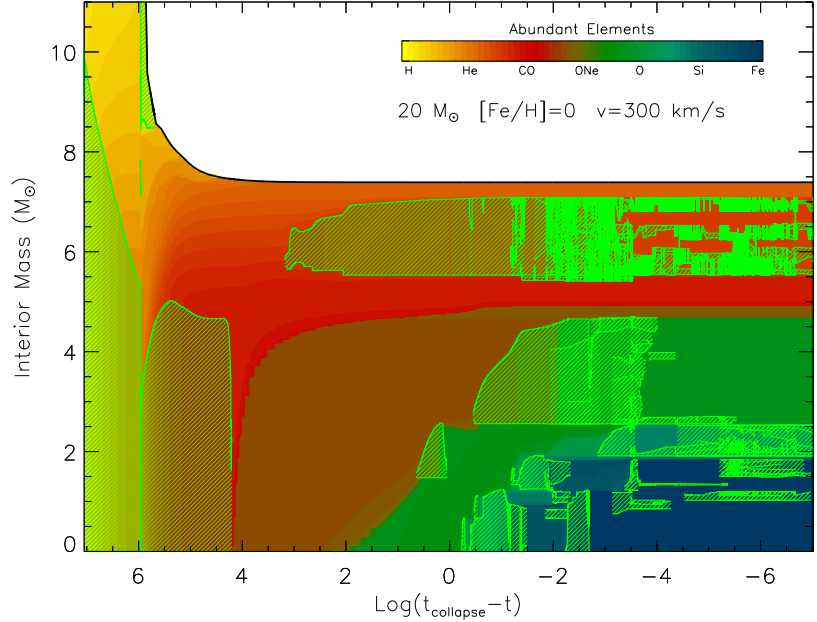}
\subfigures
\caption{Same as Figure \ref{convhista000} but for a model with initial rotation velocity $\rm v=300~km/s$.
}
\label{convhista300}       
\end{figure}


Figure \ref{convhista000} shows the chemical and convective history of a typical non rotating massive star, in this case it is a $\rm 20~M_\odot$ model with initial solar composition. In general, each burning stage begins at the center and induces the formation of a convective core. The convective core increases in mass, reaches a maximum and then disappears as the nuclear fuel is exhausted. The only exception to this general rule is core C burning that in the more massive stars occurs in a radiative environment due to the low $\rm ^{12}C$ mass fraction left by core He burning coupled to the strong neutrino losses (see section \ref{sec:features}). 
Once the nuclear fuel is exhausted at the center, the burning shifts in a shell which in general is efficient enough to induce the formation of a convective zone above it. Once the convective zone forms, the outward shift of the shell stops and the burning proceeds within the convective shell. After the nuclear fuel is depleted within the whole convective zone, the burning shell quickly shifts outward in mass and settles where the main fuel is still abundant. Then, eventually, another convective zone may form. Note that two consecutive (in time) convective shells may also partially overlap in mass - this may have some impact on the local nucleosynthesis. The details of this general behavior, i.e. number of convective zones formed in each burning stage and their overlap, depend on the mass of the CO core and its chemical composition.
Typically one to four carbon convective shells and two to three convective shell episodes for each of the neon, oxygen and silicon burning occur. In general, the number of C convective shells increases as the mass of the CO core decreases. 

The complex interplay among the shell nuclear burning, the timing and the overlap of the various convective zones determine in a direct way the mass-radius (M-R) relation \index{mass-radius relation} (i.e. the compactness) and the chemical stratification of the star at the presupernova stage.
\begin{figure}[h]
\centering
\includegraphics[scale=.25]{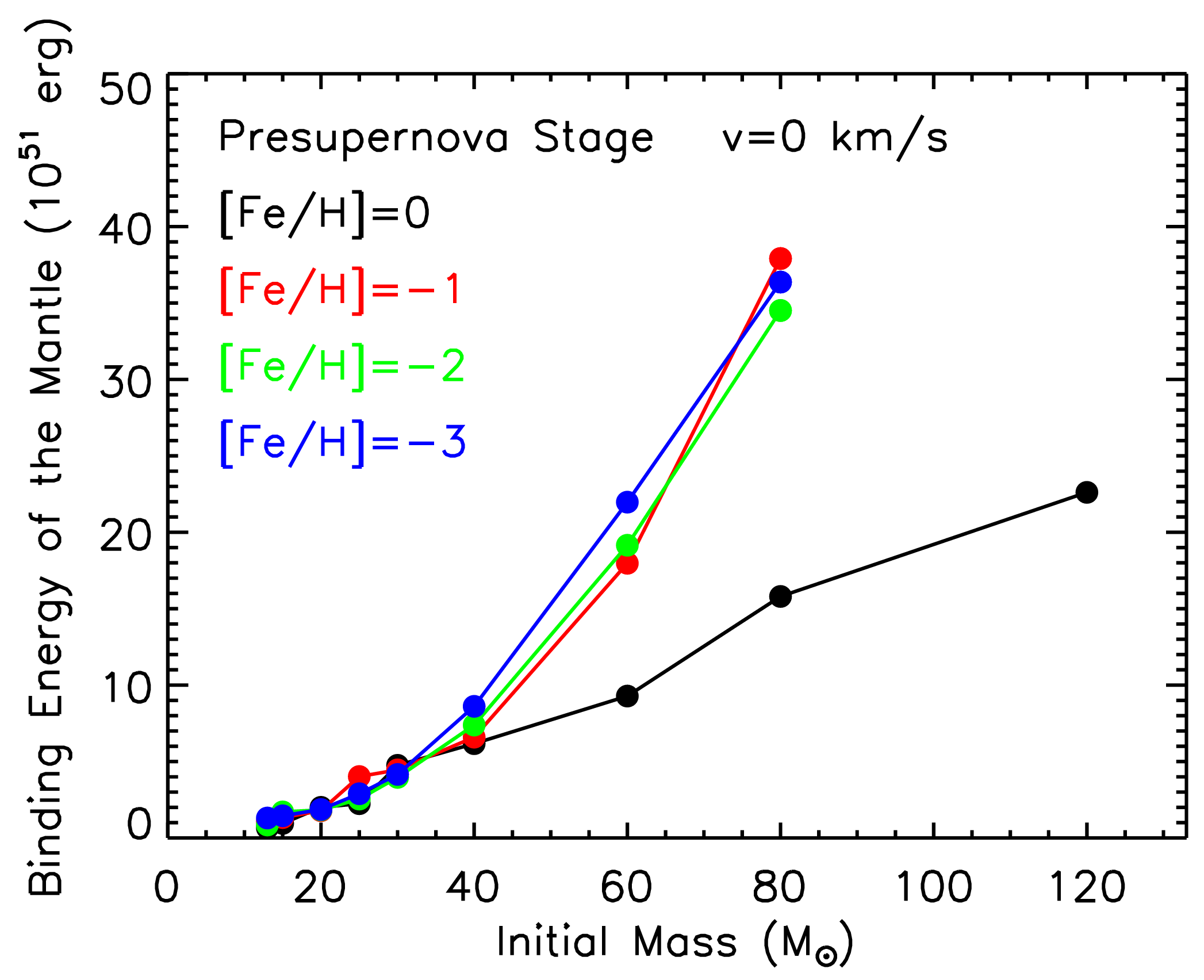}
\caption{Binding energy of the mantle (see text) at the presupernova stage for non rotating models with initial metallicities [Fe/H]=0 (black line), [Fe/H]=-1 (red line), [Fe/H]=-2 (green line) and [Fe/H]=-3 (blue line).}
\label{ebindZ}       
\end{figure} 
In general, the less efficient the C burning shell (i.e., the lower is the $\rm ^{12}C$ mass fraction left by core He burning), the smaller the number of the convective zones and the later they are formed, the higher the contraction of the CO core and the steeper the final M-R relation. This means that the higher is the mass of the CO core, the more compact is the structure of the star at the presupernova stage. The $\rm M_{CO}-M_{ini}$ relation at core He exhaustion, therefore, directly determines the scaling between the initial mass and the final M-R relation. Figure \ref{ebindZ}, that shows the binding energy of the mantle (defined as all the zones lying above the iron core) at the presupernova stage for non rotating models, clearly demonstrates that the relations between the compactness of the core at the presupernova stage and the initial mass, for the various metallicities, closely follows the corresponding $\rm M_{CO}-M_{ini}$ relations at core He exhaustion (Figure \ref{masscocoreZ}).

\begin{figure}[h]
\centering
\includegraphics[scale=.30]{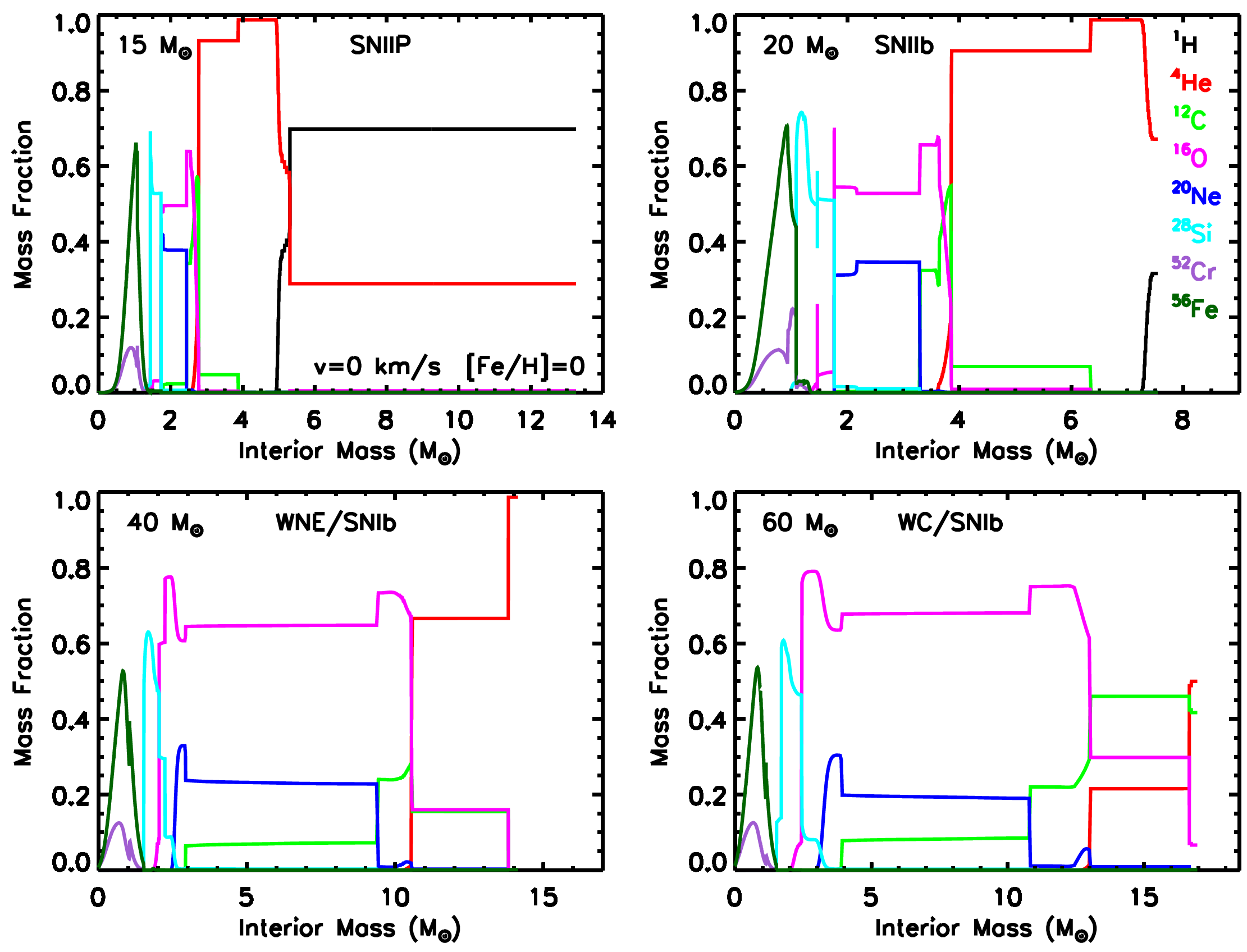}
\caption{Internal distribution of the most abundant isotopes at the presupernova stage for selected non rotating models with initial solar composition. Also shown in the figure is the supernova type that is expected for each progenitor mass (see text).}
\label{majorsolv0}       
\end{figure}

Figure \ref{majorsolv0} shows the distribution of the most abundant nuclear species at the presupernova stage for selected non rotating models of initial solar composition. As mentioned above, this chemical stratification is produced by the complex evolution of the various convective zones. In general, the presupernova star consists of an iron core \index{iron core} of mass in the range between $\sim 1.3$ and $\rm \sim 1.8~M_\odot$, depending on the initial mass of the star, which is surrounded by active burning shells located at the base of zones loaded in the main products of silicon, oxygen, neon, carbon, helium and hydrogen burning, respectively, i.e., the classical so called "onion structure". Thus, each zone of the presupernova star keeps memory of the nucleosynthesis produced by the various central and/or shell nuclear burning, occurring either in a radiative or in a convective environment. Note the effect of mass loss that reduces progressively the mass of the H-rich envelope and of the He core with increasing the initial mass.
\begin{figure}[h]
\centering
\includegraphics[scale=.30]{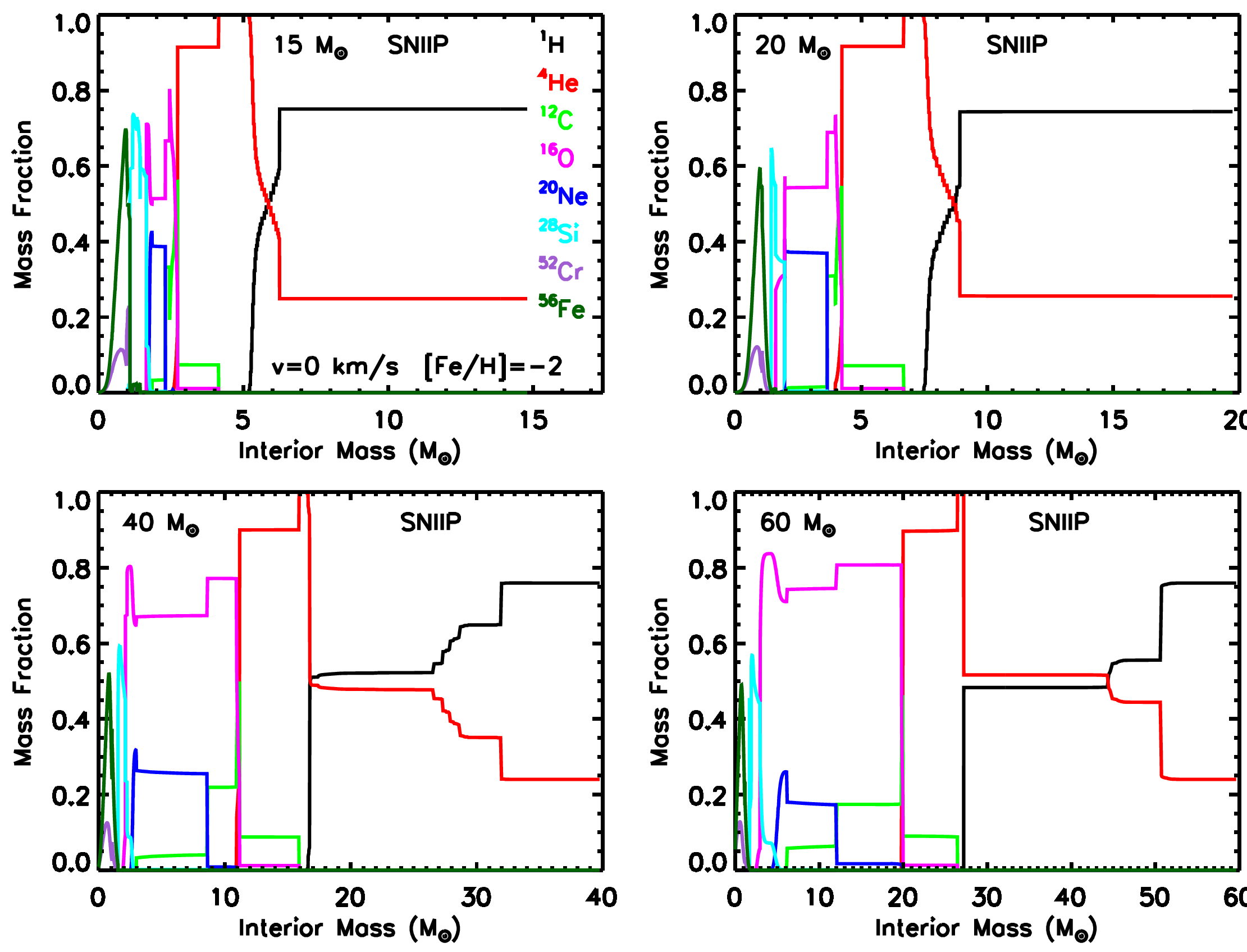}
\caption{Same as Figure \ref{majorsolv0} but for models with initial metallicity [Fe/H]=-2.}
\label{majorsolvm2}       
\end{figure}
Figure \ref{majorsolvm2} is the same as Figure \ref{majorsolv0} but for a metallicity corresponding to [Fe/H]=-2 and represents the typical behavior of non rotating stars with metallicities lower than [Fe/H]=-1. Note in this case that the overall interior properties are similar to those of the solar metallicity models. In this case all the stars retain all their H-rich envelope up to the presupernova stage because of the strong reduction of mass loss at low metallicities.

\begin{figure}[h]
\centering
\includegraphics[scale=.25]{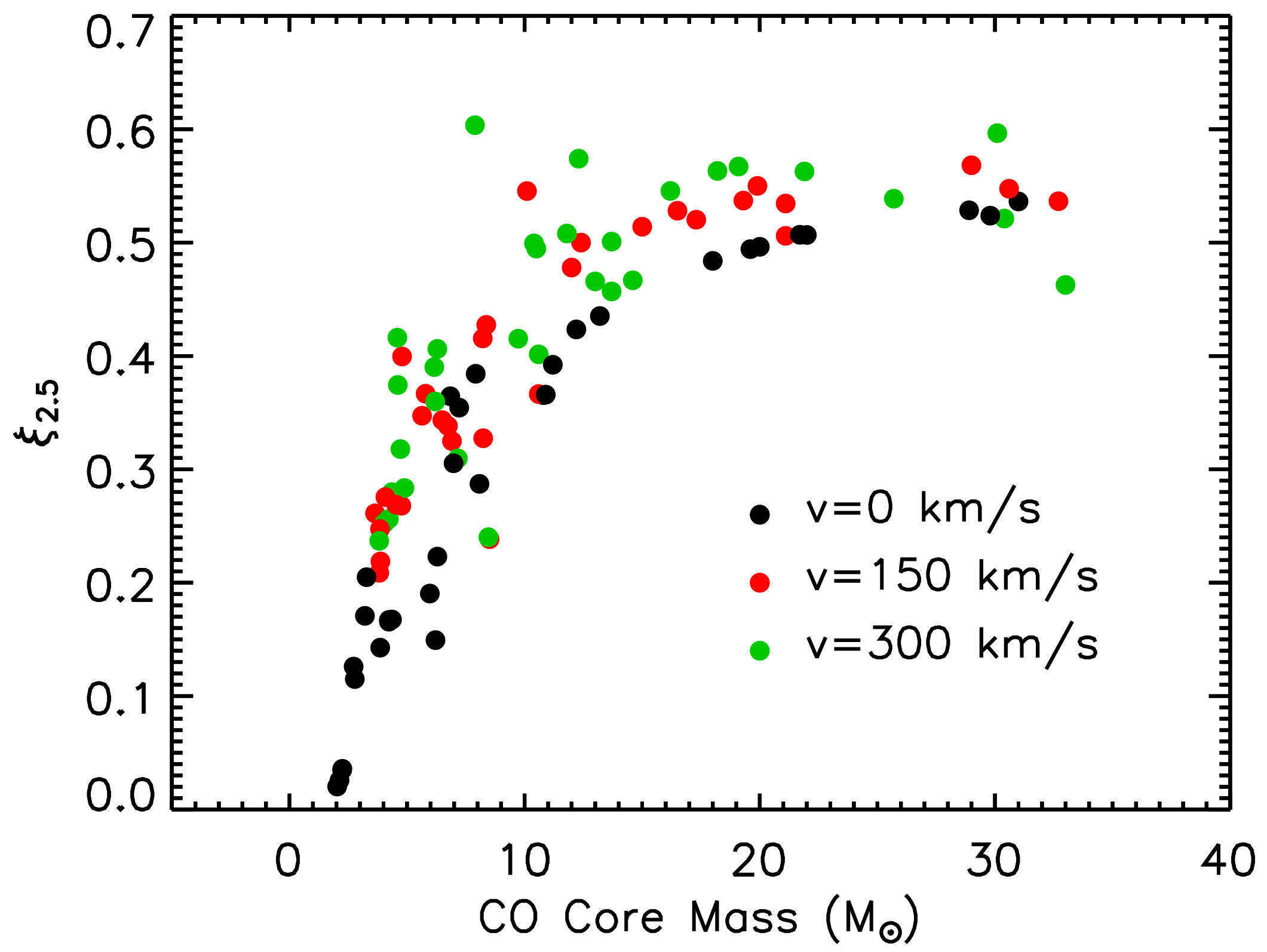}%
\caption{Compactness parameter $\rm \xi_{2.5}$ \cite{2011ApJ...730...70O} as a function of the CO core mass at the presupernova stage for all the models in the present grid.}
\label{csimco}       
\end{figure}

\begin{figure}[h]
\centering
\includegraphics[scale=.25]{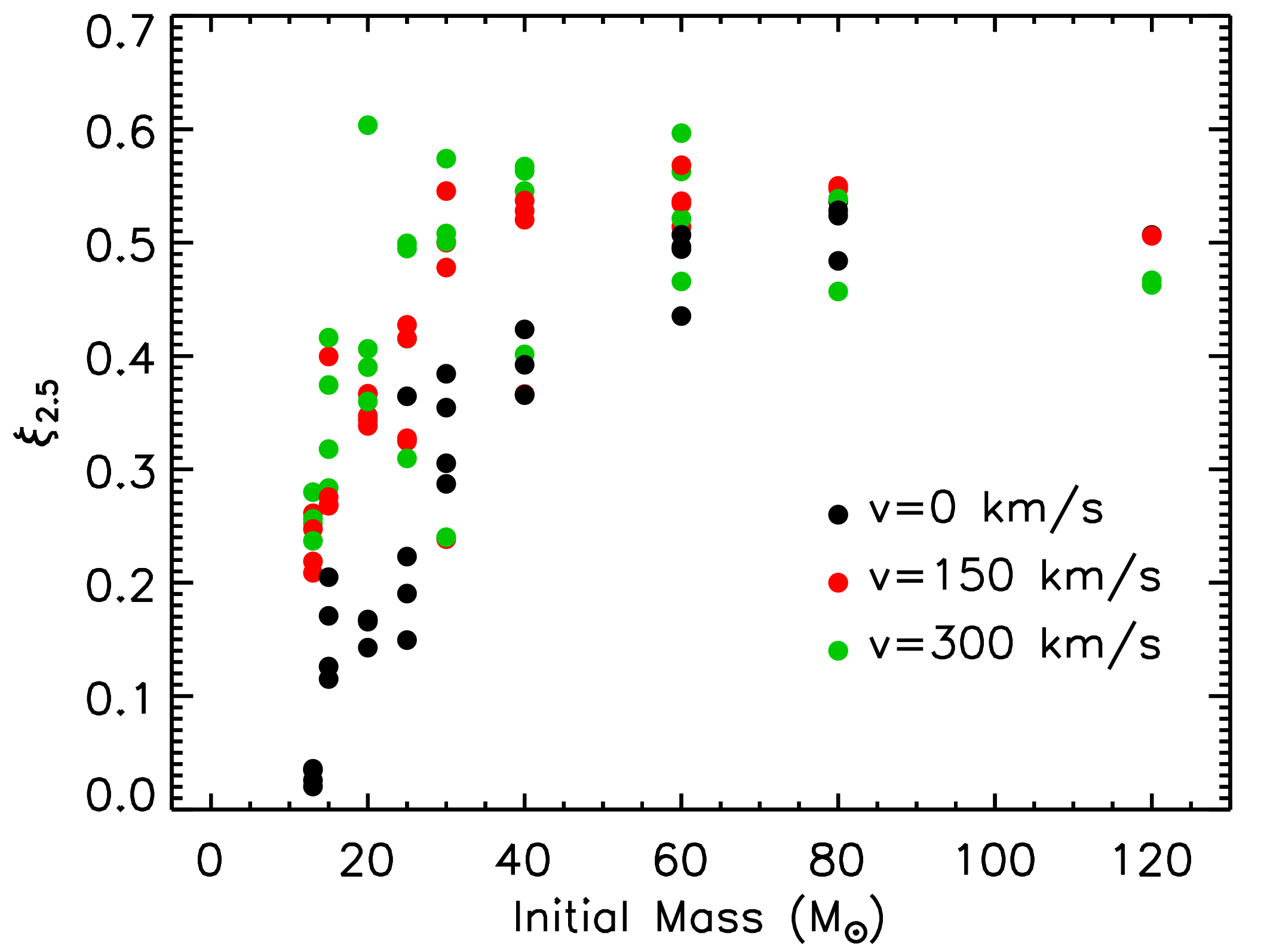}%
\caption{Compactness parameter $\rm \xi_{2.5}$ as a function of the initial mass for all the models in the present grid.}
\label{csimass}       
\end{figure}

The role of rotation on the advanced nuclear burning stages can be discussed by studying the internal variation of both the angular velocity and the degree of the deformation of the structure induced by the centrifugal force, for example, at two selected evolutionary stages after core He depletion, i.e., at core Si exhaustion and at the presupernova stage. Let us remind the reader that the deformation of the structure induced by the centrifugal force is controlled by a proper form factor parameter ($f_{\rm P}$), that enters in the hydrostatic equilibrium equation, and that varies from 0 to 1 - the higher is its value, the higher is the contribution of the centrifugal force and therefore the higher is the deformation of the star (see, e.g., \cite{1970stro.coll...20K}).
\begin{figure}[h]
\centering
\includegraphics[scale=.29]{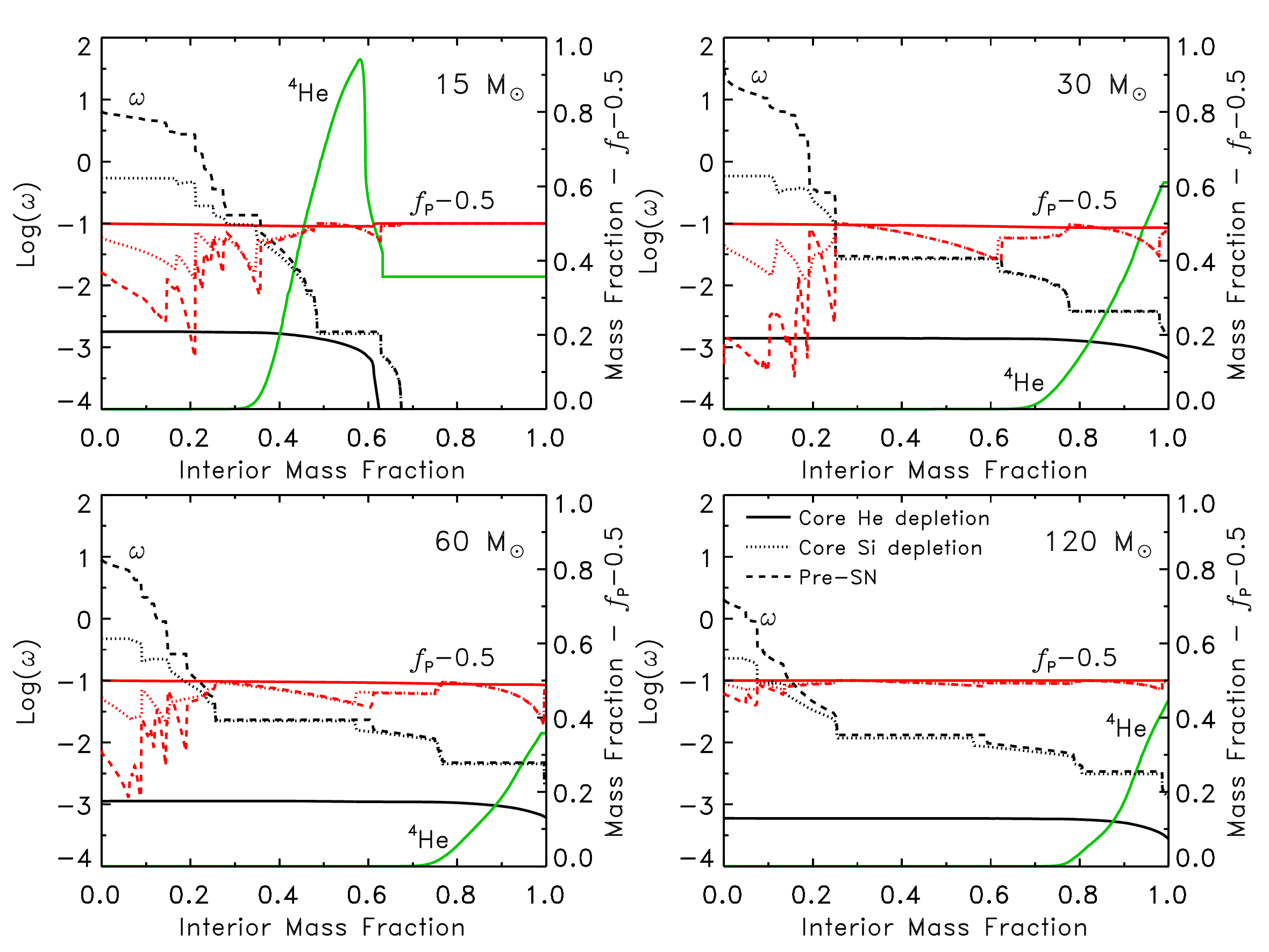}
\caption{
Angular velocity (black lines) and Òform factorÓ $f_P$ (red lines) as a function of the interior mass fraction at core He exhaustion (solid lines), core Si exhaustion (dotted lines), and presupernova stage (dashed lines), for four selected models, i.e,  $\rm 15~M_\odot$ (upper left panel), $\rm 30~M_\odot$ (upper right), $\rm 60~M_\odot$ (lower left panel), and $\rm 120~M_\odot$ (lower right panel). Also shown is the He mass fraction profile as a function of the interior mass fraction at core He exhaustion (green lines).
}
\label{rotpropsolv300}       
\end{figure}
We show in Figure \ref{rotpropsolv300}, as an example, the case of 4 selected models at solar metallicity, initially rotating at 300 km/s.
The figure shows clearly that, in spite of the spin-up of the more internal zones due to the progressive contraction of the core, the structural deformations induced by rotation start to be sizable only after Si depletion, i.e., the form factor decreases significantly below 1 only beyond this stage. This means that all the advanced evolutionary phases, from core He depletion to core Si exhaustion, are very mildly affected by rotation and hence that the final differences in the structure at the presupernova stage between rotating and non rotating models are essentially due to the differences in the CO core mass and in the central $\rm ^{12}C$ mass fraction at core He exhaustion (Figure \ref{masscocoreperc}). Let us recall again that the larger is the CO core mass - and/or the lower is the $\rm ^{12}C$ mass fraction left by core He burning - the faster is the outward shift of the C burning shell and the more compact is the core of the star. According to these general rules, for the same initial mass, rotating models behave like more massive stars and therefore they end their life with more compact structures. Figures \ref{convhista000} and \ref{convhista300} clearly show such a behavior for a rotating and non rotating 20 $\rm M_\odot$ model with initial solar composition. The non rotating model forms a CO core, at core He exhaustion, of $\rm \sim 3.1~M_\odot$ and the $\rm ^{12}C$ mass fraction left by core He burning is $\sim 0.28$; the corresponding values for the rotating models are $\rm \sim 4.5~M_\odot$ and $\sim 0.22$. The non rotating model forms a convective core during core C burning, followed by three consecutive convective shell episodes in the subsequent evolutionary phases (this is a typical behavior of the less massive stars). On the contrary, in the rotating model, core C burning occurs in a radiative core and it is followed by only two C convective shell episodes in the further evolution (typical of the more massive stars). A measure of the compactness of the core can be obtained by using the parameter $\rm \xi_{2.5}$ \cite{2011ApJ...730...70O}. Although this parameter has been originally defined at core bounce in order to predict the final fate of the star, we assume, here, that its value does not change significantly during the collapse and therefore that its evaluation at the presupernova stage may provide a similar information. Figure \ref{csimco} shows that a rather tight relation exists between $\rm \xi_{2.5}$ and the CO core mass. 
Such a tight relation becomes much more scattered if $\rm \xi_{2.5}$ is plotted as a function of the initial mass of the star (Figure \ref{csimass}) . Such a result is due to the effect of rotation.
Taking into account the effects of both the initial metallicity and the initial velocity, therefore, we find that
low metallicity fast rotating models are harder to explode, as a consequence, we expect these models to produced larger remnant masses and/or, eventually, to produce faint \index{faint supernovae} and/or "failed" supernovae \index{failed supernovae} \cite{1995ApJ...445L.129B,1999ApJ...513..780P}. 

\begin{figure}[h]
\centering
\includegraphics[scale=.25]{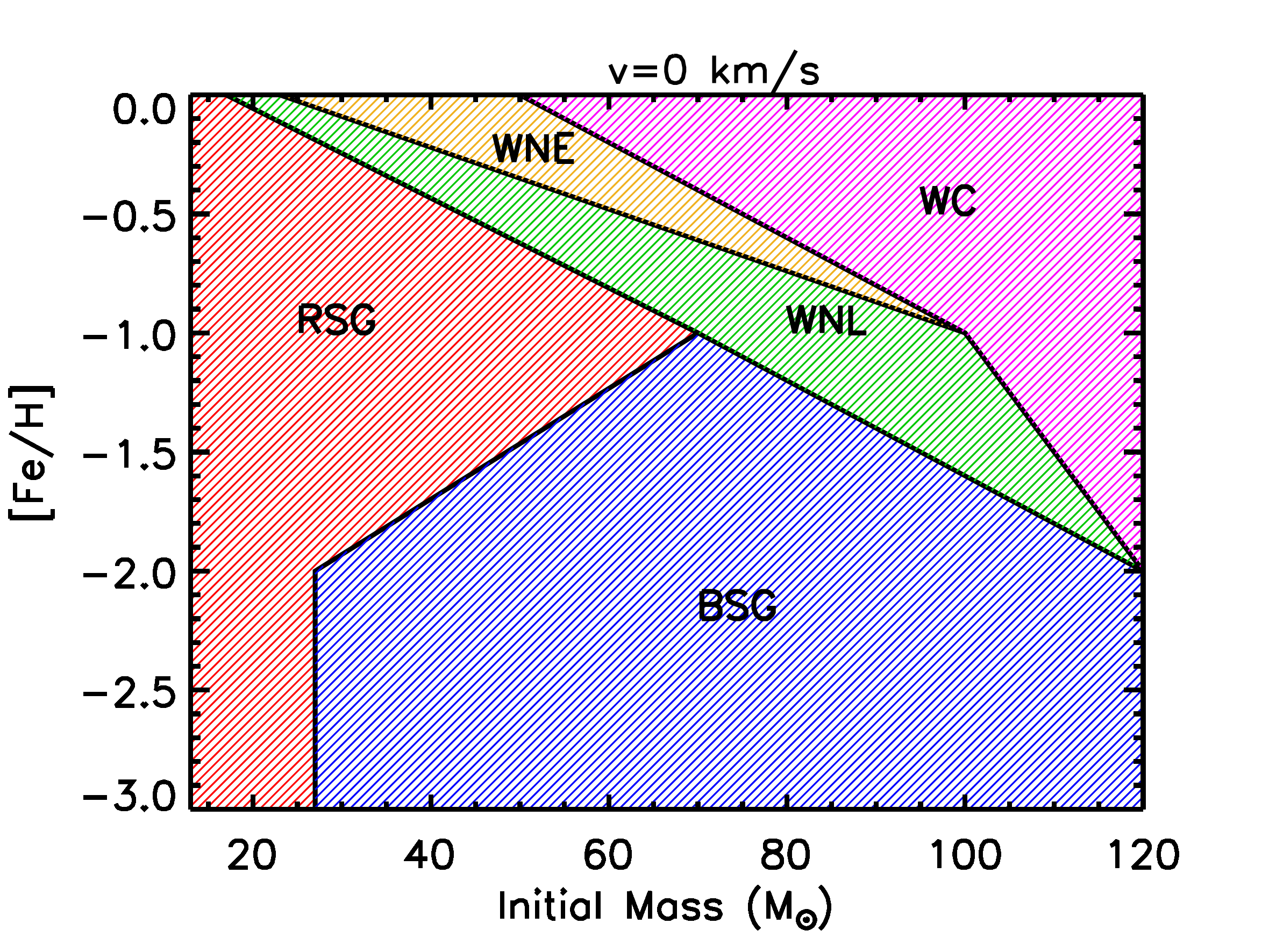}%
\subfigures
\caption{Predicted supernova progenitors for non rotating models at various metallicities. The meanings of the various labels are defined in the text.}
\label{confsnv000}       
\end{figure}

As mentioned in section \ref{sec:features}, the timescales of the advanced nuclear burning stages become progressively very short. However, in spite of such an occurrence, the location in the HR diagram after core He depletion for the majority of the models may change, even substantially. 
\begin{figure}[h]
\centering
\includegraphics[scale=.25]{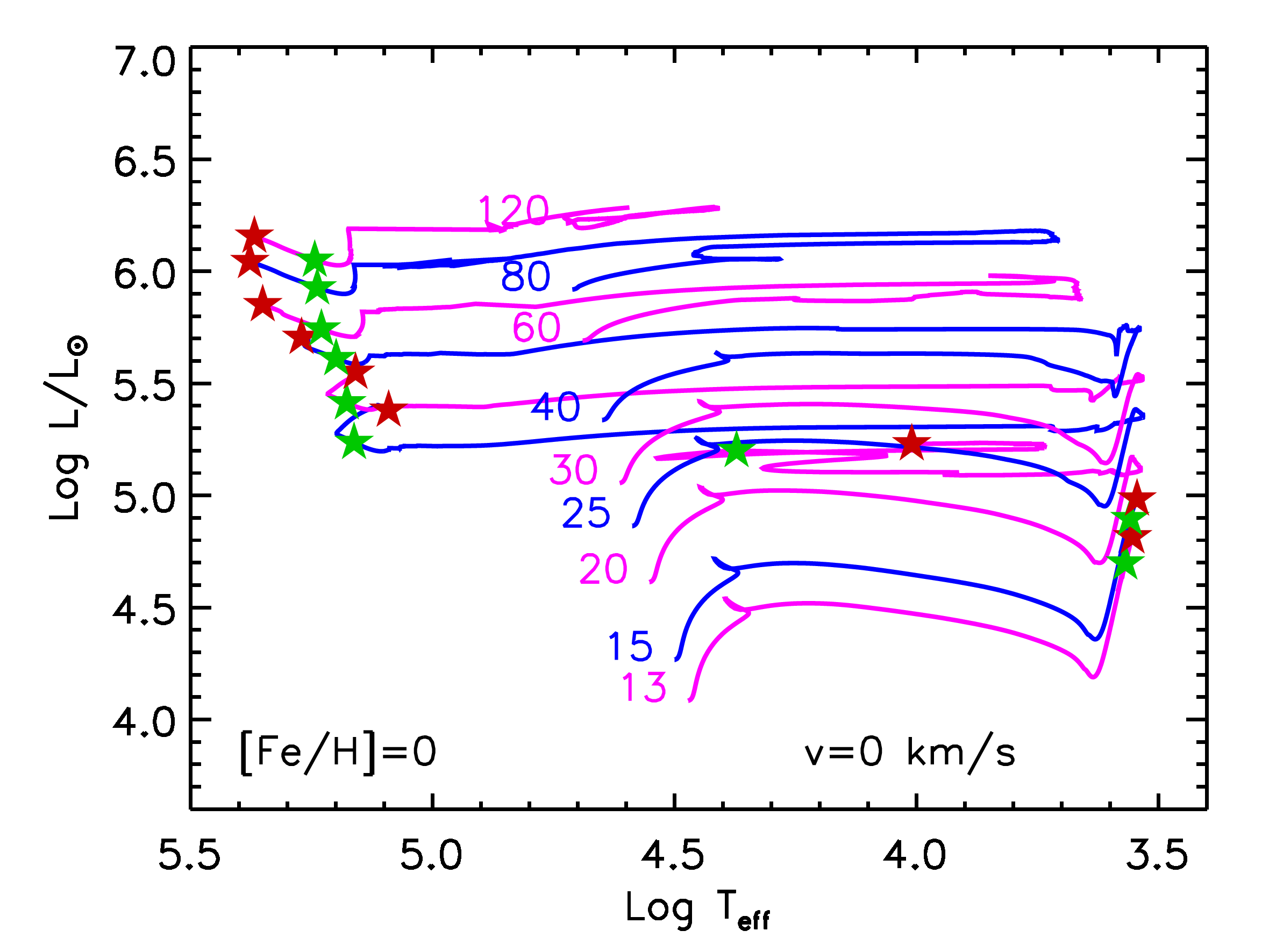}
\subfigures
\caption{HR diagram of solar metallicity non rotating models from the MS phase up to the presupernova stage. The green stars mark the core He depletion while the red stars refer to the presupernova stage.}
\label{hrpresna000}       
\end{figure}
\begin{figure}[h]
\samenumber
\centering
\includegraphics[scale=.25]{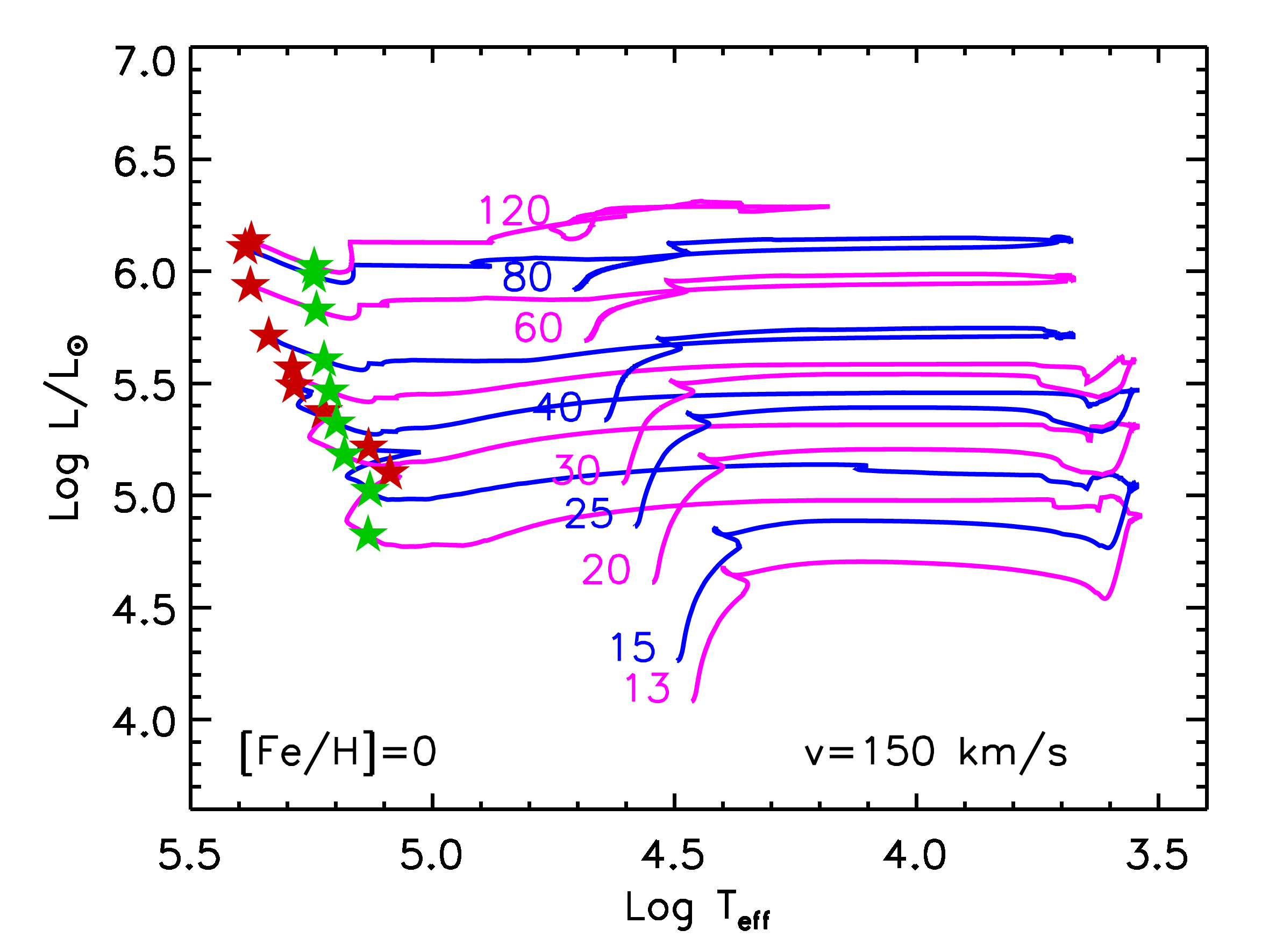}
\subfigures
\caption{Same as Figure \ref{hrpresna000} but for models with initial rotation velocity $\rm v=150~km/s$.}
\label{hrpresna150}       
\end{figure}
\begin{figure}[h]
\samenumber
\centering
\includegraphics[scale=.25]{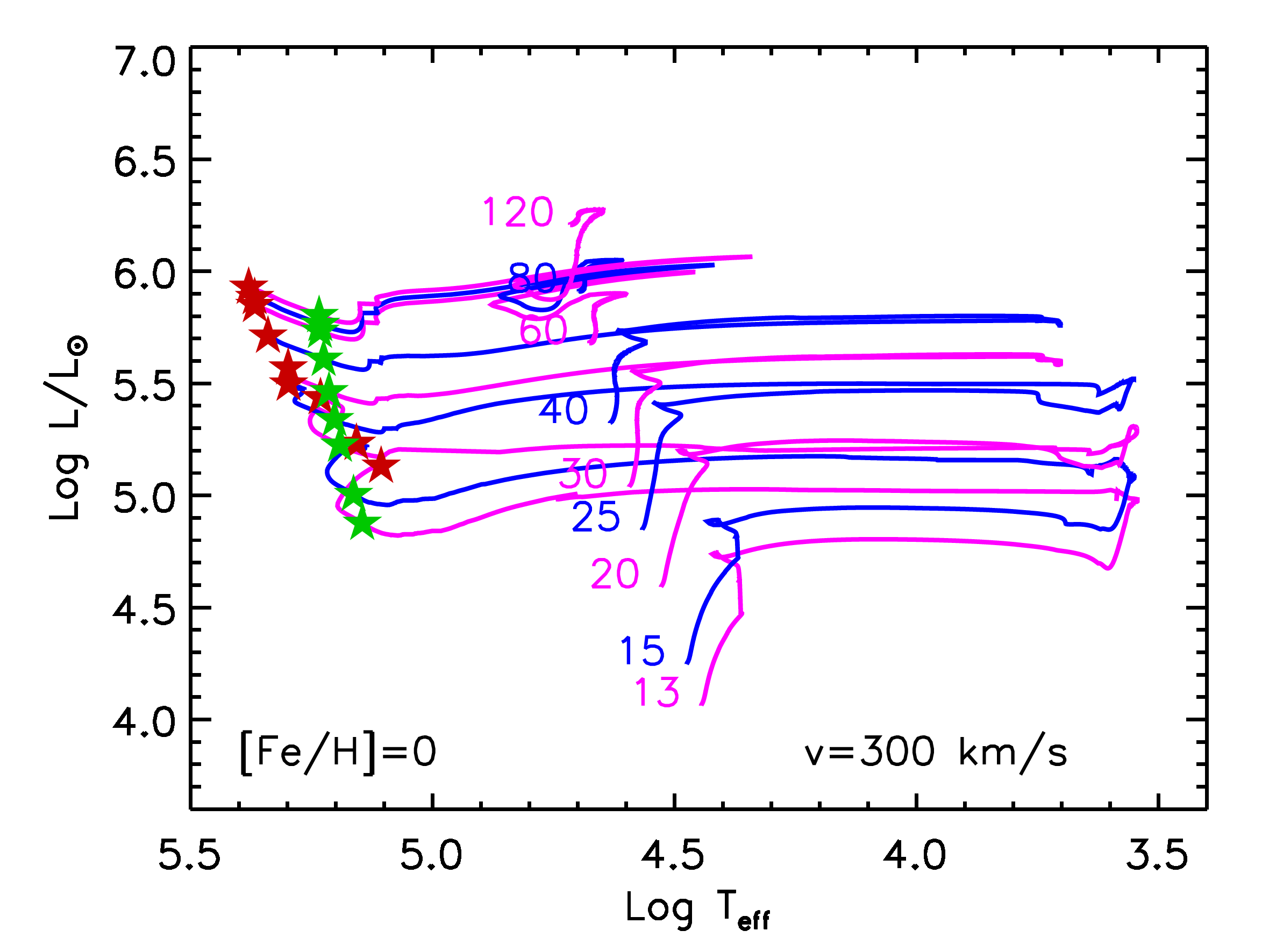}
\subfigures
\caption{Same as Figure \ref{hrpresna000} but for models with initial rotation velocity $\rm v=300~km/s$.}
\label{hrpresna300}       
\end{figure}
\begin{figure}[h]
\samenumber
\centering
\includegraphics[scale=.25]{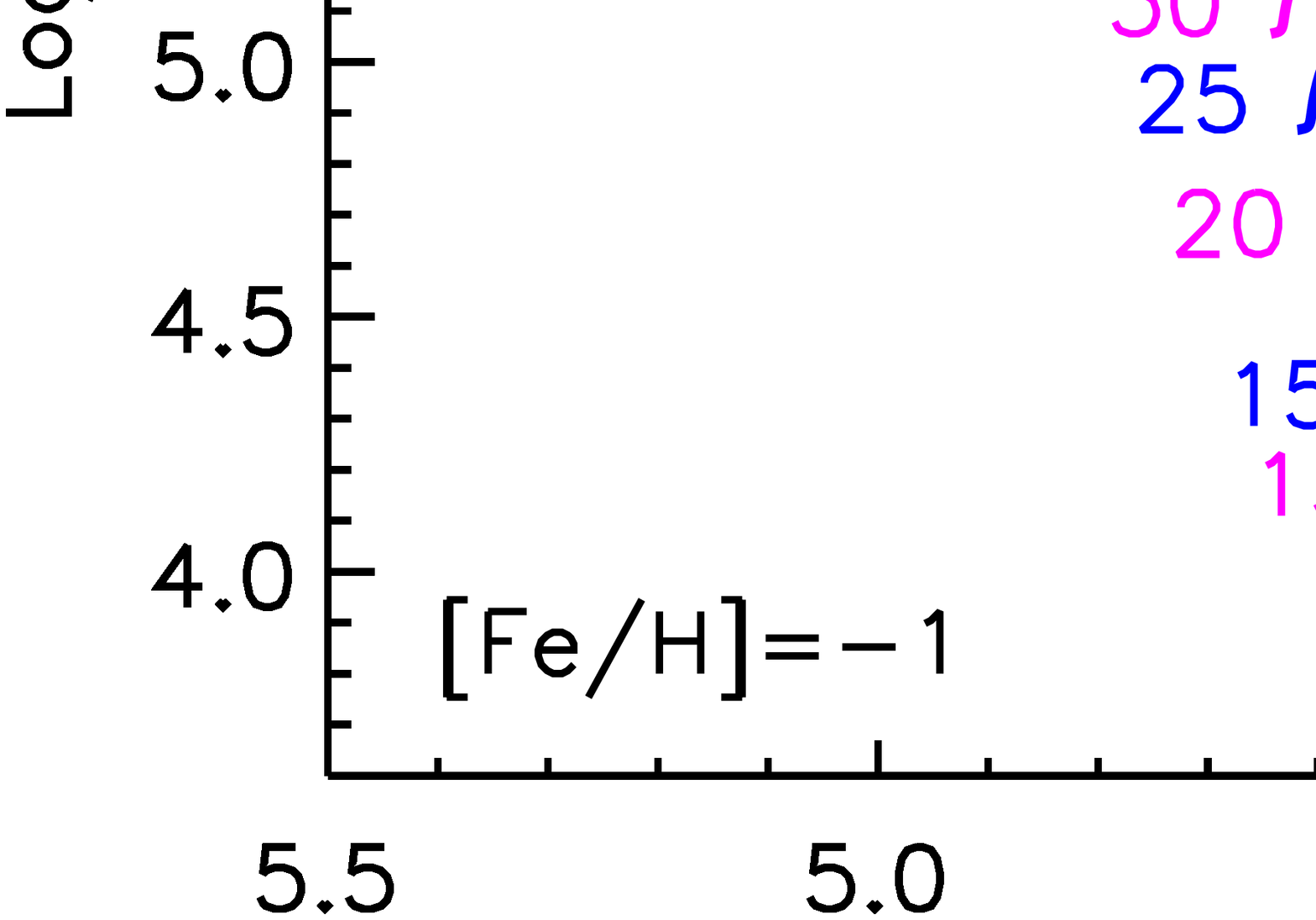}
\subfigures
\caption{Same as Figure \ref{hrpresna000} but for models with initial metallicity [Fe/H]=-1.}
\label{hrpresnb000}       
\end{figure}
\begin{figure}[h]
\samenumber
\centering
\includegraphics[scale=.25]{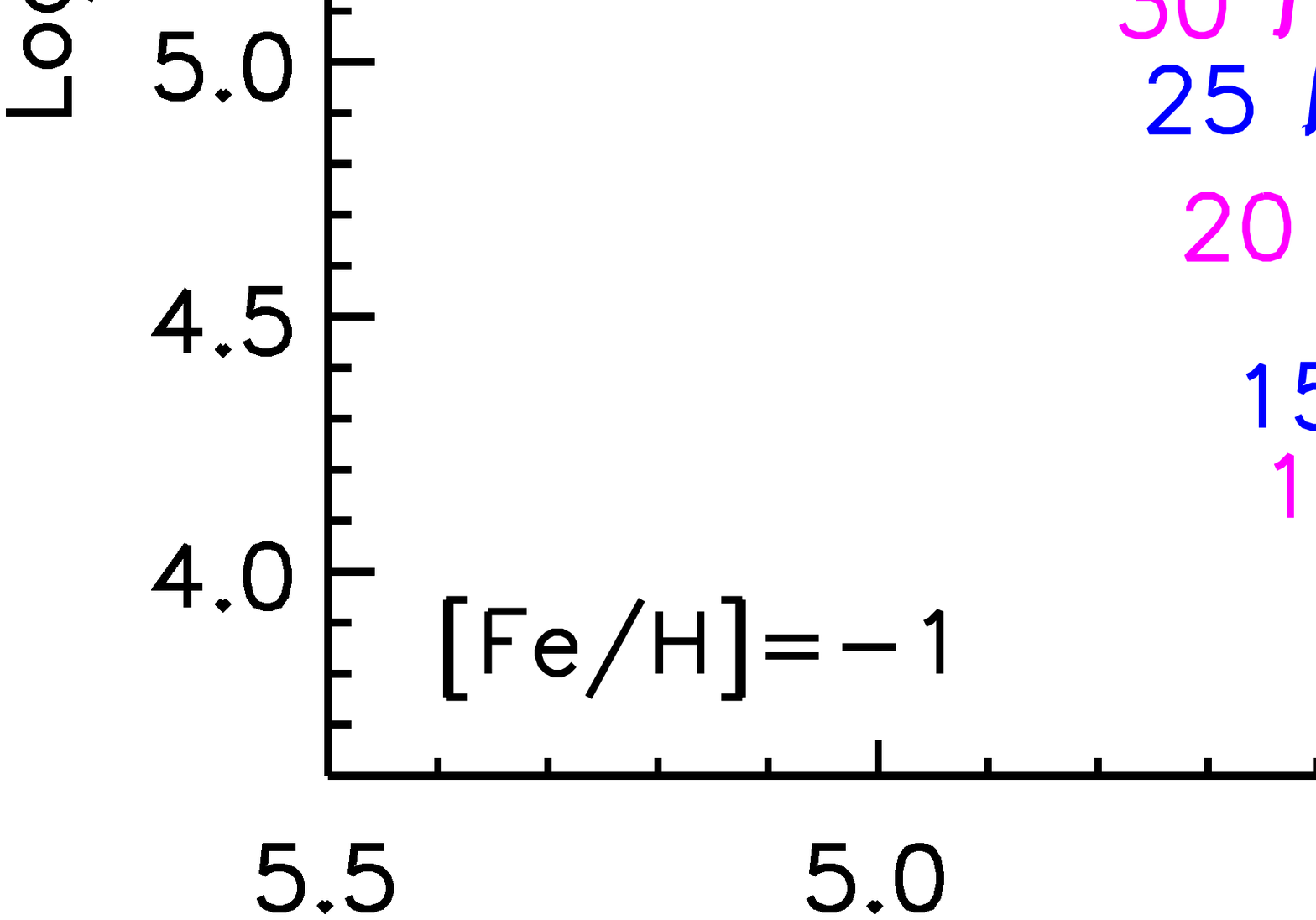}
\subfigures
\caption{Same as Figure \ref{hrpresnb000} but for models with initial rotation velocity $\rm v=150~km/s$.}
\label{hrpresnb150}       
\end{figure}
\begin{figure}[h]
\samenumber
\centering
\includegraphics[scale=.25]{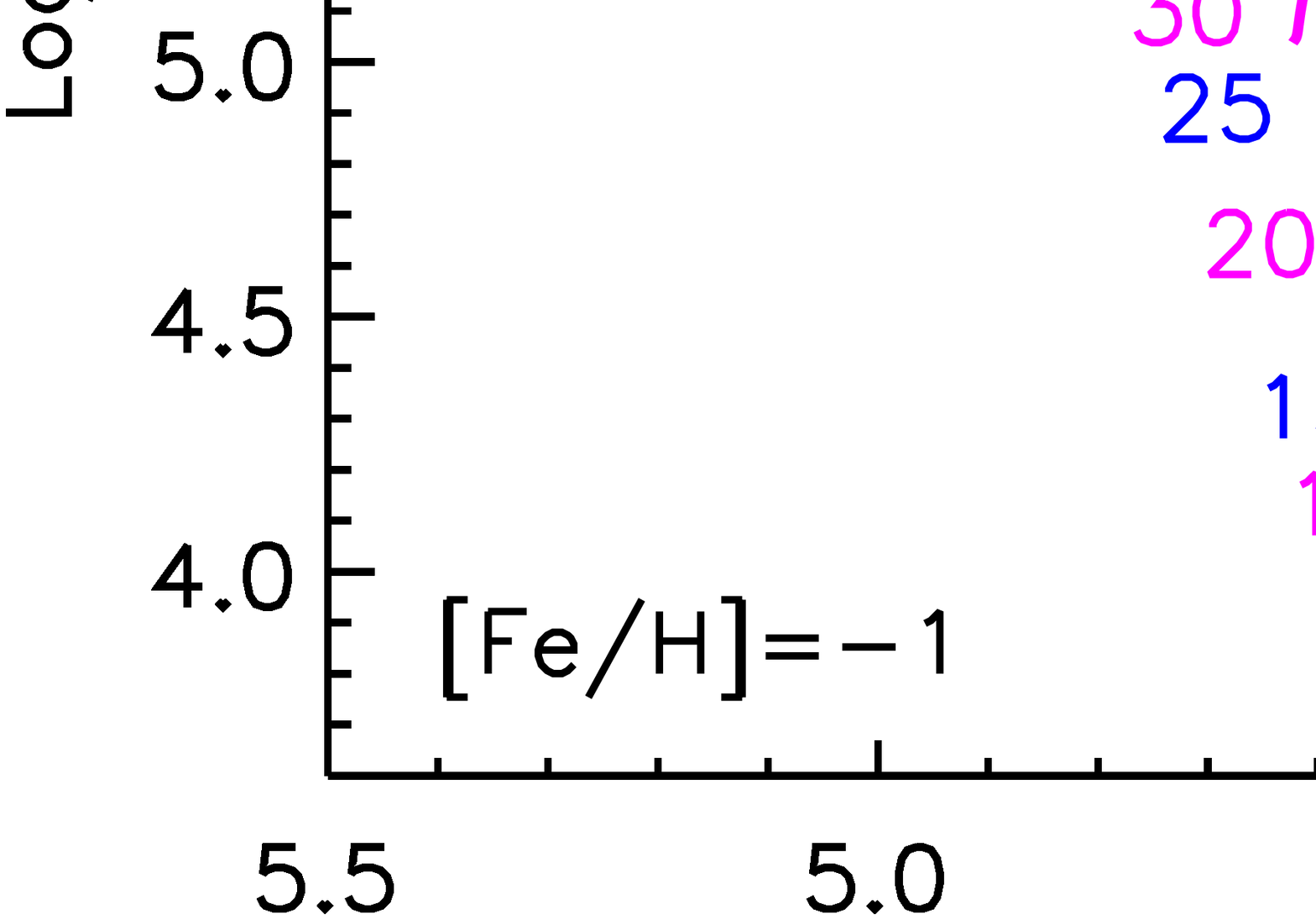}
\subfigures
\caption{Same as Figure \ref{hrpresnb000} but for models with initial rotation velocity $\rm v=300~km/s$.}
\label{hrpresnb300}       
\end{figure}
\begin{figure}[h]
\samenumber
\centering
\includegraphics[scale=.25]{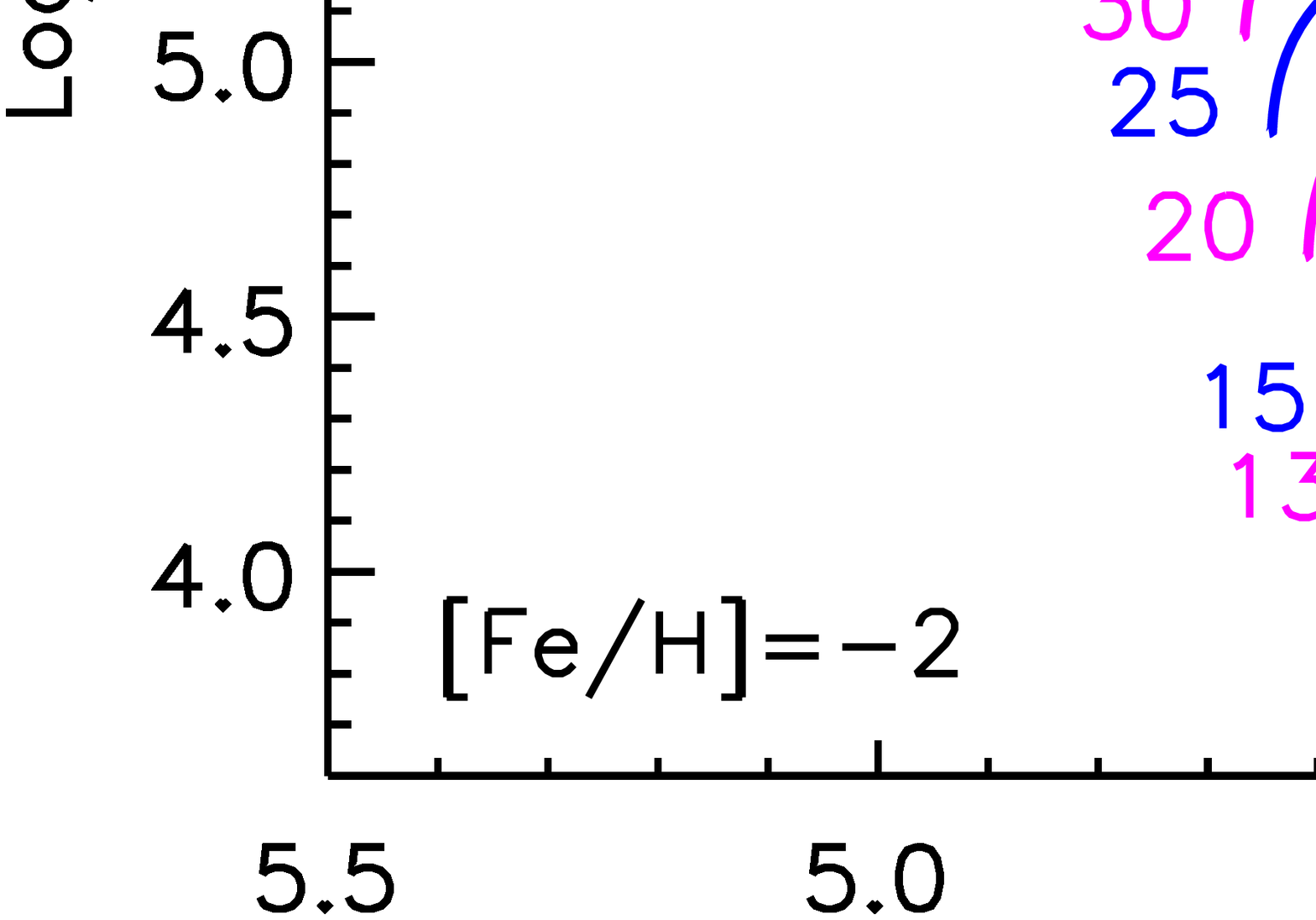}
\subfigures
\caption{Same as Figure \ref{hrpresna000} but for models with initial metallicity [Fe/H]=-2.}
\label{hrpresnc000}       
\end{figure}
\begin{figure}[h]
\samenumber
\centering
\includegraphics[scale=.25]{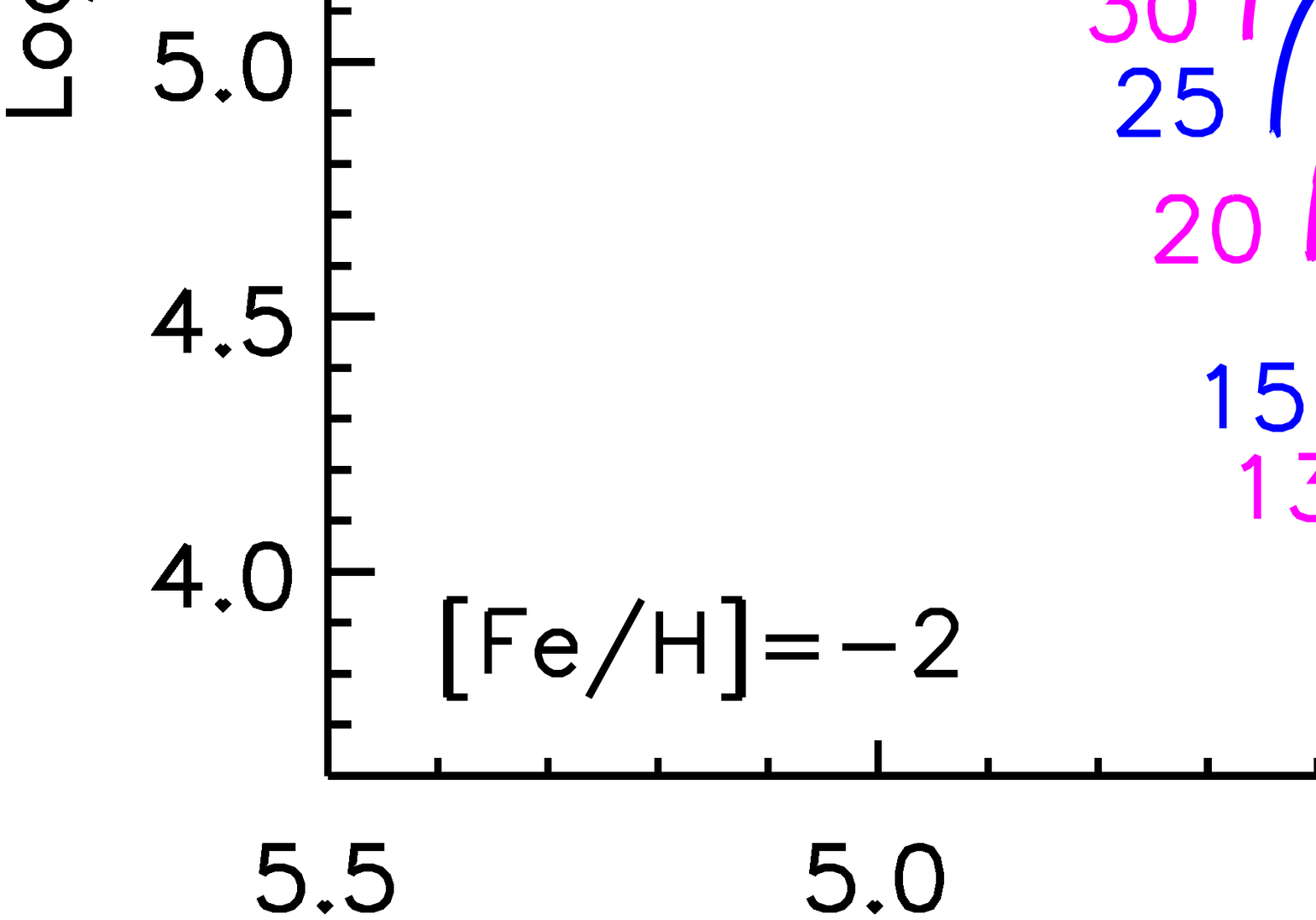}
\subfigures
\caption{Same as Figure \ref{hrpresnc000} but for models with initial rotation velocity $\rm v=150~km/s$.}
\label{hrpresnc150}       
\end{figure}
\begin{figure}[h]
\samenumber
\centering
\includegraphics[scale=.25]{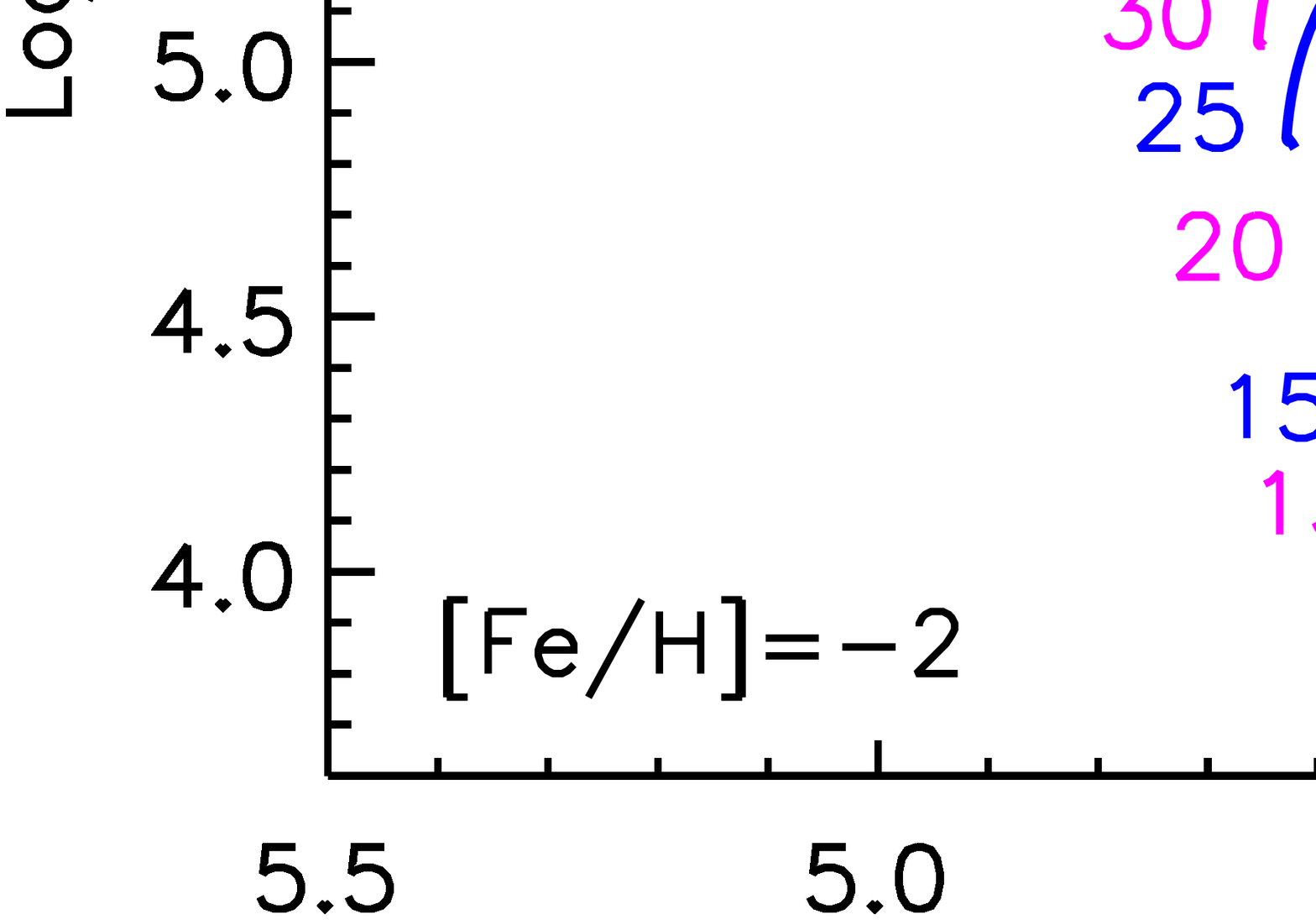}
\subfigures
\caption{Same as Figure \ref{hrpresnc000} but for models with initial rotation velocity $\rm v=300~km/s$.}
\label{hrpresnc300}       
\end{figure}
\begin{figure}[h]
\samenumber
\centering
\includegraphics[scale=.25]{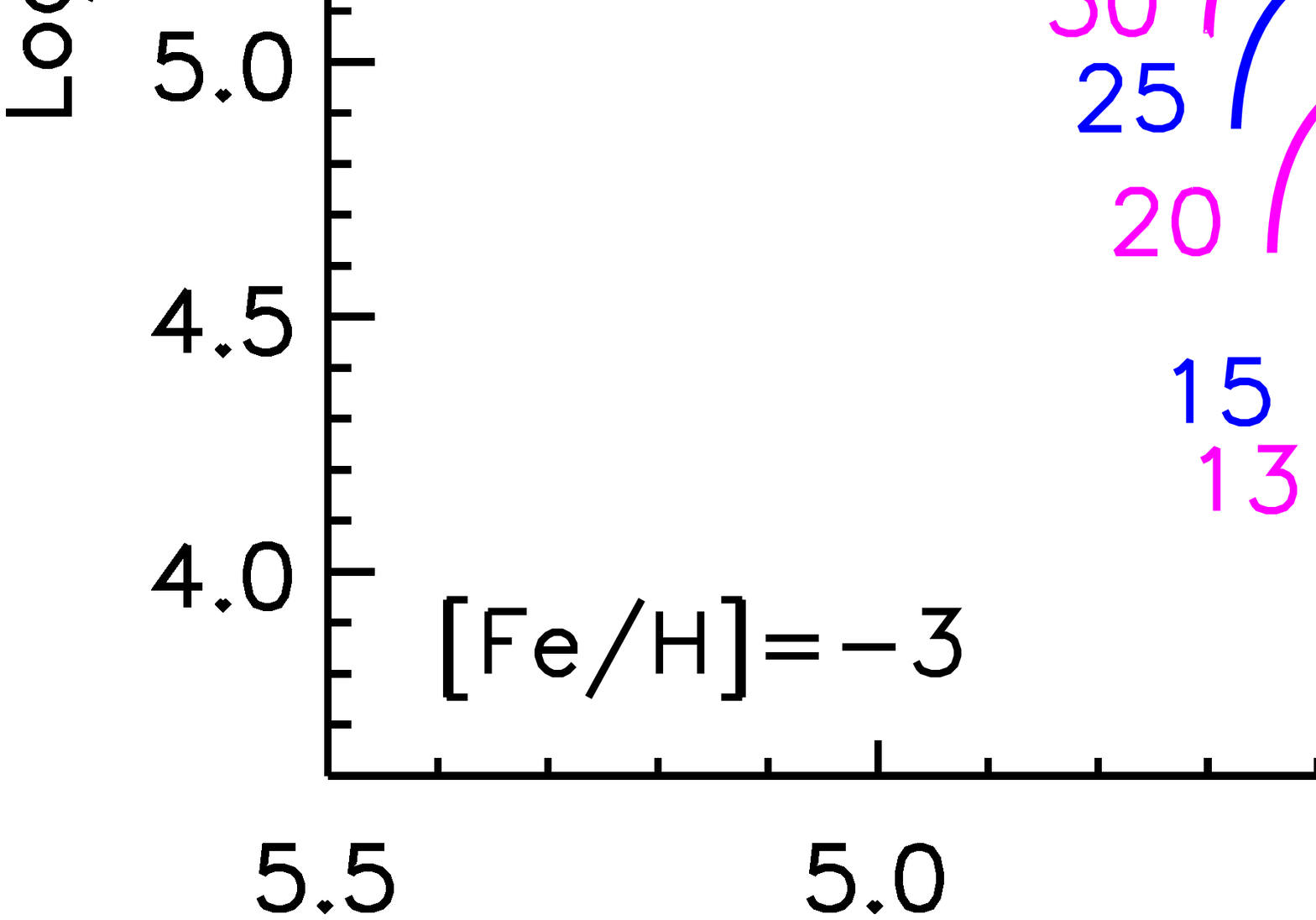}
\subfigures
\caption{Same as Figure \ref{hrpresna000} but for models with initial metallicity [Fe/H]=-3.}
\label{hrpresnd000}       
\end{figure}
\begin{figure}[h]
\samenumber
\centering
\includegraphics[scale=.25]{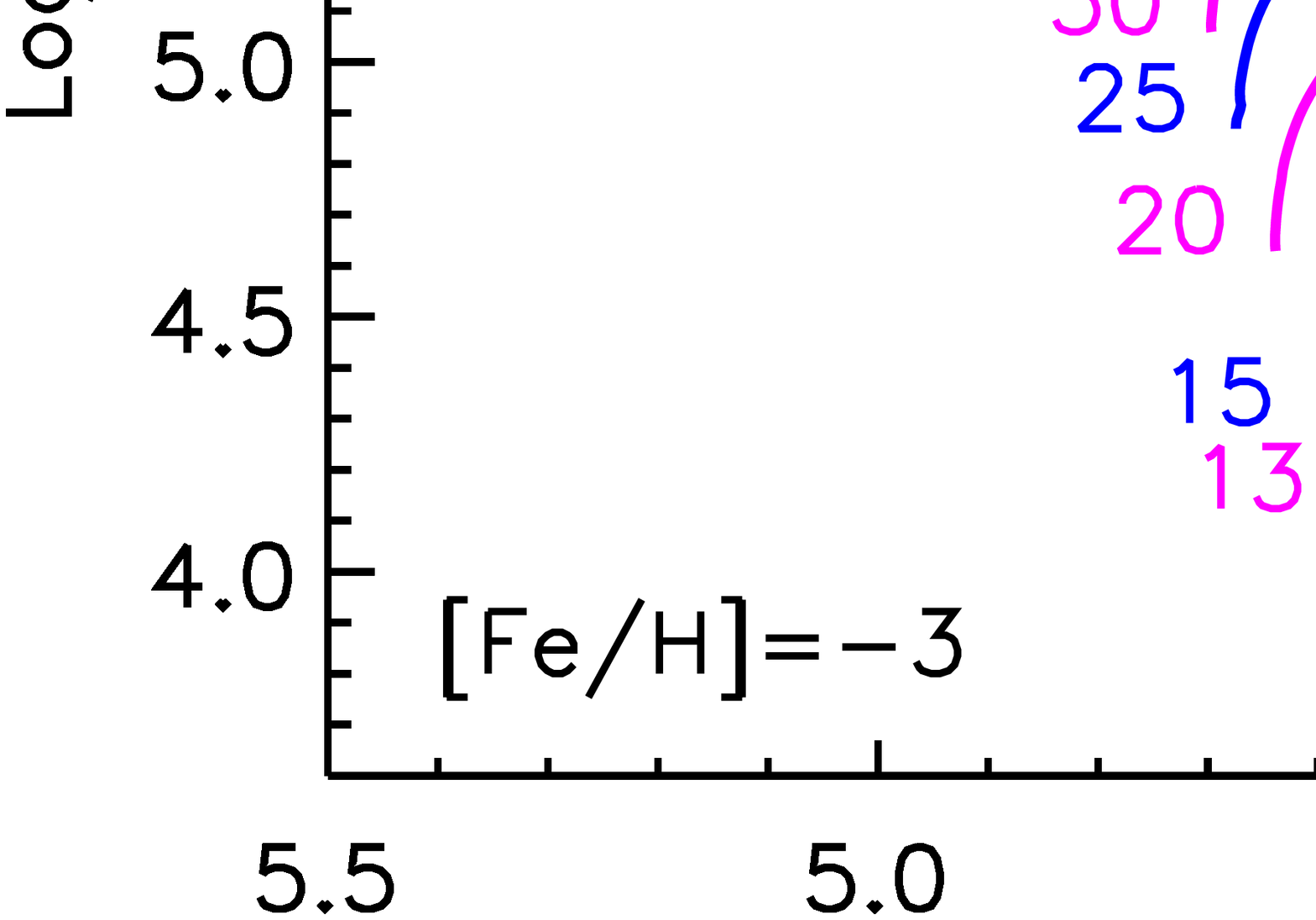}
\subfigures
\caption{Same as Figure \ref{hrpresnd000} but for models with initial rotation velocity $\rm v=150~km/s$.}
\label{hrpresnd150}       
\end{figure}
\begin{figure}[h]
\samenumber
\centering
\includegraphics[scale=.25]{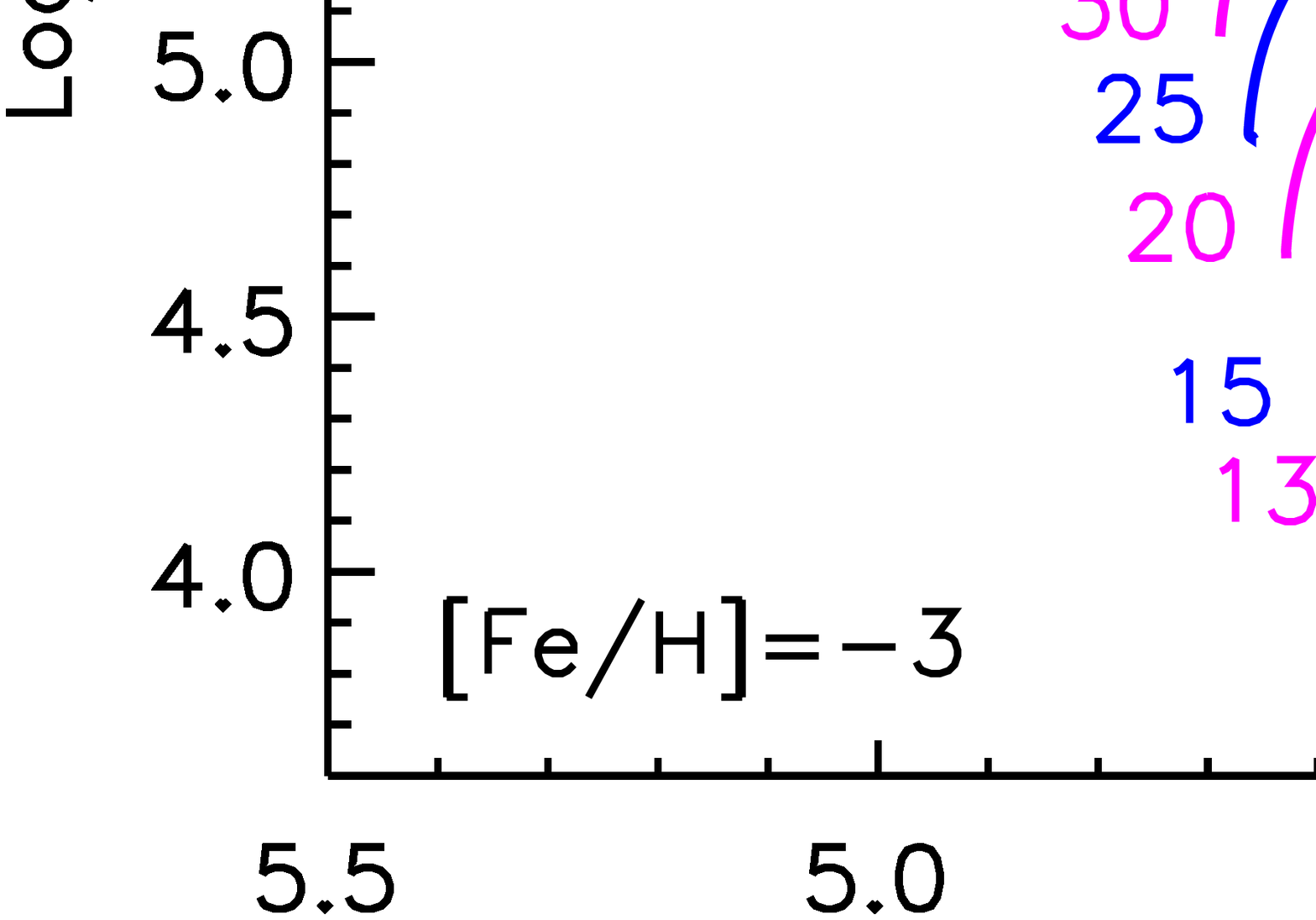}
\subfigures
\caption{Same as Figure \ref{hrpresnd000} but for models with initial rotation velocity $\rm v=300~km/s$.}
\label{hrpresnd300}       
\end{figure}
The green and red stars in Figures \ref{hrpresna000} to \ref{hrpresnd300} mark the location of each model in the HR diagram at core He depletion and at the presupernova stage, respectively, and therefore provides the final configuration of the all the stars of the present grid, at the time of the explosion, as a function of the initial mass, metallicity and rotation velocity. For non rotating models at solar metallicity, stars with initial masses below $\rm \sim 17~M_\odot$ explode as RSGs while stars with initial masses above this limit explode as WR stars. By the way, stars exploding as WRs, in general, may have a different chemical composition of the envelope and therefore may explode as WNL- \index{WNL}, WNE-, WNC- \index{WNC} or WC-WR stars. These differences result from the specific mass loss history and may have, in general, some impact on the supernova light curves and spectra \cite{2011MNRAS.414.2985D}. At solar metallicity we find a higher number of WNE and WC progenitors compared to the WNL ones. Note also the lacking of WNC progenitors in this case (as already discussed by \cite{2013ApJ...764...21C} and references therein). As the metallicity decreases to [Fe/H]=-1 the number of RSG supernova progenitors increases while the number of WR progenitors decreases because of the reduction of the mass loss. At this metallicity the limiting mass between stars exploding as RSGs and those exploding as WRs is $\rm \sim 70~M_\odot$.
Note that no WNE and WNC are expected for this metallicity.
For metallicities $\rm [Fe/H]\leq -2$ the compactness increases and the mass loss reduces substantially, therefore in these cases, the number of RSG progenitors decreases and the number WR progenitors vanishes. For these low metallicities stars with masses lower than $\rm \sim 27~M_\odot$ explode as RSGs while stars above this limit explode as BSGs with an extended H-rich envelope (Figure \ref{confsnv000}). 

\begin{figure}[h]
\samenumber
\centering
\includegraphics[scale=.25]{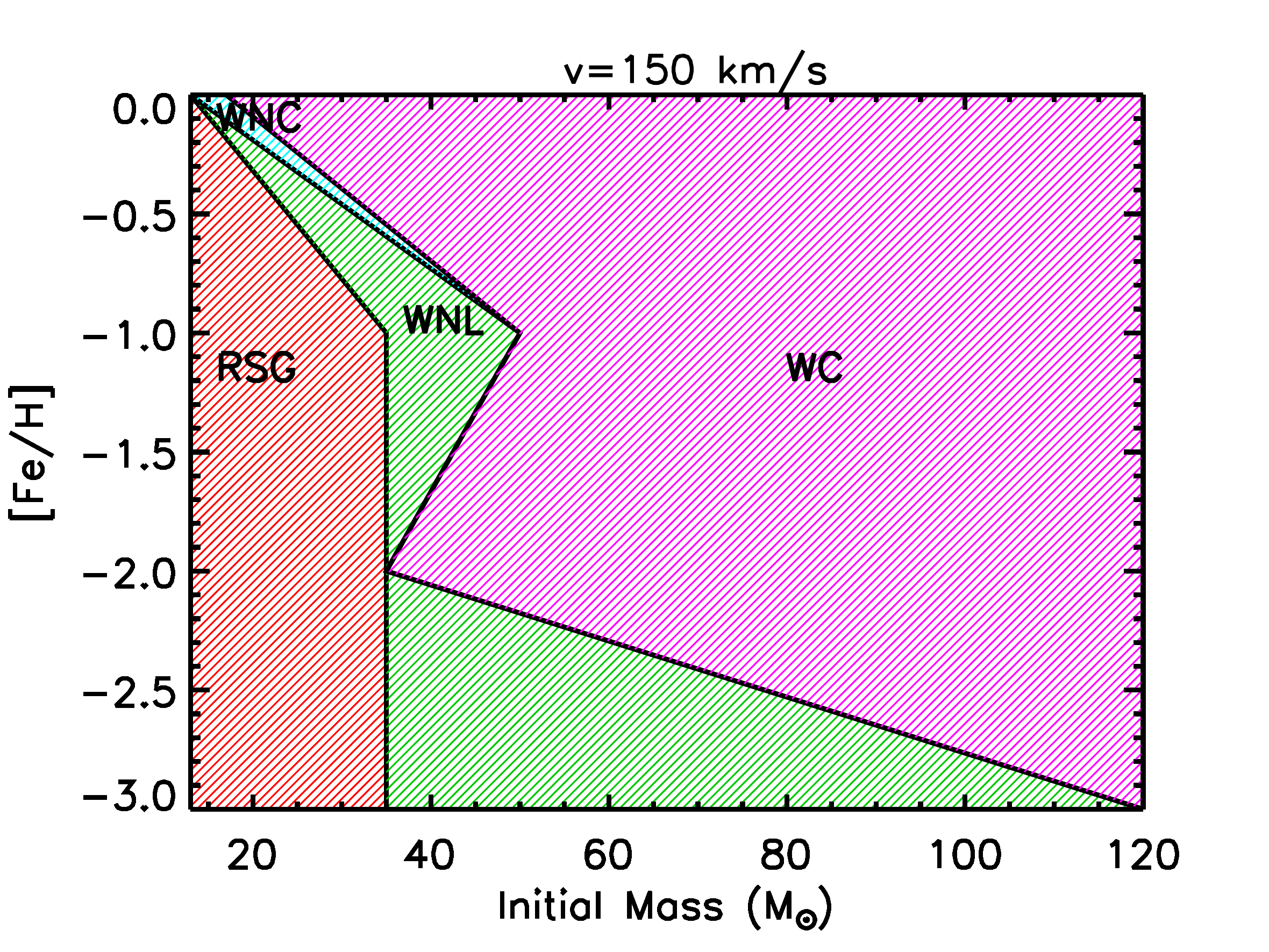}
\subfigures
\caption{Same as Figure \ref{confsnv000} but for models with initial rotation velocity of 150 km/s at various metallicities}
\label{confsnv150}       
\end{figure}

\begin{figure}[h]
\samenumber
\centering
\includegraphics[scale=.25]{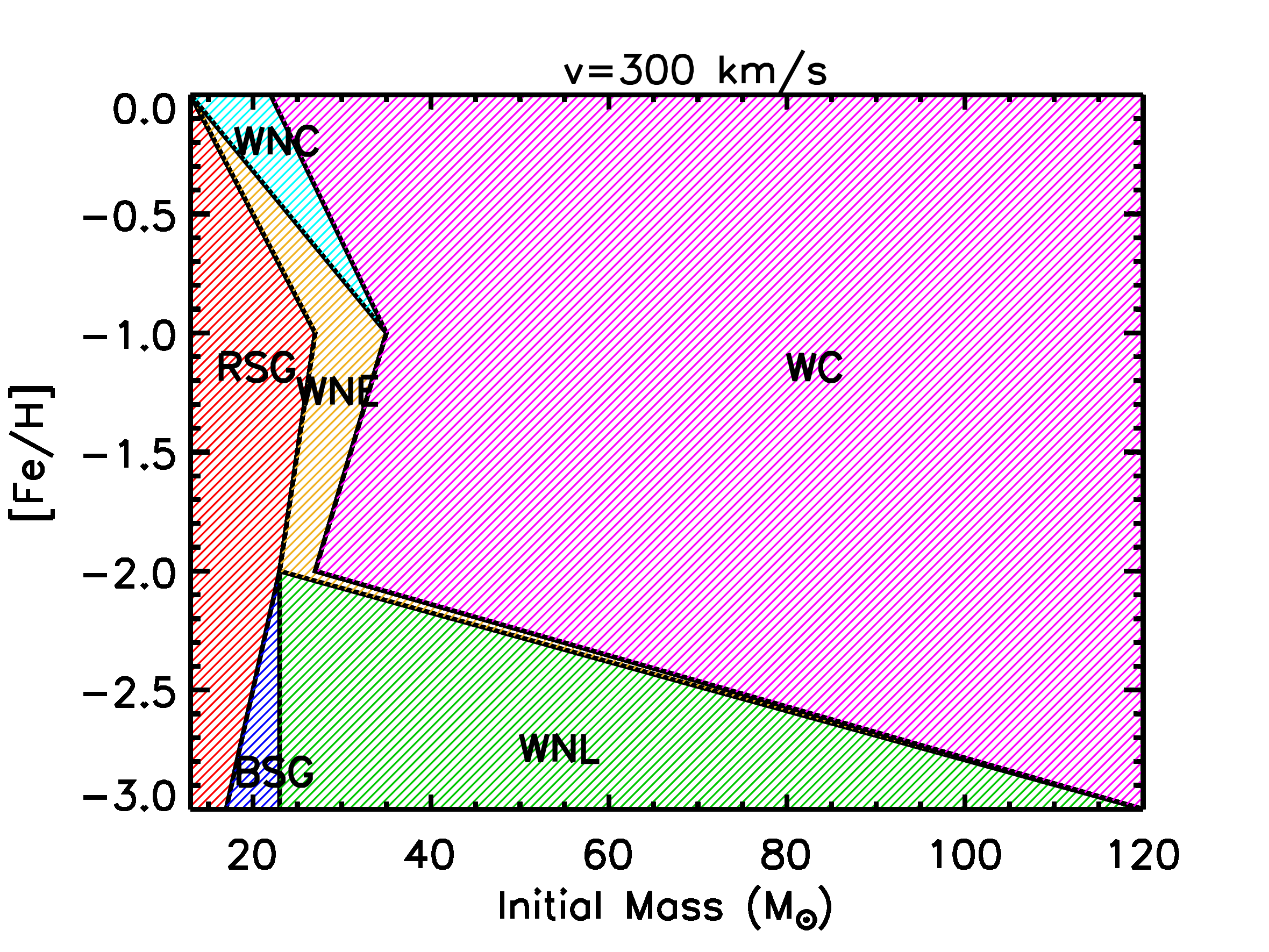}
\subfigures
\caption{Same as Figure \ref{confsnv000} but for models with initial rotation velocity of 300 km/s at various metallicities}
\label{confsnv300}       
\end{figure}
The inclusion of rotation, in general, reduces the minimum mass entering the WR stage (see above) and increases the maximum mass exploding as RSG at all the metallicities, therefore reducing the number of BSG, H-rich envelope, supernova progenitors (Figure \ref{confsnv150}, \ref{confsnv300}). Note also that rotation has a complex and non monotonic impact on the expected number of the various kinds of WR supernovae and that, in any case, allows some progenitors to explode as WNC stars \cite{2013ApJ...764...21C}.

The dramatic decrease of the lifetimes of the advanced evolutionary phases, due to the enormous neutrino energy loss, largely inhibits any angular momentum transport in the radiative zones. On the contrary,  in the convective regions the outward transport of the angular momentum is still very efficient to induce an almost flat profile of the angular velocity. It is important to note that since no convective regions crosses the mass coordinate corresponding to the CO core mass, the total angular momentum stored in the CO core will remain essentially constant up to the onset of the iron core collapse.

\section{Conclusions}

All the evolutionary properties of the present grid of models discussed so far, can be summarized in order to define a global general picture of the evolution and outcome of a generation of massive stars, in the mass range $\rm 13$ to $120~M_\odot$, as a function of the initial metallicity and initial rotation velocity. 

With the help of simple hydrodynamic simulations it is also possible to estimate which is the final remanant mass left by each supernova and therefore to estimate which is the initial mass interval of stars that give rise to neutron stars and black holes forming supernovae. These simple hydrodynamic simulations are needed because, to date, there is no well-established self-consistent 3D model that naturally obtains the explosion of core collapse supernovae with their typical observed properties \cite{2015PASA...32....9F}. These "simplified" explosions are artificially induced by injecting in the presupernova model some amount of energy in an also arbitrary mass location (typically near the edge of the iron core) and followed by means of a 1D hydro code. The amount of extra energy injected is in general calibrated in order to abtain a prefixed amount of kinetic energy of the ejecta at infinity (typically of the order of $\rm 10^{51}~erg~\equiv$ 1 foe, i.e. 1 fifty one erg). The extra energy can deposited in form of thermal energy ("thermal bomb", \index{thermal bomb} \cite{1996ApJ...460..408T,1991ApJ...370..630A}), kinetic energy ("kinetic bomb", \index{kinetic bomb} \cite{2003ApJ...592..404L}) or a piston \index{piston} \cite{1995ApJS..101..181W,1991ApJ...370..630A}. It must be emphasized, however, that since these kind of explosions are not obtained from {\em first principles}, the actual remnant mass cannot be determined with certainty of precision and therefore all the determinations of this crucial quantity available in literature and based on these kind of estimates must be taken with extreme caution.

The chemical composition of the envelope of the star at the presupernova stage determines, in a direct way, the properties of the light curves and spectra after the supernova explosion and therefore the classification of the supernova itself \cite{2015ApJ...814...63M}. \cite{2012MNRAS.422...70H} provided the values of the H and the He envelope masses ($\rm M_{H-env}$ and $\rm M_{He-env}$) corresponding to the various supernova types \cite{2012MSAIS..19...24P,2001ASSL..264..199C}. In this chapter it has been adopted the following scheme: (1) stars in which $\rm M_{H-env} \gtrsim 0.3~M_\odot$ explode as Type II Plateau SNe (SNIIP) \index{Type II Plateau Supernovae (SNIIP)}; stars in which $\rm 0.1 \lesssim M_{H-env} \lesssim 0.3~M_\odot$ explode as Type IIb SNe (SNIIb) \index{Type IIb supernovae (SNIIb)}; stars in which $\rm M_{H-env} \lesssim 0.1~M_\odot$ explode as Type Ib SNe (SNIb) \index{Type Ib supernovae (SNIb)} or Type Ic SNe (SNIc) \index{Type Ic supernovae (SNIc)} if $\rm M_{He-env} \gtrsim 0.1~M_\odot$ or $\rm M_{He-env} \lesssim 0.1~M_\odot$ respectively. On the basis of this prescription, it can be estimated, for the present set of models, the limiting masses marking the passage from one type of core collapse supernova to another one, as a function of the initial metallicity and initial rotation velocity.

\begin{figure}[h]
\centering
\includegraphics[scale=.28]{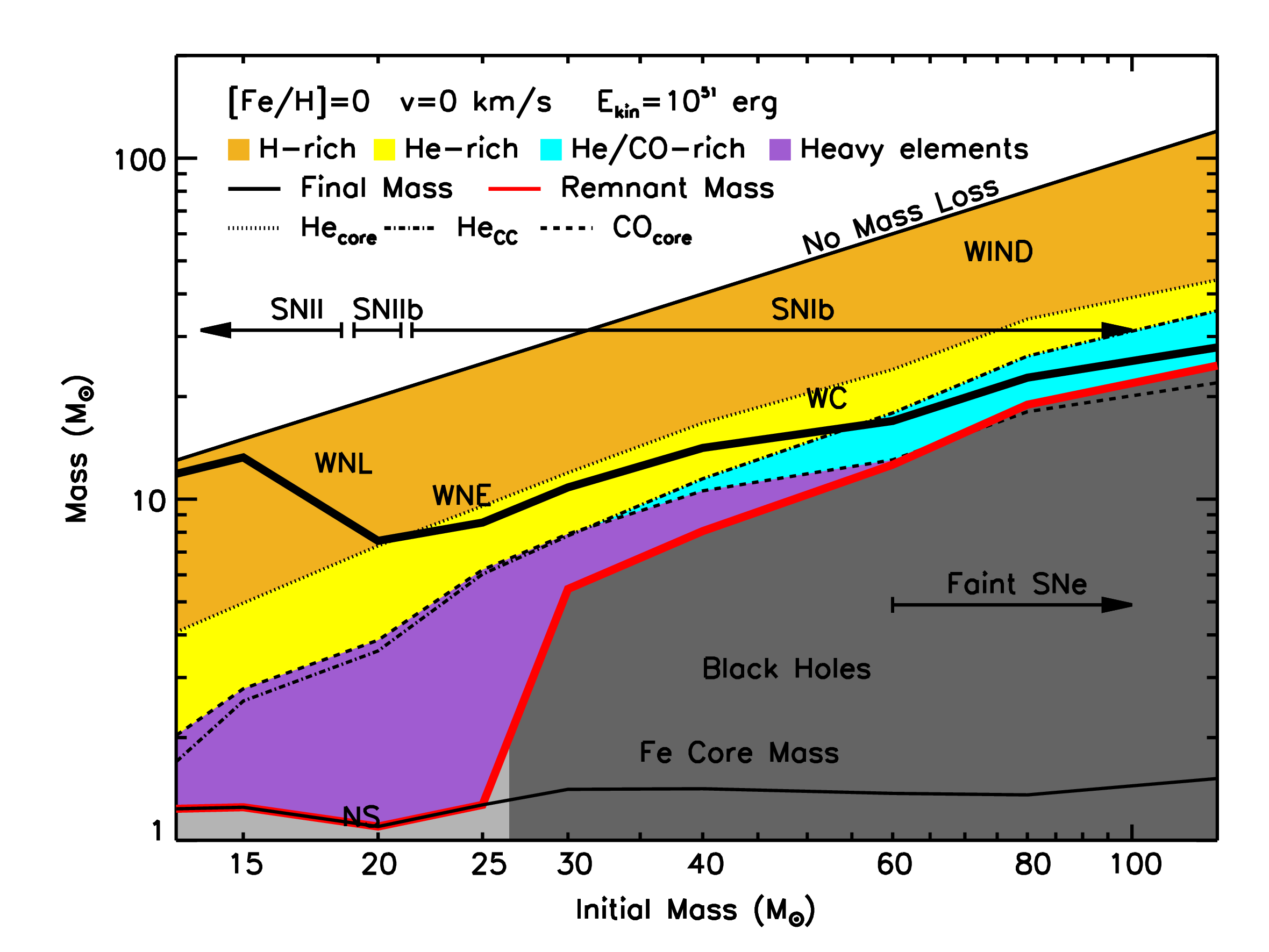}
\subfigures
\caption{Global properties of a generation of solar metallicity non rotating massive stars: the mass intervals of stars evolving through the various WR stages, the limiting masses for the various SN types, the initial mass-remnant mass relation (for 1 foe explosions), the limiting mass between neutron stars and black holes forming supernovae and, finally, the progenitor masses exploding as faint supernovae as well as those exploding as pair instability supernovae (PISN). Note that, as mentioned in the text, since we cannot determine with precision the final fate of stars with CO core masses larger than the limit for the onset of pulsation pair instabilities, for sake of simplicity we do not distinguish here between stars undergoing pulsation pair instabilities and stars entering the pair instabilities.}
\label{finalfatea000}       
\end{figure}
\begin{figure}[h]
\samenumber
\centering
\includegraphics[scale=.28]{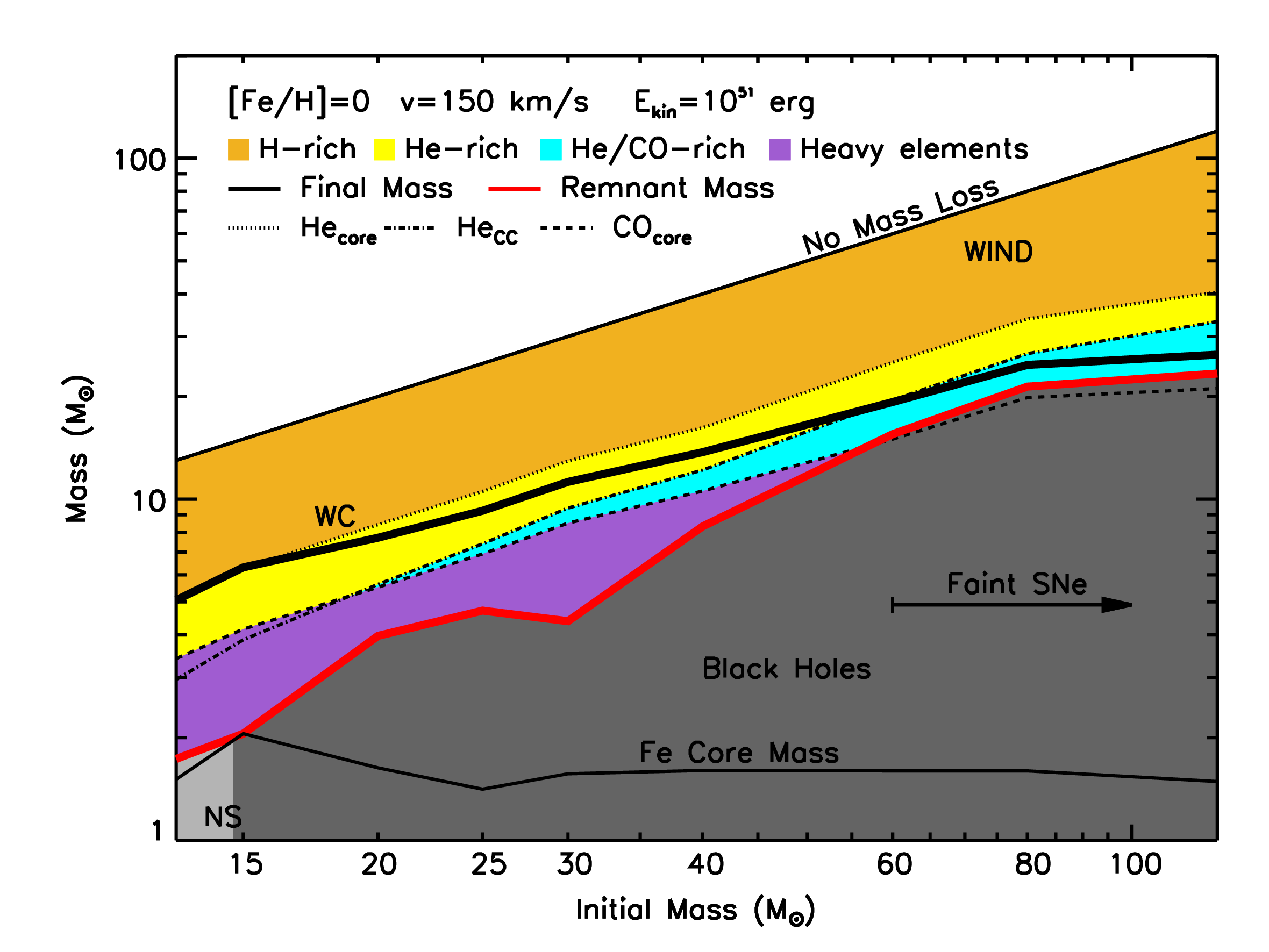}
\subfigures
\caption{Same as Figure \ref{finalfatea000} but for models with initial rotation velocity $\rm v=150~km/s$.}
\label{finalfatea150}       
\end{figure}
\begin{figure}[h]
\samenumber
\centering
\includegraphics[scale=.28]{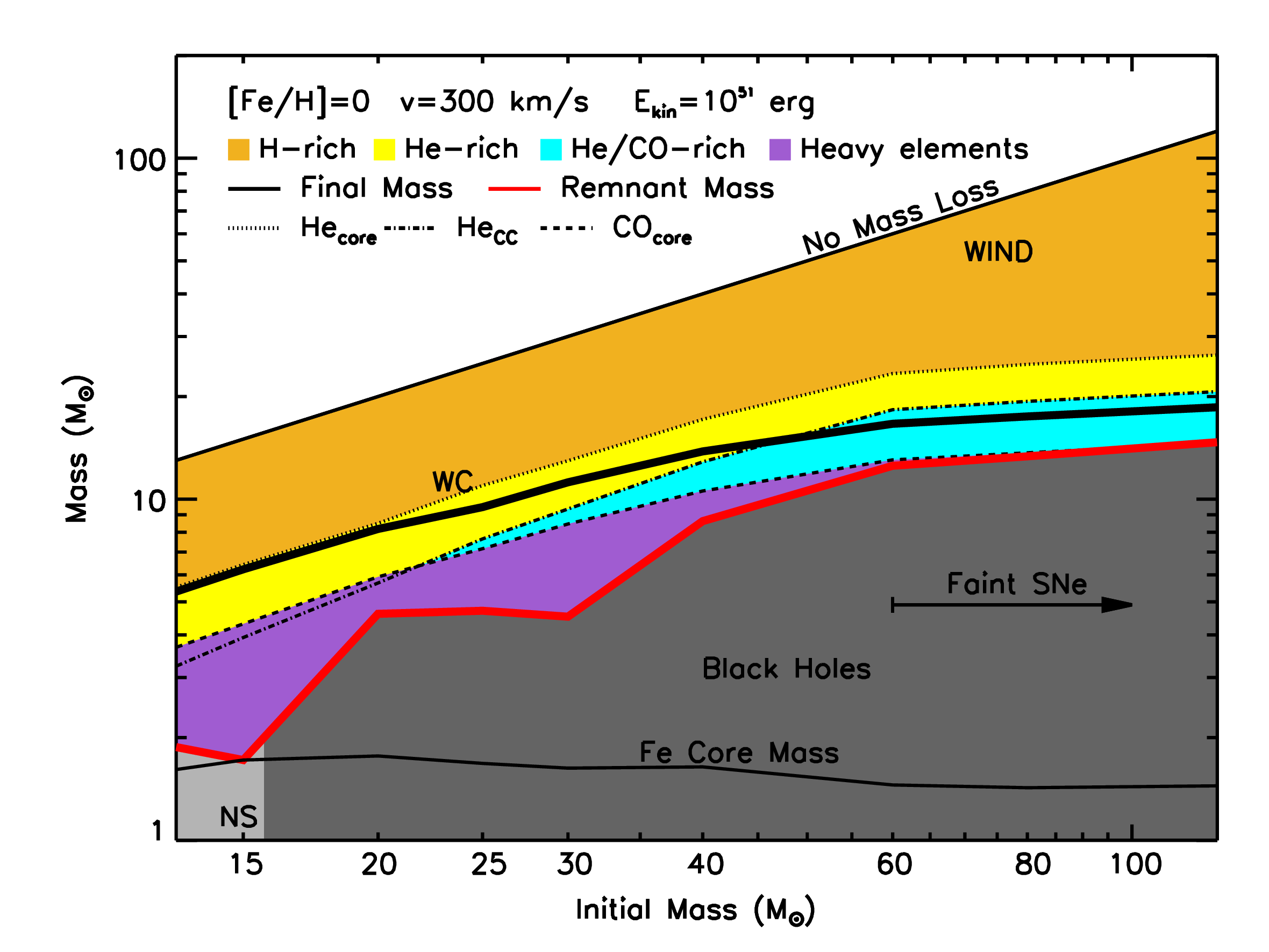}
\subfigures
\caption{Same as Figure \ref{finalfatea000} but for models with initial rotation velocity $\rm v=300~km/s$.}
\label{finalfatea300}       
\end{figure}
\begin{figure}[h]
\samenumber
\centering
\includegraphics[scale=.28]{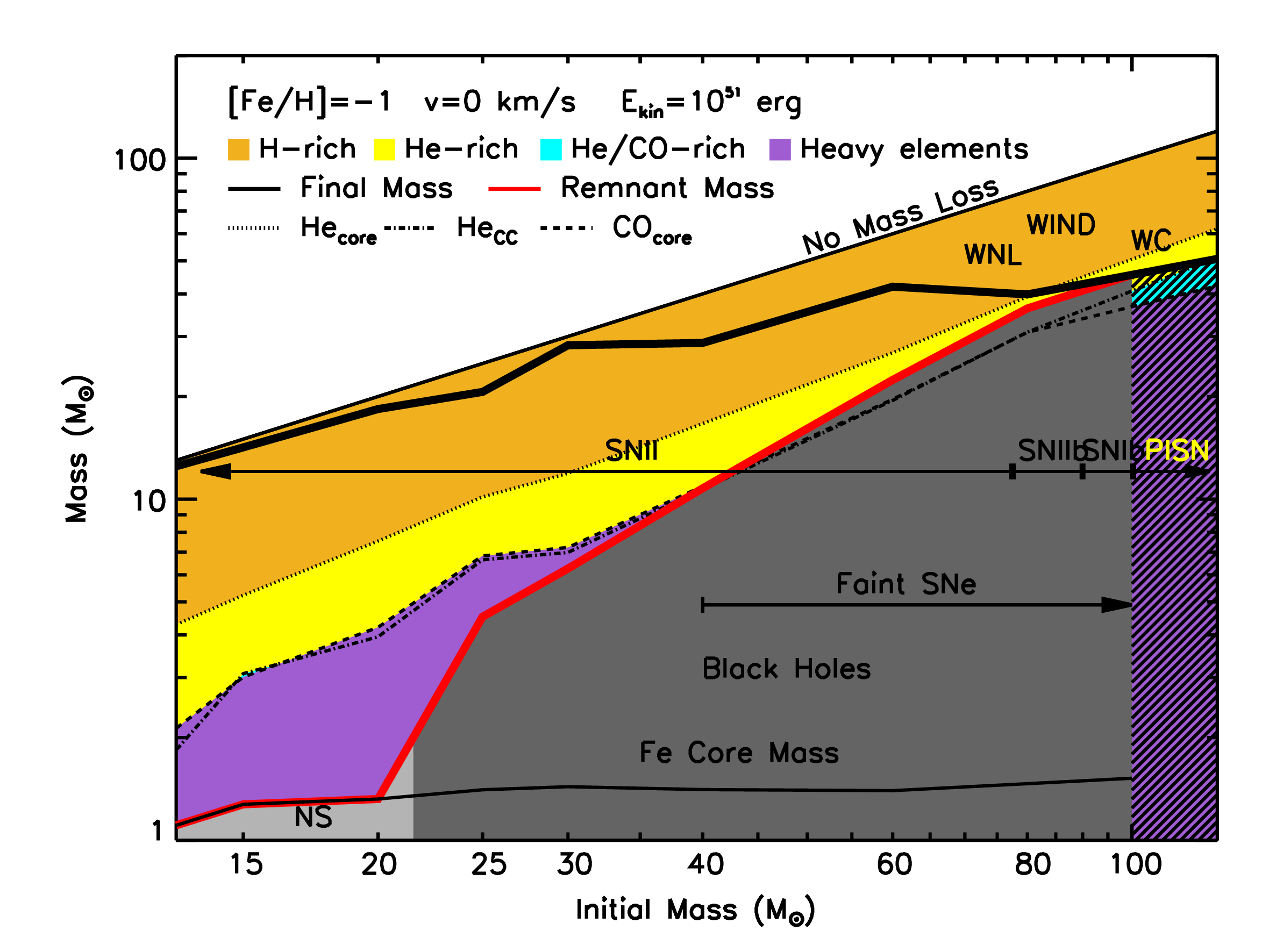}
\subfigures
\caption{Same as Figure \ref{finalfatea000} but for models with initial metallicity [Fe/H]=-1.}
\label{finalfateb000}       
\end{figure}
\begin{figure}[h]
\samenumber
\centering
\includegraphics[scale=.28]{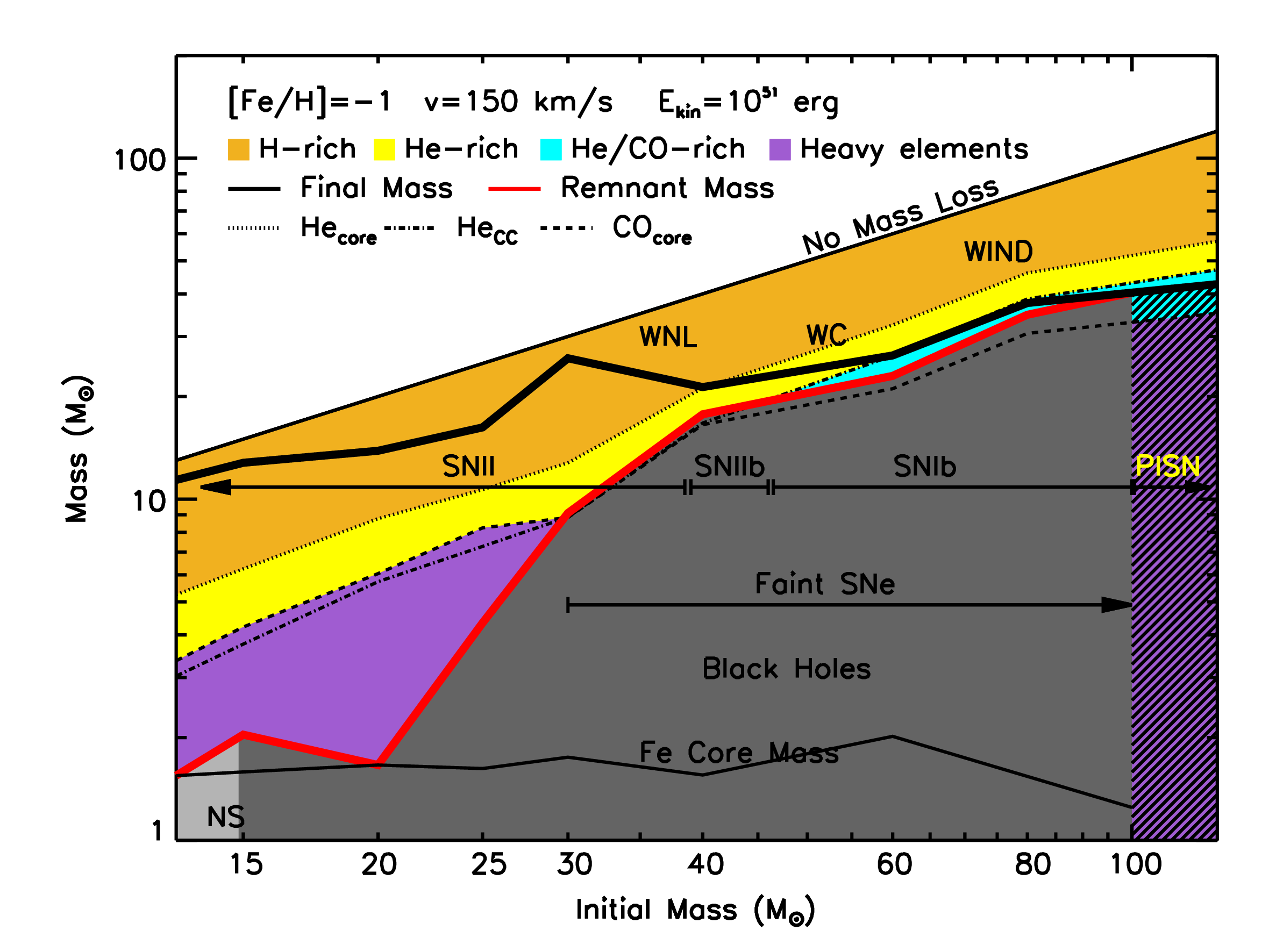}
\subfigures
\caption{Same as Figure \ref{finalfateb000} but for models with initial rotation velocity $\rm v=150~km/s$.}
\label{finalfateb150}       
\end{figure}
\begin{figure}[h]
\samenumber
\centering
\includegraphics[scale=.28]{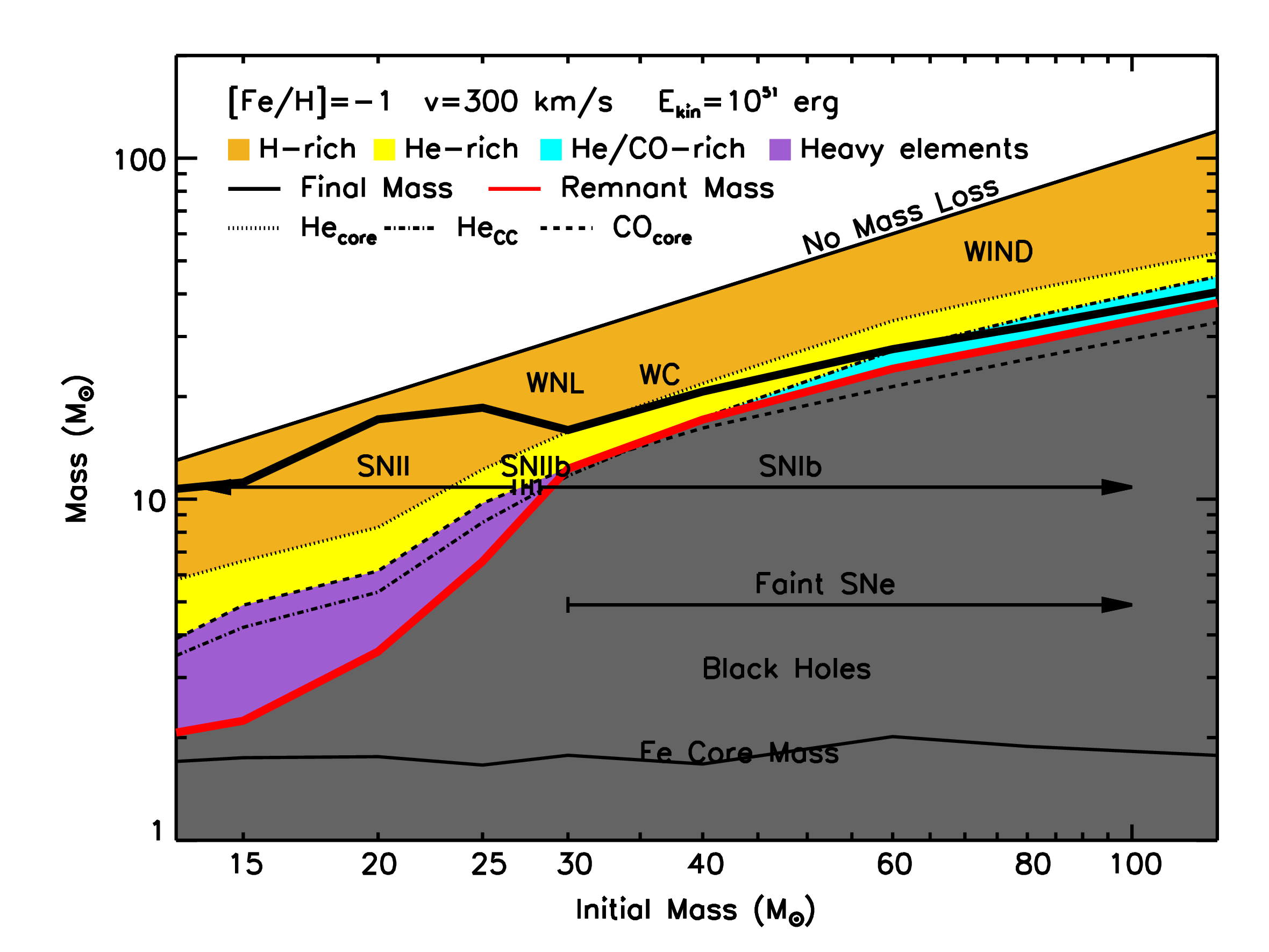}
\subfigures
\caption{Same as Figure \ref{finalfateb000} but for models with initial rotation velocity $\rm v=300~km/s$.}
\label{finalfateb300}       
\end{figure}

\begin{figure}[h]
\samenumber
\centering
\includegraphics[scale=.28]{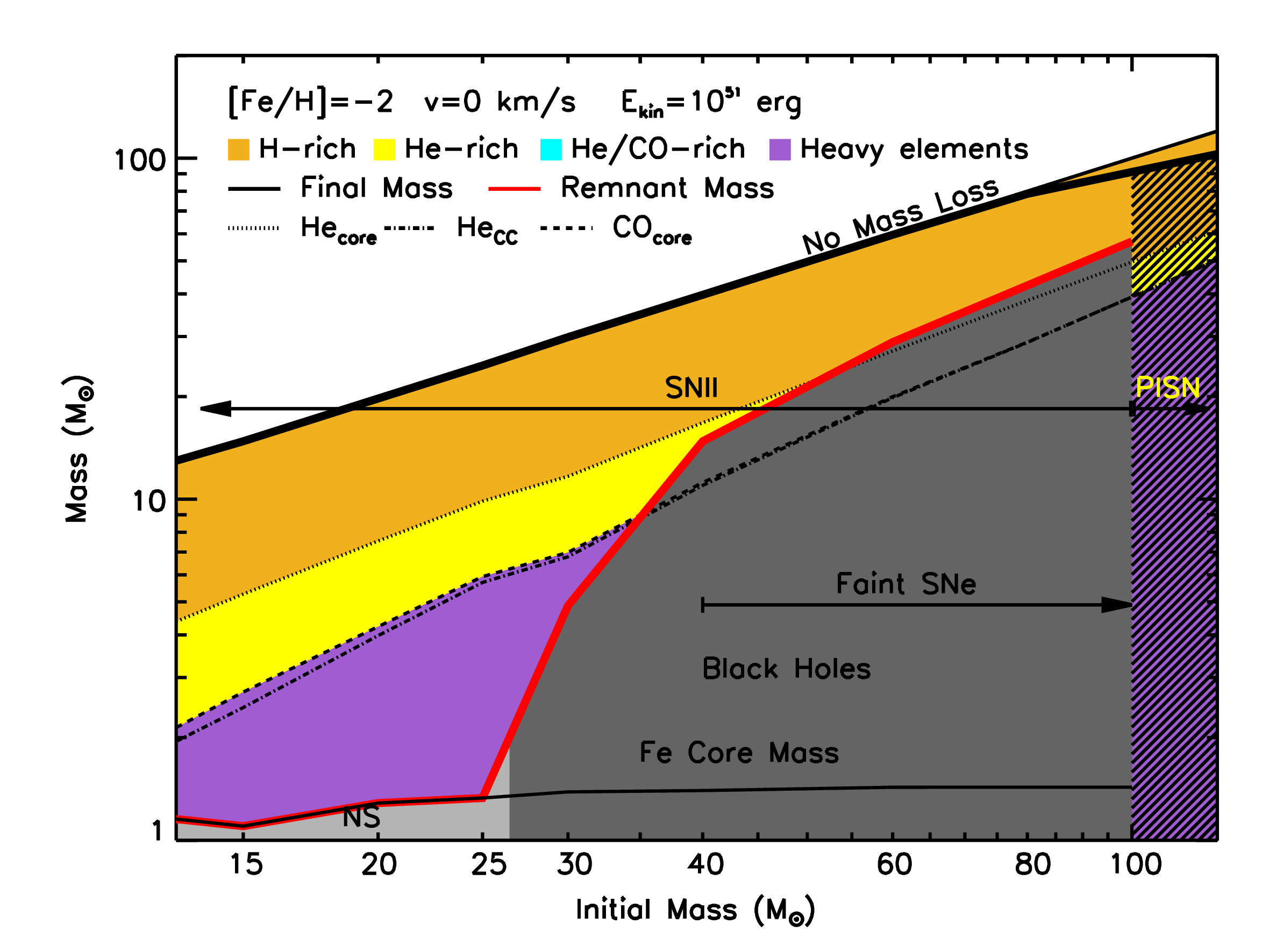}
\subfigures
\caption{Same as Figure \ref{finalfatea000} but for models with initial metallicity [Fe/H]=-2.}
\label{finalfatec000}       
\end{figure}
\begin{figure}[h]
\samenumber
\centering
\includegraphics[scale=.28]{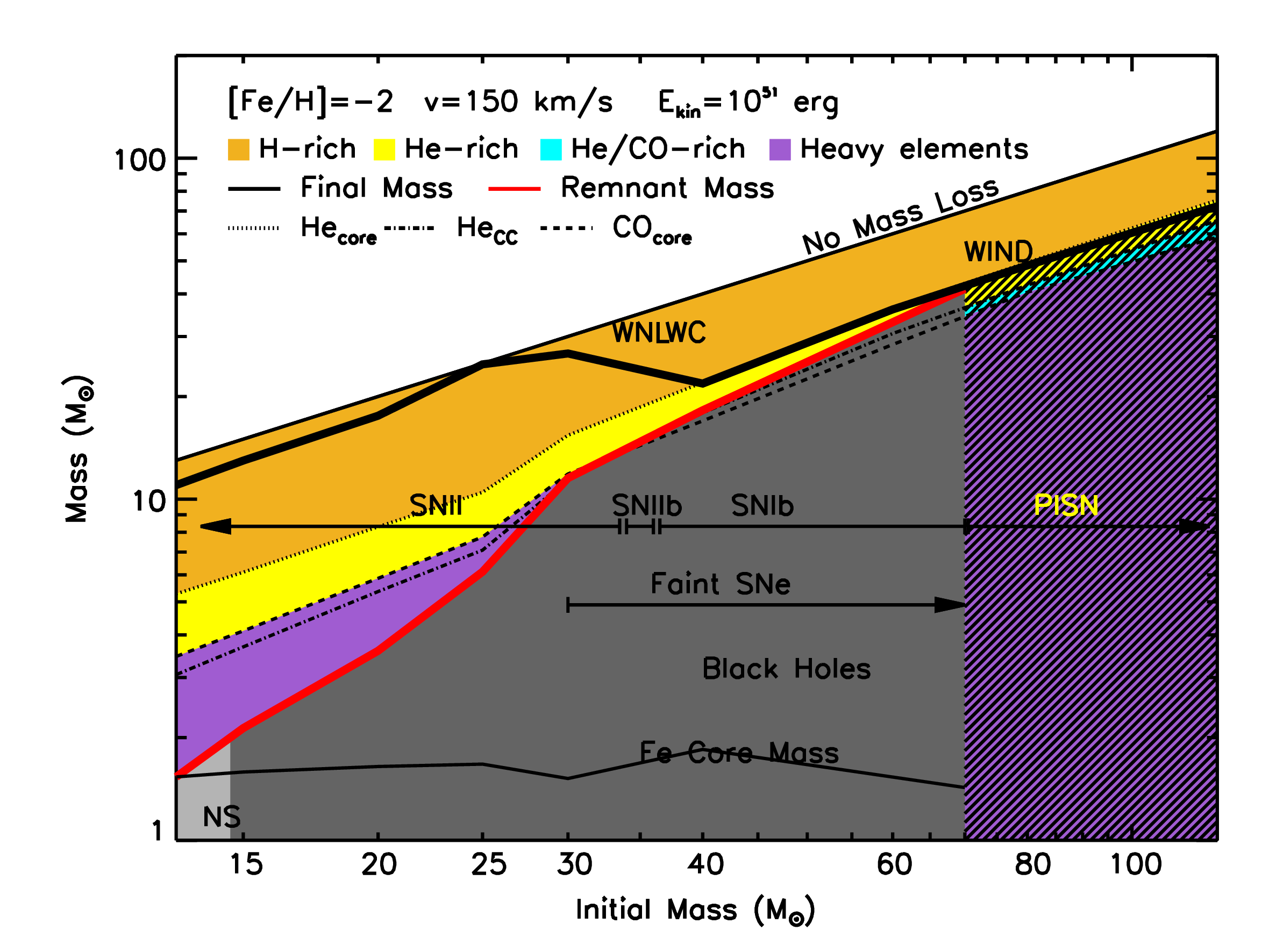}
\subfigures
\caption{Same as Figure \ref{finalfatec000} but for models with initial rotation velocity $\rm v=150~km/s$.}
\label{finalfatec150}       
\end{figure}
\begin{figure}[h]
\samenumber
\centering
\includegraphics[scale=.28]{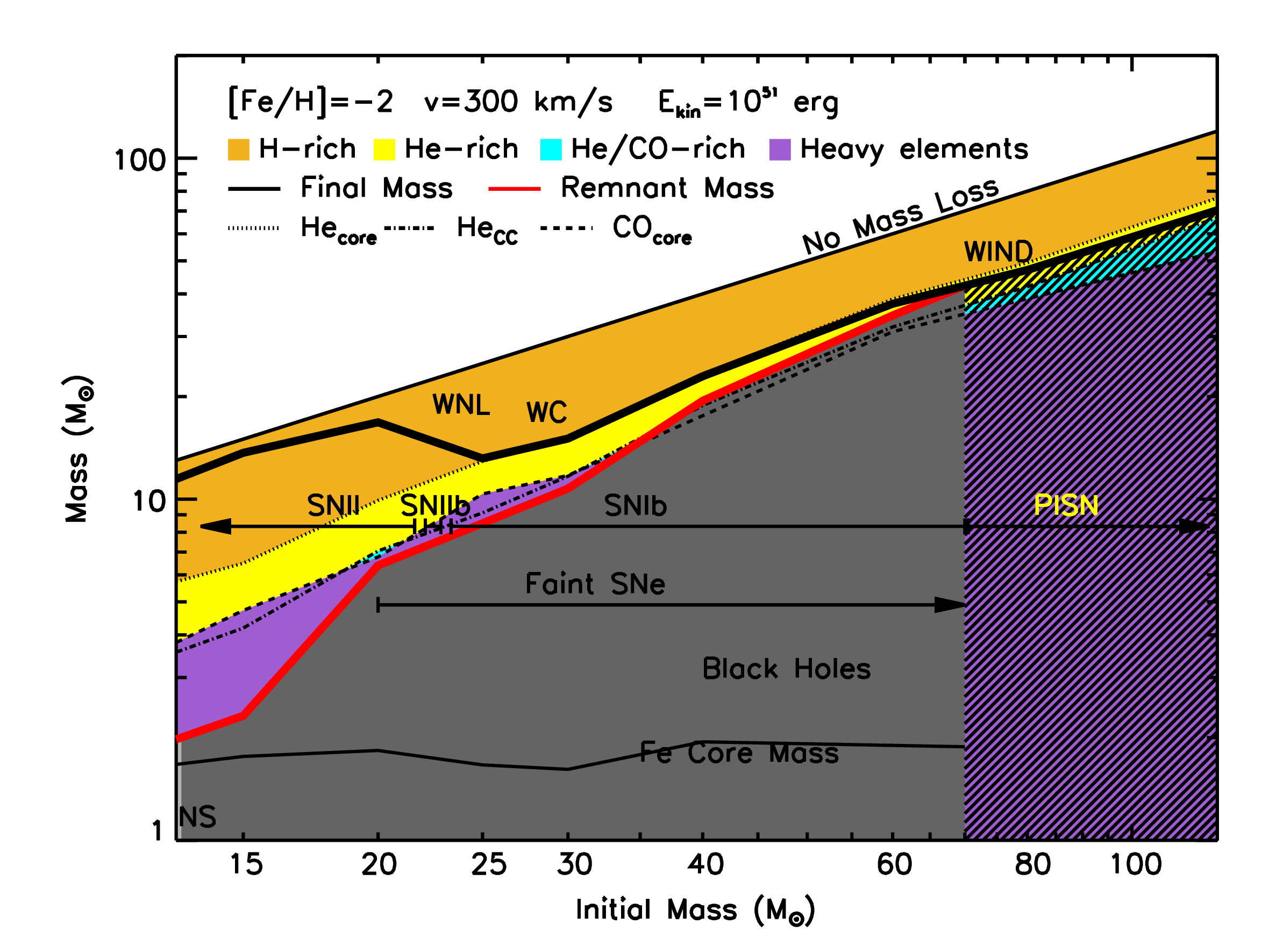}
\subfigures
\caption{Same as Figure \ref{finalfatec000} but for models with initial rotation velocity $\rm v=300~km/s$.}
\label{finalfatec300}       
\end{figure}

\begin{figure}[h]
\samenumber
\centering
\includegraphics[scale=.28]{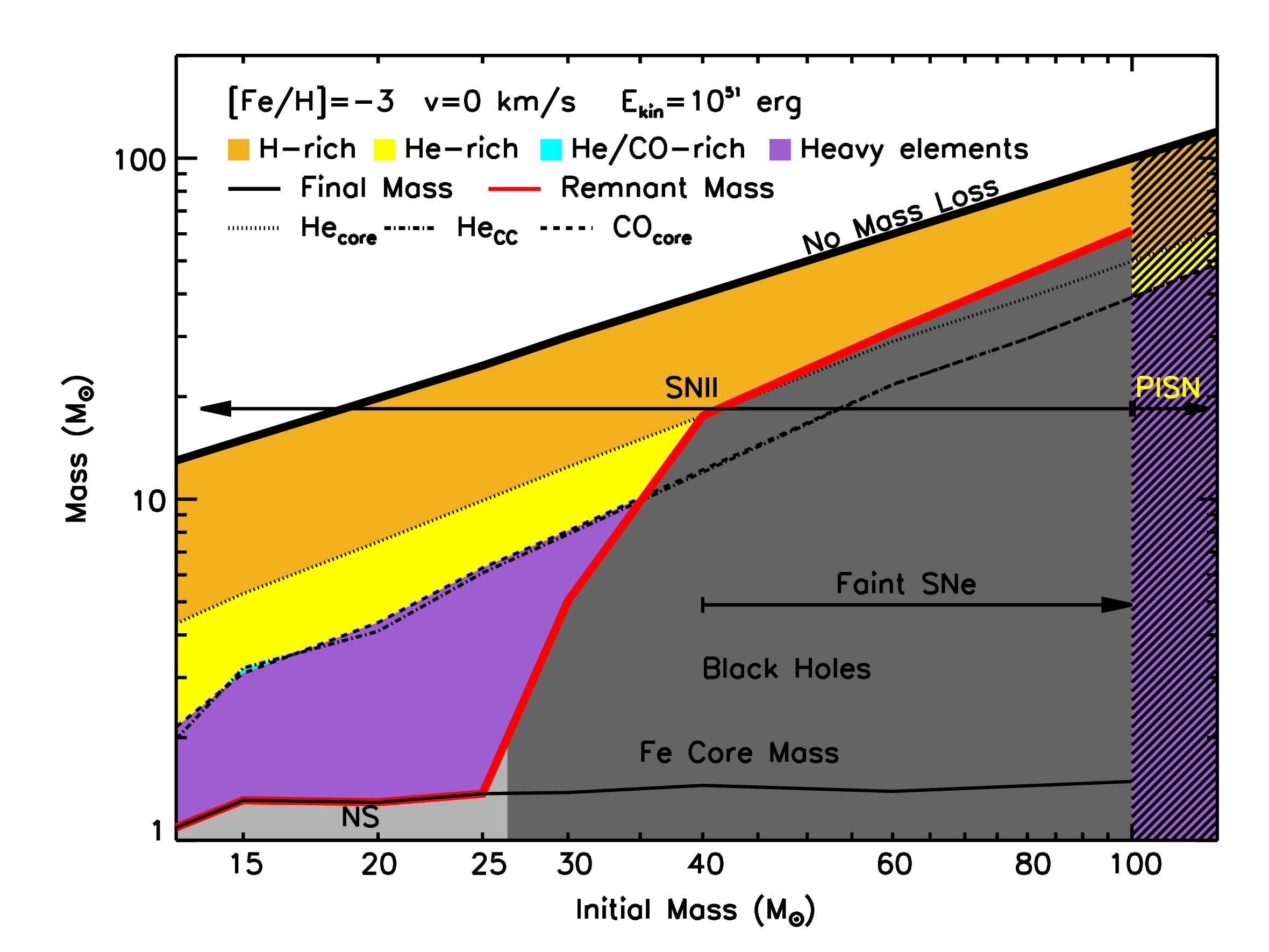}
\subfigures
\caption{Same as Figure \ref{finalfatea000} but for models with initial metallicity [Fe/H]=-3.}
\label{finalfated000}       
\end{figure}
\begin{figure}[h]
\samenumber
\centering
\includegraphics[scale=.28]{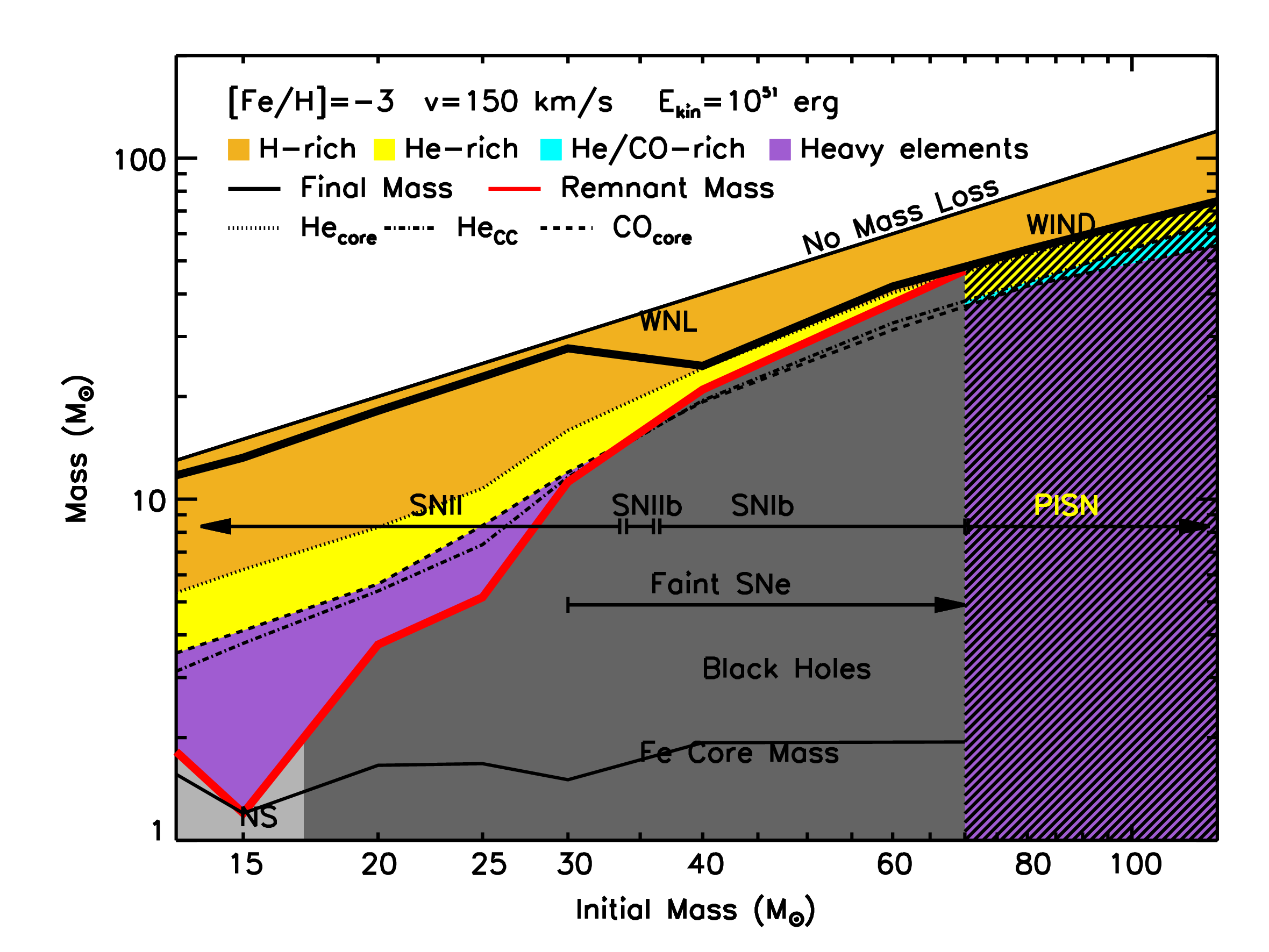}
\subfigures
\caption{Same as Figure \ref{finalfated000} but for models with initial rotation velocity $\rm v=150~km/s$.}
\label{finalfated150}       
\end{figure}
\begin{figure}[h]
\samenumber
\centering
\includegraphics[scale=.28]{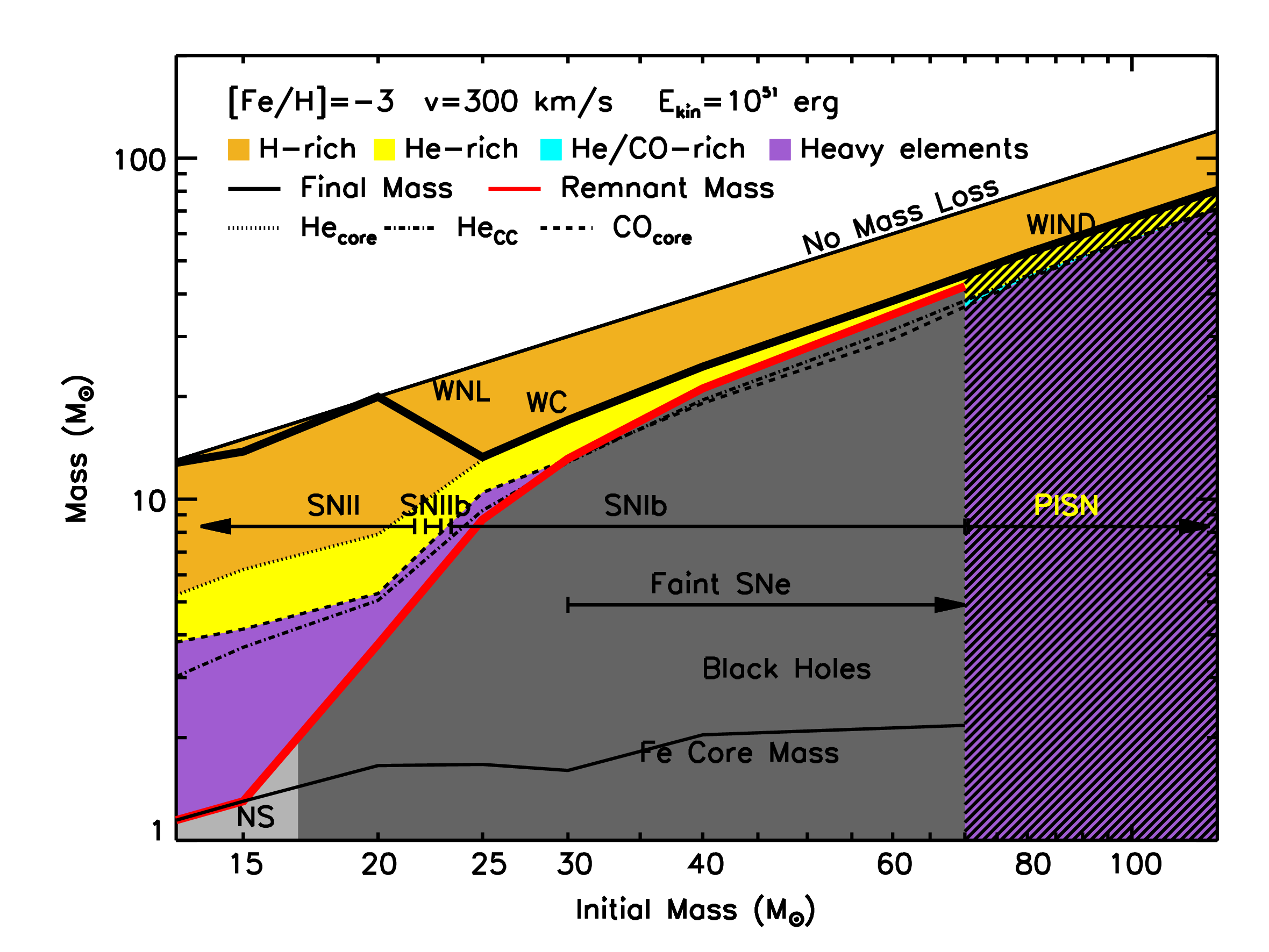}
\subfigures
\caption{Same as Figure \ref{finalfated000} but for models with initial rotation velocity $\rm v=300~km/s$.}
\label{finalfated300}       
\end{figure}

All the information discussed above, and in the present chapter, can be summarized in Figures \ref{finalfatea000} to \ref{finalfated300} that show, for any initial metallicity and initial rotation velocity, the global properties of a generation of massive stars, i.e., the mass ranges of stars that evolve through the various WR stages, the limiting masses for the various SN types, the initial mass-remnant mass relation (obtained assuming that the final kinetic energy of the ejecta is 1 foe), the limiting mass between neutron stars and black holes forming supernovae ($\rm M_{NS-BH}$) and, finally, the progenitor masses exploding as faint supernovae (defined tentatively as those supernovae producing a remnant mass corresponding to $\sim 90\%$ of the CO core mass) as well as those exploding as pair instability supernovae (PISN). By the way, since, as already mentioned above, we cannot determine with precision the final fate of stars with CO core masses larger than the limit for the onset of pulsation pair instabilities, for sake of simplicity we do not distinguish here between stars undergoing pulsation pair instabilities and stars entering the pair instabilities. For non rotating models $\rm M_{min}^{WR}$ increases with decreasing the initial metallicity - for metallicities $\rm [Fe/H]<-1$ none of the models become a WR star. As a consequence the limiting mass between SNIIP and SNIIb/SNIb ($\rm M_{IIP-IIb/Ib}$) follows the same trend, i.e., it increases with decreasing the metallicity - for metallicities $\rm [Fe/H]<-1$ all the stars explode as SNIIP. Since the CO core mass in the lower mass models is not strongly dependent on the initial metallicity $\rm M_{NS-BH}$ is essentially independent on the initial metallicity. On the contrary, the higher mass stars develop a progressively higher CO core masses with decreasing the metallicity because of the strong reduction of the mass loss, as a consequence these models enter the pair instability regime and explode as PISN - the minimum mass exploding as PISN ($\rm M_{PISN}$) is $\sim 100~M_\odot$ for metallicities lower than [Fe/H]=-1. For the same reason the minimum mass for the faint supernovae ($\rm M_{faint}$) decreases with decreasing the metallicity. The effect of rotation, for any given initial metallicity, is that of favoring the WR evolution and therefore both $\rm M_{min}^{WR}$ and $\rm M_{IIP-IIb/Ib}$ progressively reduce with increasing the initial rotation velocity.
Another effect of rotation is that of increasing the CO core mass, as a consequence $\rm M_{NS-BH}$, $\rm M_{faint}$ and $\rm M_{PISN}$ progressively reduce with increasing the initial rotation velocity.  
Inspection of Figures \ref{finalfatea000} to \ref{finalfated300} provides the approximate values for all the above mentioned critical masses as a function of the initial metallicity and initial rotation velocity. As a final comment, let us note that no progenitor star in the present set of models can produce a Type Ic SN (SNIc) because of the rather high mass of He present in the envelope at the time of the explosion. However, \cite{2011MNRAS.414.2985D} questioned about the upper limiting value of $\rm M_{He-env} \sim 0.1~M_\odot$ to produce a SNIc explosion since they find models in which the He lines (in particular the HeI 10830 line) are not excited in spite of a rather large $\rm M_{He-env}$. 

\begin{figure}[h]
\centering
\includegraphics[scale=.25]{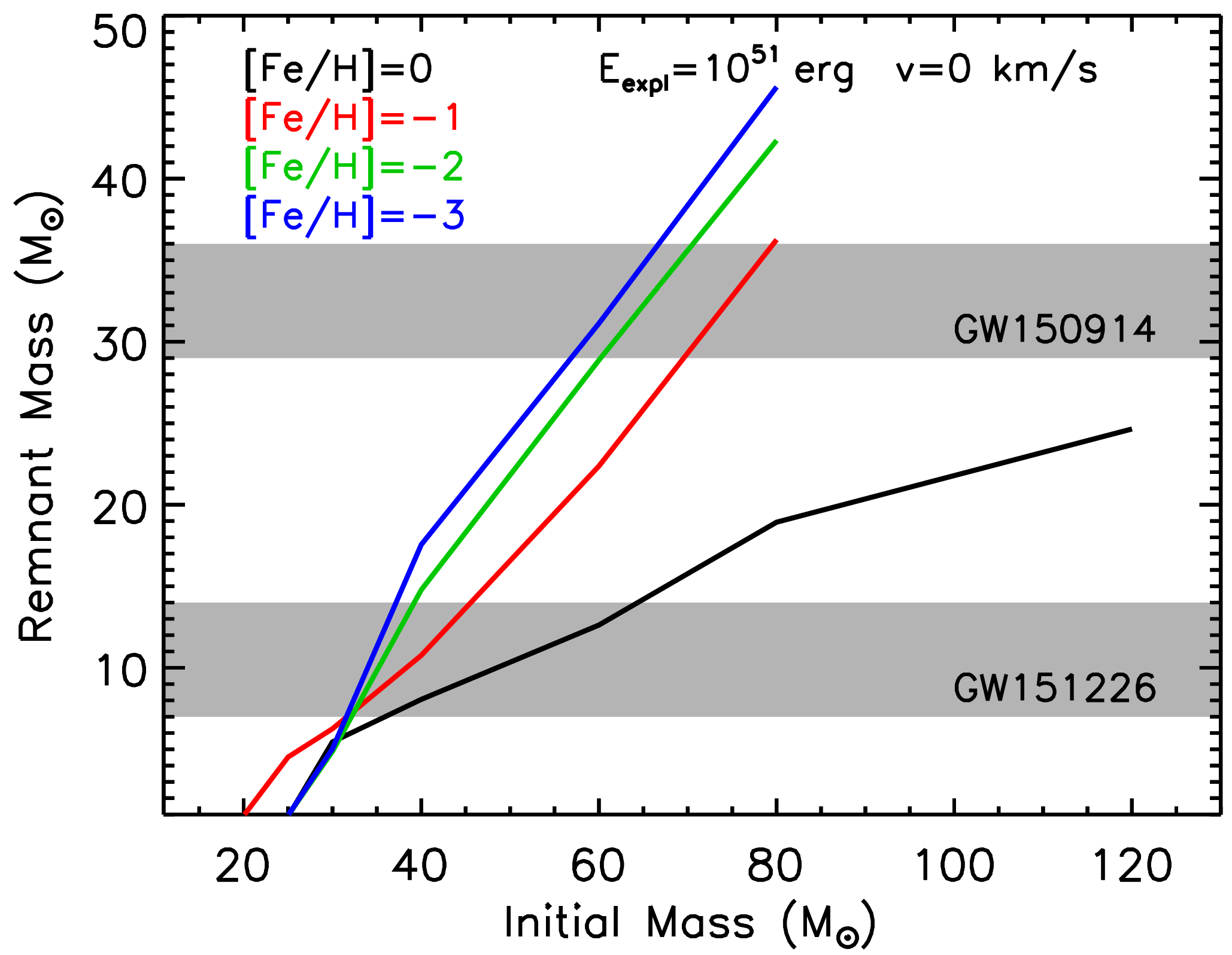}%
\caption{Mass of the compact remnant as a function of the initial mass for all the non rotating models with metallicities [Fe/H]=0 (black line), -1 (red line), -2 (green line), -3 (blue line). These results have been obtained with explosion energies of $\rm 10^{51}~erg$. The two horizontal grey strips correspond to the range of measured masses of the two black holes merging in GW150914 and GW151226.}
\label{mbh000}       
\end{figure}

\begin{figure}[h]
\centering
\includegraphics[scale=.25]{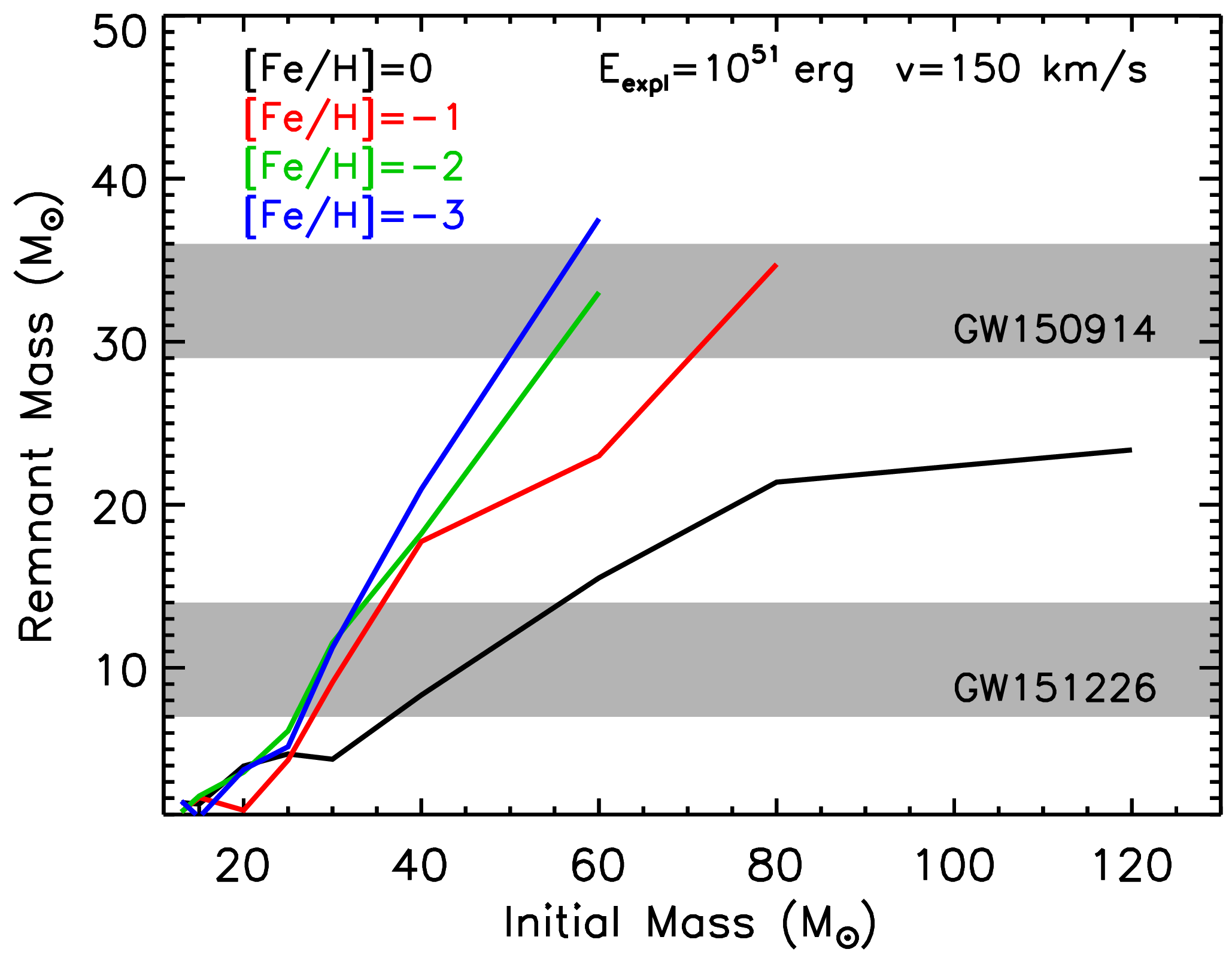}%
\caption{Same as Figure \ref{mbh000} but for models with initial rotation velocity $\rm v=150~km/s$.}
\label{mbh150}       
\end{figure}

\begin{figure}[h]
\centering
\includegraphics[scale=.25]{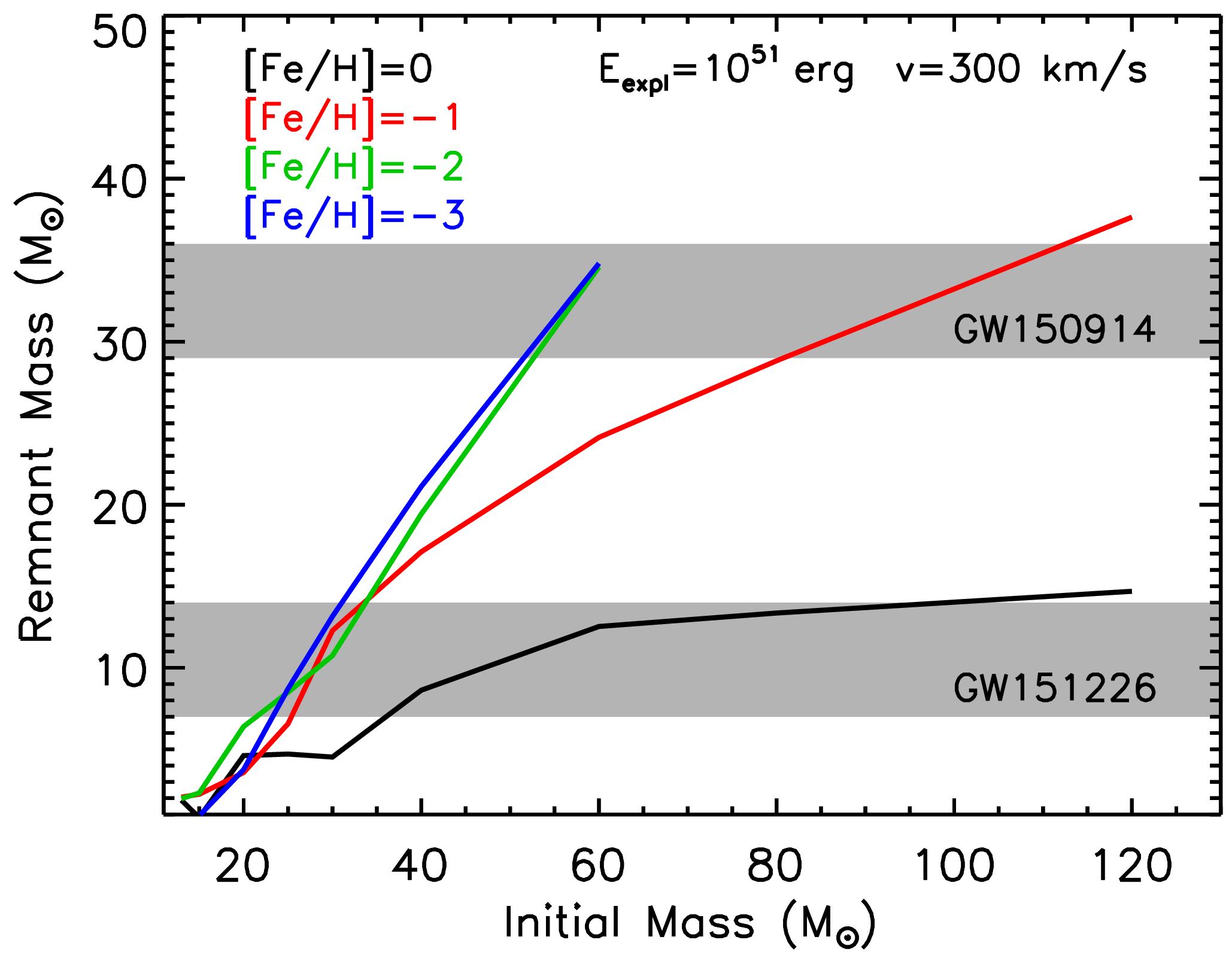}%
\caption{Same as Figure \ref{mbh000} but for models with initial rotation velocity $\rm v=300~km/s$.}
\label{mbh300}       
\end{figure}

Since the first (and only) two direct detections of gravitational waves are associated with the merger of double black hole binaries with masses $\rm 36-29~M_\odot$ (GW150914 \cite{2016PhRvL.116f1102A}) and $\rm 14-7~M_\odot$ (GW151226 \cite{2016PhRvL.116x1103A}), before closing this chapter it is worth showing the final remnant masses expected for all the models as a function of the initial metallicity for the three initial rotation velocities (Figures \ref{mbh000}, \ref{mbh150}, \ref{mbh300}). The figures show that, in the non rotating case, only low metallicity models (with $\rm [Fe/H]\leq -1$ and with initial masses in the range $\rm 50 \lesssim M/M_\odot \lesssim 80$) can produce black holes as massive as those associated to GW150914. This result is in agreement with other studies, like, e.g.  \cite{2015MNRAS.451.4086S}. On the contrary, models with initial masses lower than $\rm \sim 60~M_\odot$, at all metallicities, can easily produce black holes with the typical masses associated to GW151226. The increase of the initial rotation velocity does not change qualitatively this picture, although it lowers the mass interval compatible with the two gravitational waves events and limits, in general, the formation of massive black holes because it favors the entrance in the pair instability. 

All the models presented and discussed in this chapter can be downloaded from the web site http://orfeo.iaps.inaf.it, or obtained upon request to the author.

\begin{acknowledgement}
I am indebted to my friend and collaborator Alessandro Chieffi for enlightening discussions about several aspects of the presupernova evolution and final fate of massive stars and for many valuable suggestions. I also warmly thank my wife, Tatiana, for her continuous support and encouragement during the preparation and writing of this chapter. Finally, I am grateful to my colleague and friend Prof. Ken'ichi Nomoto for having invited me to write this chapter and for having improved the quality of the paper thanks to his revision.
\end{acknowledgement}

\noindent
{\bf{Cross-References}}
\begin{itemize}
\item Observational and Physical Classification of Supernovae
\item Hydrogen-Rich Core Collapse Supernovae
\item Hydrogen-Poor Core-Collapse Supernovae
\item Supernovae from Rotating Stars
\item Explosion Physics of Core-Collapse Supernovae
\item The Masses of Neutron Stars
\item The Core-Collapse Supernova - Black Hole Connection
\item Nucleosythesis in spherical explosion models of core collapse supernovae
\item Pre-Supernova Evolution and Nucleosynthesis in Massive Stars and their Stellar Wind Contribution
\item Nucleosynthesis in Hypernovae: Gamma Ray Bursts
\item Supernova remnants as clues to supernova progenitors
\end{itemize}

\newpage

\clearpage




\printindex

\end{document}